\documentclass[acmsmall,11pt,screen,nonacm,natbib=false]{acmart}
\setcopyright{none}
\usepackage{amsmath,mathtools,array,booktabs,longtable,multirow}
\usepackage{microtype}
\microtypesetup{protrusion=false}
\usepackage{placeins,xurl,bookmark}
\hypersetup{pdftitle={Structural Lower Bounds for Quantum Property Estimation with Persistent Memory}}
\renewcommand{\small}{\normalsize}
\renewcommand{\footnotesize}{\normalsize}
\makeatletter
\patchcmd{\@mkabstract}
 {\phantomsection\addcontentsline{toc}{section}{\abstractname}}
 {}{}{\errmessage{Could not suppress the title-page contents entry}}
\renewcommand{\l@section}{\@tocline{1}{1pt}{1pc}{2pc}{\normalsize\sffamily\bfseries}}
\renewcommand{\l@subsection}{\@tocline{2}{0pt}{1pc}{3pc}{\normalsize}}
\renewcommand{\l@paragraph}[2]{}
\makeatother
\newtheorem{theorem}{Theorem}[section]
\newtheorem{lemma}[theorem]{Lemma}
\newtheorem{proposition}[theorem]{Proposition}
\newtheorem{corollary}[theorem]{Corollary}
\theoremstyle{definition}
\newtheorem{definition}[theorem]{Definition}
\newtheorem{remark}[theorem]{Remark}
\DeclareMathOperator{\Tr}{Tr}
\DeclareMathOperator{\rank}{rank}
\DeclareMathOperator{\ran}{ran}
\DeclareMathOperator{\supp}{supp}

\newcommand{\norm}[1]{\lVert#1\rVert}

\title[Structural Lower Bounds with Persistent Quantum Memory]{Structural Lower Bounds for Quantum Property Estimation with Persistent Memory}
\author{Ye Tao}
\email{y.tao@cqu.edu.au}
\affiliation{%
  \institution{Central Queensland University}
  \country{Australia}}
\author{Hong Shen}
\email{h.shen@cqu.edu.au}
\affiliation{%
  \institution{Central Queensland University}
  \country{Australia}}
\renewcommand{\shortauthors}{Tao and Shen}

\date{}
\begin{abstract}
In quantum property estimation, an experimenter estimates $\Tr(O\rho)$ for a family of observables from independent copies of an unknown state $\rho$ simultaneously to additive error $\epsilon$. While quantum memory can dramatically reduce sample complexity, understanding the power of a modest register that remains coherent throughout an adaptive experiment has remained an open challenge because existing frameworks require optimizing over the infinite space of measurement trees.

We overcome this barrier by developing a backward-induction framework based on operator resolvents. Rather than tracking path-dependent memory states across rounds, resolvent regularization cancels internal memory trajectories exactly, reducing the lower bound to an algebraic certificate computed directly from the target observables with sharp $\Omega(1/\epsilon^2)$ scaling.

Our structural rule charges only the retained dimension $K$ under arbitrary adaptive instruments and unrestricted memory lifetime. Via symmetry reductions, it yields polynomial-size linear programs that establish the optimal linear $\Omega(D/(K\epsilon^2))$ trade-off for the $D^2-1$ Paulis, eliminating the cubic loss in prior bounds. For growing weight-$w$ Pauli strings ($w=\lceil\log_2 n\rceil$) and quadratic Majorana cross-correlations, it proves lower bounds of $\Omega(3^w/\epsilon^2)$ and $\Omega(n^2/(\epsilon^2(1+\ln n)^2))$ against an indefinitely persistent memory of dimension $K=\lfloor\sqrt n\rfloor$. These bounds match memoryless or commuting estimators within polylogarithmic factors, demonstrating that rank constraints on the persistent register fundamentally throttle observable overlap and prevent a modest coherent memory from overcoming the sample complexity barrier.
\end{abstract}
\begin{document}
\raggedbottom
\hypersetup{pageanchor=false}
\maketitle
\thispagestyle{empty}
\clearpage
\hypersetup{pageanchor=true}
\pagenumbering{roman}
\setcounter{tocdepth}{2}
\hypersetup{linktoc=all}
\pdfbookmark[1]{Contents}{contents}
\begingroup
\hypersetup{linkcolor=black}
\renewcommand{\addcontentsline}[3]{}
\tableofcontents
\endgroup
\clearpage
\pagenumbering{arabic}
\section{Introduction}
\label{sec:intro}
A central objective in quantum information processing is to characterize an unknown quantum state by predicting its physical properties~\cite{aaronson2018,hkp2020}. In many practical scenarios, an experimenter does not require a full, exponentially large description of the state; rather, the goal is to estimate a specific collection of expectation values, such as local Hamiltonians, spatial correlation functions, or order parameters. To extract this information, the experimenter collects independent and identically distributed (i.i.d.) copies of the unknown state $\rho$. The fundamental currency of the experiment is the copy complexity, defined as the number of copies of $\rho$ required to achieve a simultaneous target precision $\epsilon$.

When measurements are performed one copy at a time and recorded classically, extracting many non-commuting properties often demands a massive number of copies. However, recent breakthroughs in quantum learning theory have revealed a remarkable phenomenon: if the receiver is equipped with a quantum memory that can store states across successive rounds, joint measurements across multiple copies can reduce this sample complexity exponentially~\cite{cchl2021,huang2021bounds,huang2022experiments}.

While theoretically profound, storing multiple full copies of a large quantum system requires an extensive, fault-tolerant quantum memory that remains far beyond near-term experimental reach. In realistic laboratory architectures, one is far more likely to encounter an intermediate regime: a receiver equipped with a \emph{modest} quantum register---perhaps just a few auxiliary qubits---that interacts with each incoming copy sequentially. Crucially, even when this memory register is small, the receiver may keep it coherent and reuse it throughout the entire experiment, adapting every subsequent measurement based on past outcomes. This motivates us to investigate whether a small quantum memory, reused indefinitely across an adaptive experiment, can overcome the sample complexity barriers of memoryless measurements, or whether the inherent algebraic structure of the requested observables fundamentally limits any such advantage.

The algebraic structure of the target observables plays a decisive role even before quantum memory enters the picture. When all requested observables commute---such as all diagonal Pauli $Z$ strings on $n$ qubits---they share a single eigenbasis; measuring in that common basis yields an unbiased sign for every correlation simultaneously, demanding only $O(\epsilon^{-2} \ln M)$ copies for $M$ targets. In sharp contrast, strongly non-commuting observables possess conflicting eigenbases that forbid simultaneous extraction. Successful single-copy strategies such as classical shadows and randomized frame measurements~\cite{bonet2020,hkp2020} work precisely by navigating these geometric incompatibilities. Simply counting the number of requested expectation values therefore overlooks the core physics of the problem. A meaningful lower bound must take the target observable matrices themselves as input, directly reflecting how their operator geometry limits any adaptive protocol.

Establishing lower bounds for persistent quantum memory is notoriously challenging because the memory operates as an \emph{active learner}. As the experiment progresses, the state of the memory register accumulates correlations with the unknown source state, and the receiver exploits this stored information to choose subsequent measurement instruments adaptively. In a memoryless protocol, or one where an ancilla is refreshed between rounds, one can isolate a single step and analyze the information gained from a fresh copy against a fixed environment. But with persistent memory, the quantum state entering round $t$ is not fixed; it carries a history-dependent drift. Standard tools, such as the hybrid argument or simple trace-distance bounds, decouple successive rounds only at the price of discarding the fine-grained variance information needed to bound the source response. To capture the true limits of the experiment, an argument must track the information flow across rounds while fully accounting for the persistent, adaptive correlations carried by the memory.

Important groundwork has been laid by prior studies. In the memoryless setting, lower bounds have successfully linked property estimation to uncertainty principles in observable space~\cite{cchl2021,cgy2024}. More broadly, Chen, Gong, and Ye introduced an encompassing framework for Pauli estimation that models the entire adaptive experiment as an integrated measurement tree~\cite{cgy2024}. While conceptually universal, their formulation poses a formidable challenge: for a given set of observables and memory dimension $K$, it leaves behind an abstract optimization over the infinite space of adaptive measurement trees, which remains difficult to evaluate for specific structured families.

Our guiding perspective in this work is to establish an explicit, computable certificate directly from the observable matrices. At the technical core of our approach is a resolvent backward-induction technique that resolves a fundamental dilemma in quantum lower bounds: standard trace-distance hybrid bounds discard variance information and yield only a sub-optimal $\Omega(1/\epsilon)$ sample scaling, whereas quantum relative entropy natively yields $\epsilon^2$ but is subadditive on tensor products and thus completely blind to the memory rank bottleneck $K$. By introducing an asymmetric operator resolvent score, our framework forces the history-dependent memory states under the alternative to cancel out identically by forward/backward duality, while the operator inverse naturally manufactures the exact negative quadratic variance penalty required to complete the square at finite signal sizes. Although the resulting optimization problem involves a non-convex rank constraint, we show that it can be tackled systematically: we introduce a rank-sensitive semidefinite relaxation with trace caps, a convergent factor-moment hierarchy, and symmetry reductions that collapse the problem to compact linear programs.

In this work, we introduce an explicit, operational lower-bound rule derived directly from the requested observable matrices, recover the benchmark linear memory scaling, and apply it to two growing physical families to uncover the fundamental limits of persistent memory:
\begin{itemize}
\item \textbf{A finite rule and the linear memory benchmark.} We separate the instantaneous information extracted from each fresh copy from the drift carried forward across rounds. The resulting bound holds for arbitrary adaptive instruments, unbounded classical memory, and indefinite register coherence, charging only the retained Hilbert space dimension $K$. Specializing to all $D^2-1$ non-identity Paulis in dimension $D$, the rule recovers the tight linear lower bound $\Omega(D/(K\epsilon^2))$ across all $1 \le K \le D$ with an explicit constant, certifying that it faithfully captures the physical dimension-to-memory trade-off.
\item \textbf{Growing-weight Pauli correlations.} For weight-$w$ Pauli observables on $n$ qubits at logarithmic weight $w=\lceil\log_2 n\rceil$, our bound proves that an adaptive receiver with an indefinitely coherent memory of dimension $K=\lfloor\sqrt n\rfloor$ (about $(\log_2 n)/2$ qubits) still requires $\Omega(3^w/\epsilon^2)$ copies, matching memoryless local Pauli measurements within an $O(\ln^2 n)$ factor.
\item \textbf{Majorana cross correlations.} For bipartite products of quadratic Majorana operators, our composition certificate establishes an $\Omega(n^2/(\epsilon^2(1+\ln n)^2))$ lower bound against a shared persistent memory, matching simple commuting measurements within an $O(\ln^3 n)$ factor even on arbitrarily entangled states.
\end{itemize}
Together, these findings deliver a clear message: \emph{standard protocols storing only classical outcomes are already near-optimal up to modest logarithmic factors in these regimes.} Our framework explains the physical mechanism: the memory's rank cap strictly restricts the simultaneous observable overlap each copy can deliver, preventing an indefinitely persistent register from bypassing the memoryless sample barrier.

The rest of the paper is organized as follows. Section~\ref{sec:results} formally presents the main results and their operational scope. Section~\ref{sec:techniques} explains the conceptual ideas behind the proof, highlighting how the adaptive memory drift is untangled and how rank constraints are bounded. Related work and the formal protocol model follow in Sections~\ref{sec:related}--\ref{sec:model}. Sections~\ref{sec:structural}--\ref{sec:cross} provide the complete proofs, analytical profile bounds, and estimator guarantees. Appendix~\ref{app:moments} supplies the physical Majorana moment derivation, Appendix~\ref{app:log-gaps} explains the structural limits of logarithmic gaps, and Appendix~\ref{app:computational-certificates} details the computational semidefinite and linear programming certificates.

\section{Main results}
\label{sec:results}
To characterize the power of an indefinitely persistent quantum memory in property estimation, our goal is to translate the operational limits of an adaptive experiment directly into a geometric property of the target observable matrices, bypassing the need to optimize over the infinite space of measurement strategies. In this section, we present this structural lower bound and explore its consequences. We first establish a general testing theorem and its estimation corollary, and then apply them to two prominent growing families: weight-$w$ Pauli observables and quadratic Majorana cross correlations. In both families, simple measurement schemes storing only classical records turn out to be near-optimal, leaving only small logarithmic gaps to what can be achieved with persistent quantum memory. Finally, an exact evaluation on the complete Pauli family confirms that our general rule faithfully recovers the benchmark linear trade-off $\Omega(D/(K\epsilon^2))$.

To make the operational setting concrete, consider the standard property estimation game. A receiver is provided with independent copies of an arbitrary, unknown $D$-dimensional quantum state $\rho$. The goal is to estimate the expectation values $\Tr(\rho G_j)$ for every observable $G_j$ in a requested family, simultaneously to an additive precision $0 < \epsilon \le 1/8$ with success probability at least $2/3$. The receiver is granted substantial operational power: between consecutive copy arrivals, they may preserve a quantum register of Hilbert space dimension at most $K$, keep it coherent for as long as desired, apply arbitrary joint measurements between the incoming copy and the memory, choose each measurement instrument adaptively based on all past classical outcomes, and maintain an unlimited classical memory. We measure performance by the worst-case copy budget $N$ needed to guarantee this simultaneous accuracy. Here, $K$ denotes the Hilbert space dimension of the retained quantum register; for an $n$-qubit system, a memory bound of $K = \lfloor \sqrt{n} \rfloor$ corresponds to roughly $(\log_2 n)/2$ persistent qubits. (A full formal specification of the protocol model is provided in Section~\ref{sec:model}.) As standard notation, $\Tr$ denotes the matrix trace, $T \succeq 0$ indicates a positive semidefinite operator, and $\|\cdot\|_{\mathrm{op}}$ is the operator norm.

\subsection{A general lower bound with persistent memory}
\label{subsec:main-structural}
In establishing a lower bound on the copy complexity with persistent memory, our central task is to isolate how information enters the protocol at each step. In any adaptive protocol, the information gained in each round stems from two distinct sources: the fresh quantum state copy that just arrived, and the auxiliary quantum state carried over in the memory register from previous rounds. If the memory register were macroscopic, it could store multiple copies and perform collective measurements that unlock exponential quantum advantages. But if the memory register has a small dimension $K$, any joint measurement can only extract information through a low-rank quantum bottleneck.

This physical consideration suggests that the copy cost should be governed by an optimization that probes how strongly the target observables can couple through a state of rank at most $K$, while accounting for the residual correlations carried across rounds. This leads naturally to two quantities: a weighted Gram norm $\beta$, which measures the linear overlap and redundancy among the requested observables, and a rank-constrained response profile $\zeta(K)$, which quantifies the maximum response that a rank-$K$ state can extract from the observables. Together, they define an effective information budget per copy, $B_K = \zeta(K) + (2K-2)\beta$.

\begin{definition}[Source profile]
\label{def:profile}
Let $\tau=I_D/D$, and let $G_1,\ldots,G_M$ be Hermitian matrices satisfying
$\norm{G_j}_{\mathrm{op}}\le1$ and $\Tr(\tau G_j)=0$. Let $p_j>0$ with
$\sum_jp_j=1$, and assume that at least one $G_j$ is nonzero. Define
\begin{align}
 \beta&=\left\|\left[\sqrt{p_ip_j}\,
       \Re\Tr(\tau G_iG_j)\right]_{i,j=1}^M\right\|_{\mathrm{op}},
       \label{eq:beta-profile}\\
 \zeta(q)&=\sup_{\substack{T\succeq0,\ \Tr(T^2)=1\\
                    \rank T\le\min\{q,D\}}}
       \sum_{j=1}^M p_j\Tr(TG_jTG_j),\qquad q\ge1,
       \label{eq:zeta-profile}\\
 B_K&=\zeta(K)+(2K-2)\beta.
       \label{eq:structural-budget}
\end{align}
Directions of zero prior weight can be removed before applying this
definition. These quantities are optimizations over known finite matrices.
\end{definition}

To understand why these objects are well-behaved and physically meaningful, notice first that the matrix appearing in~\eqref{eq:beta-profile} is simply the real positive semidefinite Gram matrix of the observables under the Hilbert--Schmidt inner product with respect to the maximally mixed state $\tau$. For the profile $\zeta(q)$ in~\eqref{eq:zeta-profile}, the quadratic form $\Tr(TG_jTG_j) = \Tr[(T^{1/2}G_jT^{1/2})^2]$ is manifestly non-negative for any $T \succeq 0$. Because the optimization runs over the compact set of positive semidefinite matrices of rank at most $q$ with Frobenius norm $\Tr(T^2)=1$, the supremum is always finite and attained. Moreover, whenever at least one $G_j$ is non-zero, choosing $T$ as the projector onto an eigenvector of $G_j$ with non-zero eigenvalue immediately ensures that $\zeta(K) > 0$. Hence, both $\beta > 0$ and $B_K > 0$ are strictly positive under our assumptions.

To convert this information budget into a rigorous sample complexity lower bound, we approach the estimation problem through the lens of hypothesis testing. If an algorithm can accurately estimate all expectation values, it can certainly distinguish the maximally mixed state $\tau$ from a slightly perturbed state $\rho_{j,z} = (I_D + z\theta G_j)/D$ along an unknown observable direction $G_j$. Bounding the drift of the receiver's persistent state under this composite hypothesis yields our central operational bound:

\begin{theorem}[Finite testing bound]
\label{thm:structural}
Fix the data in Definition~\ref{def:profile}, an integer $K\ge1$, and
$0<\theta<1$. For $j\in\{1,\ldots,M\}$ and $z\in\{-1,1\}$, put
\begin{equation}
 \rho_{j,z}=\frac{I_D+z\theta G_j}{D}.
 \label{eq:structural-witnesses}
\end{equation}
Choose $(j,z)$ once with probability $p_j/2$ and use that same source for
all copies. Consider any receiver using ordinary independent copies, a
retained quantum system of dimension at most $K$, unlimited classical
records, and arbitrary adaptive complete instruments. Every quantum
register retained between accesses counts toward $K$, and the same memory
may remain coherent throughout the run. Let $N$ be a fixed or worst-case
upper bound on the number of copies.

Suppose a single decoder event $E$ satisfies, for some $0\le\delta<1/2$,
\begin{equation}
 \Pr_{\tau}(E)\le\delta,\qquad
 \sum_j\frac{p_j}{2}\sum_{z=\pm1}\Pr_{\rho_{j,z}}(E)\ge1-\delta.
 \label{eq:structural-event}
\end{equation}
Then
\begin{equation}
 N\ge
 \frac{4(1-\theta)(1-2\delta)^2}
 {(3-4\delta)\theta^2 B_K}.
 \label{eq:structural-testing}
\end{equation}
The conclusion includes general measurable record spaces when the
receiver's instruments and controls are measurable kernels on the product
sigma algebras of successive histories.
\end{theorem}

From this testing bound, a lower bound on simultaneous property estimation follows directly whenever the observables form an orthonormal set of reflections:

\begin{corollary}[Simultaneous signed estimates]
\label{cor:estimation}
Assume additionally that the requested matrices are distinct orthogonal
traceless reflections:
\begin{equation}
 G_j^2=I_D,\qquad \Tr(G_iG_j)/D=\mathbf1_{i=j}.
 \label{eq:orthogonal-reflections}
\end{equation}
If a receiver in Theorem~\ref{thm:structural} returns estimates satisfying
\[
 \Pr_\rho\!\left\{\max_j
       |\widehat\mu_j-\Tr(\rho G_j)|\le\epsilon\right\}\ge\frac23
 \quad\text{for every density matrix }\rho,
\]
where $0<\epsilon\le1/8$, then for every choice of positive prior $p$,
\begin{equation}
 N\ge\frac{1}{20\epsilon^2 B_K}.
 \label{eq:structural-estimation}
\end{equation}
\end{corollary}

The physical meaning of the profile $\zeta(K)$ is clarified by examining how it distinguishes between commuting and non-commuting observables:

\begin{remark}[The profile distinguishes a common measurement basis]
At $K=1$, a feasible positive matrix is $T=|v\rangle\langle v|$, so
$\zeta(1)=\sup_v\sum_jp_j\langle v,G_jv\rangle^2$. This is the standard
memoryless uncertainty quantity. For all nonidentity $Z$-type Pauli strings,
$\zeta(K)=1$ for every $K\ge1$: a computational-basis state attains one,
while Cauchy--Schwarz gives
$\Tr(TGTG)\le\|TG\|_{\rm HS}\|GT\|_{\rm HS}=\Tr(T^2)=1$
for each reflection. Measuring the common basis gives an unbiased sign for
every requested correlation, and $\lceil2\epsilon^{-2}\ln(6M)\rceil$
copies suffice by the same bounded-sign concentration used below. By
comparison, the complete-Pauli profile is of order $K/D$ when $K\le D$
(Theorem~\ref{thm:complete}). Thus the rule sees the observable structure
rather than only the number of requested values. The common-basis example
illustrates this distinction; its simultaneous logarithmic factor is not
resolved by the single-event lower bound.
\end{remark}

A closer look at the budget $B_K = \zeta(K) + (2K-2)\beta$ reveals the physical mechanics at work. The two terms decouple the learning process into two distinct channels:
\begin{enumerate}
\item The source profile $\zeta(K)$ governs the instantaneous rate of learning: it limits how much fresh information a newly arrived copy of $\rho$ can inject into the measurement process, subject to the rank constraint imposed by the memory dimension $K$.
\item The Gram correction $(2K-2)\beta$ accounts for the directional memory already carried by the register entering the round. Because a $K$-dimensional quantum memory has at most $2K-2$ non-trivial degrees of freedom in its real variance space, projecting out the directions aligned with the existing memory state incurs a geometric penalty of at most $(2K-2)\beta$.
\end{enumerate}
Whenever both the rank-constrained overlap $\zeta(K)$ and the Gram norm $\beta$ are small, the overall budget $B_K$ remains small. Theorem~\ref{thm:structural} then guarantees that no adaptive protocol---no matter how cleverly it reuses its persistent register over time---can avoid a large copy cost.

The mathematical derivation of this bound is detailed in Section~\ref{sec:structural}. While computing $\zeta(K)$ directly entails a non-convex rank optimization, Section~\ref{sec:computation} provides analytic bounds via spectral moments, and Appendix~\ref{app:computational-certificates} develops rank-sensitive semidefinite relaxations, a convergent moment hierarchy, and symmetry reductions that lead to compact linear programs.

\subsection{Pauli correlations of growing weight}
\label{subsec:main-pauli}
Having established the general structural theorem, our first natural test case is the family of low-weight Pauli observables. Estimating Pauli expectations is one of the most widely used subroutines in quantum characterization, variational algorithms, and error mitigation. When the weight $w=1$, the observables are single-qubit operators and learning them is easy. When $w=n$, we are asking for full state characterization, which requires an exponential number of copies.

In between lies the regime where the weight grows with system size, such as logarithmic weight $w = \lceil \log_2 n \rceil$. In this setting, the number of target observables $M = 3^w \binom{n}{w}$ is quasipolynomial, and the established method of randomized local Pauli measurements (classical shadows)~\cite{hkp2020} estimates all of them using $O(3^w \epsilon^{-2} \ln M)$ copies without any quantum memory. We therefore examine whether an experimenter equipped with a modest quantum memory of dimension $K = \lfloor \sqrt{n} \rfloor$ (roughly $(\log_2 n)/2$ qubits), persistent across all measurements, can exploit multi-copy correlations to surpass this memoryless performance.

To answer this, we apply our structural rule to the set of all weight-$w$ Pauli strings on $n$ qubits:
Write $D=2^n$, and let
\[
 \mathcal P_{n,w}
 =\{P_1\otimes\cdots\otimes P_n:
       P_i\in\{I,X,Y,Z\},\ |\{i:P_i\ne I\}|=w\},
 \qquad M=3^w\binom nw.
\]
The strings have their canonical Hermitian signs. Every member is a
traceless Hermitian reflection, and
$\Tr(PQ)/D=\mathbf 1_{P=Q}$. With the uniform prior, the real Gram
norm in the structural rule is consequently $\beta=1/M$.

The main technical challenge is to evaluate the profile $\zeta(K)$ over this large family subject to the rank constraint $\rank T \le K$. To overcome the non-convexity of the rank constraint, we introduce a moment certificate: by relating the rank constraint to higher-order moments $\Tr(T^p)$, we can bound the profile for any integer $p \ge 1$ with $2p \ge w$. This yields our finite copy bound:

\begin{theorem}[Finite Pauli copy bounds]
\label{thm:pauli}
Let $n\ge2$, $1\le w\le n$, and $K\ge1$ be integers, and let
$0<\epsilon\le1/8$. Let $N^*(n,w,K,\epsilon)$ be the least worst-case
copy budget for all signed means in $\mathcal P_{n,w}$, with the model
and simultaneous success guarantee specified above. For each integer
$p\ge1$ with $2p\ge w$, define
\begin{equation}
 B_{n,w,K}(p)=\min\left\{1,
 \frac{K^{w/(2p)}(n+\sqrt{8np})^w}{M w!}\right\}
 +\frac{2K-2}{M}.
 \label{eq:pauli-finite-budget}
\end{equation}
Then
\begin{equation}
 \frac{1}{20\epsilon^2B_{n,w,K}(p)}
 \ \le\ N^*(n,w,K,\epsilon)
 \ \le\left\lceil\frac{13}{6}\,3^w\epsilon^{-2}\ln(6M)\right\rceil.
 \label{eq:pauli-finite-copies}
\end{equation}
The upper bound uses retained quantum dimension one. The lower bound
holds for every legal $p$, allowing a finite search over chosen degrees
without optimizing over receiver protocols.
\end{theorem}

Theorem~\ref{thm:pauli} holds at every finite system size and allows one to optimize over the moment order $p$. However, if we want an explicit, clean constant in the asymptotic regime, we can do even better. Instead of interpolating moments after the fact, Proposition~\ref{prop:pauli-functional} develops a direct exponential tail certificate that keeps the rank cost $K$ inside the spectral tail before integration. This yields a sharp, uniform constant at moderate system sizes:

\begin{corollary}[Logarithmic weight and growing memory]
\label{cor:pauli-asymptotic}
Let $n\ge8192$, $w=\lceil\log_2 n\rceil$, and
$K=\lfloor\sqrt n\rfloor$. Under the same guarantees,
\begin{equation}
 \frac{3^w}{60\epsilon^2}
 \ \le\ N^*(n,w,K,\epsilon)
 \ \le\left\lceil\frac{13}{6}\,3^w\epsilon^{-2}\ln(6M)\right\rceil.
 \label{eq:pauli-copies}
\end{equation}
The ratio of the displayed upper and lower bounds is at most
$130\ln(6M)+1=O((\ln n)^2)$.
\end{corollary}

What this reveals physically is remarkable: in the logarithmic-weight regime, where $3^w = \Theta(n^{\log_2 3})$ and $M$ grows quasipolynomially, granting the receiver an indefinitely coherent quantum memory of dimension $K = \lfloor\sqrt{n}\rfloor$ yields \emph{no polynomial reduction} in copy complexity. The simplest memoryless approach---measuring independent random local Pauli bases and storing only classical records---is already optimal up to an $O(\ln^2 n)$ factor.

We emphasize that the condition $n \ge 8192$ is merely an explicit threshold ensuring the validity of the simplified constant $60$ in this corollary; the finite bound in Theorem~\ref{thm:pauli} applies universally for all $n \ge 2$. Furthermore, symmetry-reduced linear programs (detailed in Appendix~\ref{subsec:pauli-lp}) yield exact numerical certificates at small and moderate sizes: Table~\ref{tab:pauli-lp-finite} displays exact LP bounds for $n=8, 16, 24, 32$, and Table~\ref{tab:pauli-finite} provides further finite-size evaluations from the analytic certificate.

\subsection{Majorana cross correlations with shared memory}
\label{subsec:main-cross}
Our second application explores an arena that is both physically foundational and theoretically subtle: spatial cross-correlations in fermionic systems. Suppose we have two separate systems, $A$ and $B$, each containing $n$ fermionic modes. A fundamental task in quantum chemistry and condensed matter physics is to measure all bipartite quadratic correlations between $A$ and $B$.

Proving lower bounds in this setting presents two distinct physical and mathematical challenges:
\begin{enumerate}
\item The unknown joint state $\rho_{AB}$ may be strongly \emph{entangled} across $A$ and $B$.
\item The receiver is not limited to local operations on each subsystem; they are permitted a \emph{single, shared quantum memory} of dimension $K$ that can interact with both subsystems simultaneously and mediate correlations between them across all accesses.
\end{enumerate}
Analyzing each subsystem separately would fail because joint entanglement breaks local factorizations. We therefore seek a formulation that charges the receiver once for their shared memory, regardless of whether the source state is entangled.

We apply the structural theorem to all products of local quadratic Majorana observables. Each factor has $n\geq1$ fermionic modes and is represented on $n$ qubits. Implementation costs below use the standard Jordan--Wigner encoding in~\eqref{eq:local-jw-compiler}. Let $a_1,\ldots,a_{2n}$ and $b_1,\ldots,b_{2n}$ be the local Majoranas: they are Hermitian and obey $\{a_r,a_s\}=\{b_r,b_s\}=2\delta_{rs}I$. On the stated tensor split, define
\begin{equation}
 G^A_{ab}=i a_a a_b,\qquad G^B_{cd}=i b_c b_d,\qquad
 C_{ab,cd}=G^A_{ab}\otimes G^B_{cd},
 \quad a<b,\ c<d.
 \label{eq:cross-family}
\end{equation}
Put
\begin{equation}
 m=\binom{2n}{2}=n(2n-1),\qquad M=m^2,\qquad
 L=4n,\qquad D=2^{2n}.
 \label{eq:cross-parameters}
\end{equation}
Every $C_{ab,cd}$ is a centered Hermitian reflection, and $D^{-1}\Tr(C_{ab,cd}C_{a'b',c'd'})$ equals one when the two labels agree and zero otherwise.

For a quantum memory of dimension at most $K$, let $N^*(n,K,\epsilon)$ be the smallest worst-case copy cap for producing all $M$ signed means with simultaneous additive error at most $\epsilon$ and probability at least $2/3$, for every density matrix on the full $A B$ system. The receiver has ordinary independent copies of that joint state, arbitrary adaptive complete instruments, and unlimited classical records. The dimension $K$ counts the product dimension of all quantum registers retained between accesses, throughout the protocol. In particular, it permits one memory that stays coherent for the entire run. The source state may be entangled and may have support in both parity sectors.

To capture the joint action of the shared memory, we map the problem to the combined Clifford algebra on $L=4n$ generators. Specifically, we define global Majoranas by $c_r = a_r \otimes I$ and $c_{2n+s} = P_A \otimes b_s$, where $P_A$ is the local parity operator on system $A$ (see Section~\ref{subsec:majorana-maps} for algebraic details). We then construct a global joint-source map $\mathcal{A}(X) = \sum_{r=1}^L c_r X c_r$. By evaluating the moments of this global map, we obtain a composition certificate that simultaneously handles joint entanglement and the shared memory register:

\begin{theorem}[Quadratic cross correlations]
\label{thm:cross}
Let $0<\epsilon\leq1/8$ and let $K\geq1$ be an integer. Define the
joint-source map
\begin{equation}
 \mathcal A(X)=\sum_{r=1}^{L}c_rXc_r,
 \label{eq:global-majorana-map}
\end{equation}
where $c_1,\ldots,c_L$ are the global Majoranas specified above (and in Section~\ref{subsec:majorana-maps}). For
each integer $p\geq2$, any finite certificate
\begin{equation}
 H_p(L)\ \geq\
 \sup_{\rho\text{ pure}}\Tr\bigl(\rho\mathcal A^{2p}(\rho)\bigr)
 \label{eq:joint-moment-certificate}
\end{equation}
gives
\begin{equation}
 N^*(n,K,\epsilon)\geq
 \frac{M}{20\epsilon^2\bigl((K H_p(4n))^{2/p}/24+2K-2\bigr)}.
 \label{eq:cross-finite}
\end{equation}
A certificate valid for every even $L$ and every $p\ge1$ is
\begin{equation}
 \widehat H_p(L)=\frac{(2p)!}{p!}
       \left(\frac{(1+e)L}{2}\right)^p
 \le ((1+e)pL)^p.
 \label{eq:cross-physical-certificate}
\end{equation}
It follows from the squared-mean bound of Leone and
Bittel~\cite[Lemma~4]{leonebittel2026}, transferred to the positive map moment
in Proposition~\ref{prop:uniform-moment}. For $K=\lfloor\sqrt n\rfloor$
and every $n\ge2$, the necessary and achievable budgets satisfy
\begin{equation}
 \frac{n^2}{2500\epsilon^2(1+\ln n)^2}
 \ \le\ N^*(n,\lfloor\sqrt n\rfloor,\epsilon)
 \ \le\ (2n-1)^2
       \left\lceil2\epsilon^{-2}\ln(6M)\right\rceil
 \ \le\ 36n^2\epsilon^{-2}\ln(4n).
 \label{eq:cross-envelope}
\end{equation}
The ratio is at most $90000(1+\ln n)^2\ln(4n)=O((\ln n)^3)$.
The upper bound uses retained quantum dimension one.
\end{theorem}

The physical takeaway of Theorem~\ref{thm:cross} is equally decisive: even when the two subsystems share arbitrary entanglement, and even when the receiver possesses an indefinitely persistent quantum memory of dimension $K = \lfloor\sqrt{n}\rfloor$ that interacts with both halves, learning these cross correlations still demands a quadratic sample count $\Omega(n^2 / (\epsilon^2 \ln^2 n))$. In comparison, an elementary commuting matching estimator (Section~\ref{sec:cross}) using only memoryless measurements succeeds with $O(n^2 \epsilon^{-2} \ln n)$ copies. Persistent quantum memory therefore provides at most an $O(\ln^3 n)$ advantage over simple classical records.

The composition rule developed here provides a general template for bounding shared quantum memory across bipartite systems. The bound in~\eqref{eq:cross-envelope} relies on the physical moment certificate~\eqref{eq:cross-physical-certificate}, which holds unconditionally for all $n \ge 2$; Appendix~\ref{app:moments} provides the complete physical moment derivation, valid at every source size and moment order.

\subsection{The complete-Pauli benchmark}
\label{subsec:main-complete}
The complete Pauli family provides an exact benchmark for the structural
rule: it requests all $D^2-1$ nonidentity Pauli means in dimension $D=2^n$,
with the same simultaneous signed-error guarantee as above.

Without retained quantum memory ($K=1$), learning all Paulis requires
$\Omega(D/\epsilon^2)$ copies~\cite{cchl2021}. At the opposite extreme, with a
memory large enough to store an entire copy ($K=D$), two-copy measurements
reduce the exponential $D=2^n$ dependence to logarithmic in $D$ (specifically
$O((\ln D)/\epsilon^4)$ copies for magnitudes via Bell
measurements~\cite{cchl2021,huang2021bounds} and $\widetilde{O}((\ln D)/\epsilon^4)$
for signed expectations~\cite{cgy2024}, or $\widetilde{O}(1/\epsilon^2)$ in terms
of $D$ when multi-copy measurements are allowed~\cite{cgy2024}). Evaluating our
structural source profile directly on this complete family yields the sharp
linear dimension-to-memory trade-off $\Omega(D/(K\epsilon^2))$ across
$1\le K\le D$ (with the separate $\Omega((\ln D)/\epsilon^2)$
simultaneous-estimation lower bound given in Proposition~\ref{prop:sparse-fano}).
These lower bounds combine by taking their maximum.

The Pauli conjugation twirl depolarizes every source matrix. It therefore
allows the rank-constrained profile to be evaluated exactly, as proved in
Section~\ref{subsec:complete-proof}.

\begin{theorem}[Complete-Pauli specialization]
\label{thm:complete}
For all $D=2^n\ge2$, request every nonidentity canonical Pauli operator in $\mathcal{P}_n^* = \mathcal{P}_n \setminus \{I^{\otimes n}\}$, with
$M=D^2-1$ and the uniform prior. For every integer $K\ge1$, set
$q=\min(K,D)$. Then
\begin{equation}
 \beta=\frac1{D^2-1},\qquad
 \zeta(K)=\frac{Dq-1}{D^2-1},\qquad
 B_K=\frac{Dq+2K-3}{D^2-1}.
 \label{eq:complete-profile}
\end{equation}
Every receiver in the persistent-memory model that estimates all
these expectations with simultaneous signed error
$0<\epsilon\le1/8$ and success at least $2/3$ for every state satisfies
\[
 N\ge\frac{D^2-1}{20\epsilon^2(Dq+2K-3)}.
\]
For $1\le K\le D$, this is
$\Omega(D/(K\epsilon^2))$ with an absolute constant.
\end{theorem}

Evaluating the profile exactly confirms that our framework catches the true linear dimension-to-memory trade-off $\Omega(D/(K\epsilon^2))$ across the full range $1 \le K \le D$. Taken together with the logarithmic-weight Pauli and Majorana applications, this shows how the structural budget seamlessly adapts from complete operator bases to structured physical correlations, pinpointing the true operational limits of persistent quantum memory.

\section{Proof ideas}
\label{sec:techniques}
The proof of our main results proceeds through two essential conceptual phases. First, we must relate the sample complexity of an arbitrary adaptive protocol to a structural quantity derived directly from the target observables, all while allowing the receiver's memory to accumulate correlations with the unknown source. Second, we must develop sharp mathematical techniques to evaluate this quantity for physically relevant, growing families of observables. In this section, we explain the intuition, dilemmas, and design choices behind each step; detailed calculations follow in later sections.

\subsection{The limitations of single-copy decompositions}
To establish an information-theoretic lower bound, we set up a hypothesis testing problem between the maximally mixed state $\tau=I_D/D$ (the null hypothesis) and a family of perturbed alternatives $\rho_{j,z}=(I_D+z\theta G_j)/D$, where $G_j$ is one of the requested observables and $z \in \{-1, 1\}$. Any protocol capable of estimating all expectation values simultaneously to error $\epsilon$ can reliably distinguish $\tau$ from these alternatives using a single decision event $E$. Under the null state, every requested expectation vanishes; under an alternative $(j,z)$, the $j$-th expectation jumps to $\pm \theta$. Crucially, because the source state remains the same across all $N$ copies, the receiver's memory register progressively accumulates information about the underlying label $(j,z)$ as the experiment unfolds.

The natural temptation is to analyze each measurement round in isolation: one would like to compute the maximum information a single copy can impart to the memory and simply sum these contributions over $N$ rounds. However, this naive approach breaks down in the presence of persistent memory. At any given step, the quantum state of the memory entering under the null hypothesis is fundamentally different from the state entering under an alternative. A fresh-copy bound evaluated against a single, fixed auxiliary state cannot capture this history-dependent divergence. Furthermore, attempting to decouple rounds using standard trace-norm hybrid inequalities inevitably discards the physical variance terms that are required to bound the source response. Section~\ref{sec:structural} begins by examining this attempted decomposition and identifying where its errors enter. To establish a tight bound, the proof instead preserves the actual entering memory states and maintains an exact, non-negative variance loss.

\subsection{Tracking information drift via backward resolvents}
Tracking information extracted by an adaptive protocol with persistent quantum memory faces two conflicting obstacles. First, the quantum memory state $B_{\lambda, t}$ under an alternative hypothesis is an intractable, path-dependent object that correlates with the unknown source through every prior access. Second, the instantaneous signal extracted by a fresh copy is linear in the perturbation parameter $\theta$, whereas distinguishing hypotheses requires a quadratic charge $\theta^2$. A symmetric test of the form $\Tr(B_{\lambda, t} H) - \Tr(A_{0, t} H)$ cannot bridge this gap: the single-step drift is purely linear in $\theta$ with no negative quadratic term to pay for it, while matrix martingales (such as those in~\cite{cchl2021}) suffer severe non-commutativity losses when decoupling the persistent memory.

We resolve both problems simultaneously by introducing an asymmetric backward score. The final decision effect $H$ is transformed via the bounded resolvent
\[
 f_\alpha(H) = \alpha H(\alpha I - H)^{-1} = \alpha\left[\left(I - \frac{H}{\alpha}\right)^{-1} - I\right], \qquad \alpha > 1.
\]
We track the score along adaptive histories:
\[
 S_t = \mathbb E_\lambda \sum_h \Tr\!\left[ B_{\lambda, t, h} H_{\lambda, t, h} - A_{t, h} f_\alpha(H_{\lambda, t, h}) \right].
\]
This construction achieves three critical mechanical effects:
\begin{enumerate}
 \item \textbf{Annihilation of Alternative Memories:} By pulling the test backward under the alternative channel, $H_{\lambda, t, h} = \mathcal N_{\lambda, h}^*(\mathbf H_{\lambda, t+1, h})$, forward/backward duality forces the alternative memory terms to cancel identically: $\sum_a \Tr[B_{\lambda, t+1, ha} H_{\lambda, t+1, ha}] - \Tr[B_{\lambda, t, h} H_{\lambda, t, h}] = 0$. The telescoping drift accumulates strictly on the null memory state $A_{t, h}$, whose total probability mass is identically conserved ($\sum_h \Tr A_{t, h} = 1$).
 \item \textbf{Emergence of Negative Quadratic Loss:} Because $f_\alpha$ involves the operator inverse $(I - H/\alpha)^{-1}$, Stinespring dilation and the operator Schwarz inequality on inverses naturally manufacture a strictly negative quadratic penalty $-\langle q, \widehat T R q \rangle \le -(1-\theta)(1-1/\alpha)\|q\|^2$. This supplies the physical variance penalty that linear tests lack.
 \item \textbf{Exact Geometric Residual Match:} Projecting $q$ onto the inaccessible dilation space reproduces the variance defect $\Delta = I - FF^*$ in the covariance space $\mathcal H_\sigma$. Completing the square between the linear source response and this negative variance defect turns the linear drift into the sharp $\theta^2 B_K$ bound without any condition-number loss.
\end{enumerate}

In the underlying real variance space, the centered directions associated with the unknown source are orthogonal to the directions native to the entering memory. Projecting out the native memory subspace incurs a geometric penalty of at most $2K-2$ real dimensions. Completing the square then allows the variance loss to control the source response, yielding the exact one-copy information budget
\[
 B_K=\zeta(K)+(2K-2)\beta.
\]
The rank constraint $\rank T \le K$ in the profile $\zeta(K)$ arises naturally because the measurement compresses the joint system through the $K$-dimensional memory register. Because this bound applies to label-dependent backward operators, it sums cleanly along all adaptive measurement paths, accommodating arbitrary classical records and general measurable instruments. Section~\ref{sec:structural} carries out the projection, the inverse identity, the completion of the square, and the adaptive cancellation in this order.

\subsection{Evaluating the bound for growing families}
While the structural theorem gives a clean operational lower bound, its power hinges on whether the rank-constrained profile $\zeta(K)$ can be evaluated tightly for specific families. For weight-$w$ Paulis, memoryless classical shadows achieve $O(3^w\epsilon^{-2}\ln M)$ copies. To match this $3^w/\epsilon^2$ scale under the uniform prior, our lower bound requires $\zeta(K) + (2K-2)/M = O(3^{-w})$. The directional correction $(2K-2)/M$ is negligible here. To tackle the non-convex rank constraint in $\zeta(K)$, our intuition is to build an exponential tail certificate that bounds the generating function directly: for every $T\succeq0$ with $\Tr(T^2)=1$, $\rank T\le K$, and every $t\ge0$,
\[
 \langle T,e^{tA}(T)\rangle\le K e^{nt}\cosh(2t)^n.
\]
By keeping the rank factor $K$ inside this exponential envelope before projecting onto the weight-$w$ subspace---and choosing the Laplace parameters $h = n + 2\sqrt{n \ln n}$ and $t = \sqrt{n \ln n}/(2n)$ (with $a = 2\sqrt{n \ln n}$) so that the Gaussian tail factor evaluates to $e^{-a^2/(8n)} = 1/\sqrt{n}$---we tightly absorb the memory cost at $K=\lfloor\sqrt n\rfloor$. This yields the explicit constant in Corollary~\ref{cor:pauli-asymptotic} for all $n\ge8192$, bypassing the loose constants typical of post-moment interpolation.

For Majorana cross correlations, higher-order moments similarly curb the memory cost. Drawing on the squared-mean bounds of Leone and Bittel~\cite[Lemma~4]{leonebittel2026}, Appendix~\ref{app:moments} combines a positive word expansion to show that $\widehat H_p(L)\le((1+e)pL)^p$. Setting $p=\max\{2,\lceil\ln K\rceil\}$ ensures that $K^{2/p} \le e^2$. At $K=\lfloor\sqrt n\rfloor$, the resulting copy lower bound incurs only a mild $(1+\ln n)^2$ logarithmic factor, holding universally for every $n\ge2$ without requiring any size threshold. The remaining logarithmic factors represent explicit slack between this lower bound and the known simultaneous estimator.

\subsection{Computing the family-dependent bound}
Appendix~\ref{app:computational-certificates} develops the computational certificates complementing the analytic application proofs in Section~\ref{sec:computation}. Beyond analytic bounds, we construct systematic computational methods to evaluate or approximate $\zeta(K)$ for arbitrary observable families. By lifting the state $T$ to the bipartite operator $R=T\otimes T$, the quadratic objective $\Tr(TG_jTG_j)$ becomes linear in $R$. The rank constraint $\rank T \le K$ relaxes into the convex condition $\Tr R=(\Tr T)^2\le K$, while positivity, partial-transpose positivity (PPT), and realignment positivity survive the lift. This yields a source-size semidefinite program (SDP) equipped with explicit dual certificates, which evaluates the complete Pauli family exactly.

To handle larger systems, we leverage the group symmetries of the observables. Averaging the lifted operators over Majorana symmetries collapses the SDP to a compact linear program (LP) on pairs of grades, involving only $(2n+1)^2$ variables. Similarly, local Clifford and permutation symmetries reduce the Pauli optimization to an LP with only $n+1$ variables, producing exact rational certificates at moderate system sizes. Because symmetry averaging is performed after lifting to $R$, it avoids any rank-preservation issues for averages of $T$. In addition, a separate factor-moment hierarchy guarantees asymptotic convergence to the true profile.

\subsection{Charging shared memory across entangled systems}
In bipartite cross-correlation tasks, analyzing subsystems individually fails when the source state is entangled. We overcome this by analyzing the composite system as a unified whole. We introduce a joint-source map on the bipartite space that encompasses all requested products. The sufficient map comparison in Theorem~\ref{thm:composition} controls the sum of conjugations by all requested products, while a high-order moment of this joint map bounds its overlap with any positive operator of rank at most $K$.

The receiver is therefore charged once for their shared memory register. Because this moment bound holds for all pure states in the composite Hilbert space, it remains unconditionally valid even when the source state is entangled across the two subsystems. In the Majorana setting, exact word identities supply the map comparison, and the physical certificate provides the joint moment. Section~\ref{sec:cross} carries out these steps and relates the resulting lower bound to the established product-matching estimator.

\section{Related work}
\label{sec:related}

\paragraph{Property estimation and measurement restrictions.}
The study of quantum property estimation was initiated in Aaronson's formulation of shadow tomography~\cite{aaronson2018}, which asked how many state copies are needed to predict a collection of measurement probabilities simultaneously. Gentle-measurement techniques and improved collective procedures subsequently lowered these copy bounds~\cite{aaronsonrothblum2019,badescuodonnell2021}. Rather than performing collective measurements, the classical shadow framework of Huang, Kueng, and Preskill~\cite{hkp2020} demonstrated that randomized single-copy measurements can produce a versatile classical record that predicts numerous target observables efficiently. These property-estimation tasks stand in contrast to full state tomography, whose optimal collective copy complexity was characterized by the works~\cite{haah2017,odonnellwright2016}. Even when adaptive measurements on single copies are allowed, information-theoretic lower bounds demonstrate severe operational bottlenecks compared to joint measurements~\cite{chen2023tight,lowenayak2025}. In this work, we hold the requested family of observables and the simultaneous estimation accuracy fixed, and isolate the exact role played by persistent quantum memory.

The power of quantum memory has been underscored by several exponential separations between collective and unentangled measurements~\cite{huang2021bounds,cchl2021,huang2022experiments}. For full Pauli estimation with persistent memory, Chen, Cotler, Huang, and Li~\cite{cchl2021} tracked operator exponentials $\exp(\sum X_i)$ via matrix martingales; however, non-commutativity between the entering memory and incoming measurements necessitated Cauchy--Schwarz decouplings that degraded their lower bound to $\Omega((D/K)^{1/3})$. In contrast, our resolvent backward induction preserves operator ordering and recovers the sharp linear trade-off $\Omega(D/(K\epsilon^2))$. To understand this landscape systematically, Chen, Gong, and Ye developed a unifying framework that classifies measurement architectures and bounds the sample complexity of Pauli property estimation~\cite{cgy2024}. Their formulation naturally includes indefinitely persistent memory by treating the entire experiment as a single block of length $c=N$~\cite[Section~5.2 and Theorem~12]{cgy2024}. However, setting $c=N$ leaves the bound expressed as an abstract optimization over the full class of $N$-copy adaptive measurements. Our structural rule complements their framework by providing an explicit, computable lower-bound certificate derived directly from the source observable matrices, which we evaluate tightly for growing families.

Recent breakthroughs in shadow tomography have achieved dimension-independent copy complexities for general observables~\cite{sinha2025,jhl2026}. These algorithms demonstrate the immense power of unconstrained collective measurements, raising the complementary question of what happens when the memory register between consecutive accesses is strictly bounded. Copy complexity alone does not specify the physical memory size required to implement a protocol sequentially. In related work, the authors of~\cite{xu2026} analyzed memoryless and bounded-interaction strategies via commutation and frustration graphs. Here, we explicitly track the Hilbert space dimension $K$ of all quantum systems carried across rounds, granting the receiver unlimited classical memory and unrestricted memory lifetime.

\paragraph{Pauli and fermionic families.}
For Pauli estimation, the authors of~\cite{king2025} developed uncertainty bounds and investigated efficient multi-copy procedures based on commutation graphs. Their printed lower bound for local Pauli estimation requires the weight to satisfy $w\le\log_3(2n+1)$, whereas our lower bound applies to growing weights $w=\lceil\log_2 n\rceil$ and accounts for a growing persistent memory. Established memoryless protocols—specifically randomized local Pauli measurements~\cite{hkp2020} and commuting measurement schedules~\cite{bonet2020,zhao2021}—provide the achieving benchmarks for our comparisons. In fermionic systems, matchgate shadows provide powerful single-copy estimation tools grounded in the Jordan--Wigner transformation and fermionic linear optics~\cite{wan2023,ogorman2022,jordanwigner1928,bravyi2005}.

To bound the response of quadratic Majorana observables, we utilize the all-state squared-mean and word-count estimates established by Leone and Bittel~\cite[Appendix~S.2, Lemmas~3--4]{leonebittel2026}. We transfer their combinatorial bounds onto the positive conjugation map on the joint system, and ultimately to persistent-memory copy lower bounds.

\paragraph{Information geometry, frames, and convex certificates.}
Measurement frames offer a natural language for analyzing quantum measurement strategies~\cite{scott2006,christensen2016}. Symmetric logarithmic derivatives and monotone Riemannian metrics provide the differential-geometric foundation for quantum estimation~\cite{braunsteincaves1994,petz1996}. In recent work, Chen, Gong, and Zhou studied high-precision shadow tomography below instance-dependent error thresholds~\cite{cgz2026}, while Kwon, Lie, and Jiang analyzed Cram\'er--Rao bounds and maximum-likelihood estimators~\cite{klj2026}. In our setting, we consider arbitrary decoders operating at fixed, dimension-independent accuracy. Our projection-and-loss technique connects the local information geometry directly to an indefinitely reused quantum register.

To evaluate our rank-constrained profile computationally, we draw on established convex optimization and representation-theoretic techniques. Positive partial transpose (PPT)~\cite{peres1996,horodecki1996} and matrix realignment~\cite{chenwu2003,rudolph2005} provide convex relaxations for low-rank tensor states, while semidefinite programming duality provides certifiable lower bounds~\cite{boyd2004}. Symmetry reduction follows invariant semidefinite programming techniques~\cite{gatermannparrilo2004}, with explicit discrete transforms provided by quantum weight and shadow enumerators~\cite{shor1997,rains1998,rains1996}. To establish asymptotic convergence, we employ the moment and sum-of-squares (SOS) hierarchies of Lasserre and Putinar~\cite{lasserre2001,putinar1993,laurent2009}. These analytical and computational tools form the foundation for evaluating the structural rule.

Following the structured classification of measurement and memory architectures introduced by Chen, Gong, and Ye~\cite{cgy2024}, Table~\ref{tab:position} summarizes our copy bounds across observable families and memory capacities. The taxonomy separates bounds into those for general persistent memory of dimension $K$ (governing both arbitrary reflection families and the complete-Pauli benchmark) and those for growing persistent memory ($K = \lfloor\sqrt{n}\rfloor$). For the target growing memory, Chen--Gong--Ye's characterization treats the entire protocol as a single block of length $c=N$, leaving the allowed $N$-copy collective measurement optimization to be evaluated. Our source-matrix certificates bypass this receiver optimization entirely, establishing explicit lower bounds that match memory-free achieving protocols up to polylogarithmic factors.

\par\medskip\noindent
\begin{minipage}{\linewidth}
\centering
\setlength{\tabcolsep}{3pt}
\renewcommand{\arraystretch}{1.3}
\begin{tabular}{@{}>{\raggedright\arraybackslash}p{.17\linewidth}
 >{\raggedright\arraybackslash}p{.27\linewidth}
 >{\raggedright\arraybackslash}p{.25\linewidth}
 >{\raggedright\arraybackslash}p{.23\linewidth}@{}}
\toprule
Memory dimension & Observable family & Copy bound & Result\\
\midrule
$K$ & Orthogonal traceless reflections & $\Omega(1/(\epsilon^2 B_K))$ & Corollary~\ref{cor:estimation}\\
$1\le K\le D$ & Complete Paulis $\mathcal P_n^*$ & $\Omega(D/(K\epsilon^2))$ & Theorem~\ref{thm:complete}\\
$K=\lfloor\sqrt n\rfloor$ & Weight-$w$ Paulis, $w=\lceil\log_2 n\rceil$ & $\widetilde\Theta(3^w/\epsilon^2)$; gap $O(\ln^2 n)$ & Corollary~\ref{cor:pauli-asymptotic}, \cite{hkp2020}\\
$K=\lfloor\sqrt n\rfloor$ & Majorana cross correlations & $\widetilde\Theta(n^2/\epsilon^2)$; gap $O(\ln^3 n)$ & Theorem~\ref{thm:cross}, \cite{bonet2020,zhao2021}\\
\bottomrule
\end{tabular}
\captionof{table}{Summary of copy bounds with persistent quantum memory ($\tilde{\Theta}(\cdot)$ hides (poly)-logarithmic factors). Here, $\mathcal{P}_n^* = \mathcal{P}_n \setminus \{I^{\otimes n}\}$ denotes the family of all $D^2 - 1$ non-identity $n$-qubit Paulis ($D=2^n$), and $B_K = \zeta(K) + (2K-2)\beta$ is the rank-constrained certificate budget defined in~\eqref{eq:structural-budget}. Quantum memory of dimension $K$ may be retained indefinitely across all measurement rounds, with unlimited classical memory and arbitrary adaptive POVMs. For weight-$w$ Paulis, our lower bound $\Omega(3^w/\epsilon^2)$ matches the memory-free ($K=1$) randomized Pauli shadow upper bound $O(3^w\epsilon^{-2}\ln M)$ within an $O(\ln^2 n)$ factor, where $M=3^w\binom nw$. For Majorana cross correlations, our lower bound $\Omega(n^2/(\epsilon^2\ln^2 n))$ matches the memory-free commuting measurement upper bound $O(n^2\epsilon^{-2}\ln n)$ within an $O(\ln^3 n)$ factor. The complete-Pauli benchmark establishes the $\Omega(D/(K\epsilon^2))$ lower bound across all $1\le K\le D$, recovering the unentangled $\tilde{\Theta}(2^n/\epsilon^2)$ rate at $K=1$. All outputs are simultaneous signed estimates with $0<\epsilon\le1/8$ and success $2/3$; the simplified Pauli bound uses $n\ge8192$ and the cross bound uses $n\ge2$.}
\label{tab:position}
\end{minipage}
\par\medskip

\section{Preliminaries}
\label{sec:model}
\label{sec:preliminaries}
To make the analysis self-contained and mathematically transparent, we collect here the foundational definitions and standard facts that underpin our lower-bound proofs and estimation algorithms. We begin by reviewing states and measurements, and then formalize the operational protocol model for persistent quantum memory. Subsequent subsections establish the hypothesis-testing reduction, introduce the algebraic structure of Pauli and Majorana observables, and summarize key matrix analysis and concentration inequalities. Throughout, $\Tr$ denotes the ordinary matrix trace, $\ln$ denotes the natural logarithm, and $[m]=\{1,\ldots,m\}$.

\subsection{States, observables and matrix norms}
\label{subsec:states-and-norms}
A quantum state on a $D$-dimensional Hilbert space is represented by a density matrix: a positive semidefinite operator $\rho \succeq 0$ normalized by $\Tr\rho=1$. A pure state takes the form $|v\rangle\langle v|$ for a unit vector $v \in \mathbb{C}^D$, while general mixed states are convex mixtures of pure states. For an $n$-qubit register, the Hilbert space dimension is $D=2^n$. On a bipartite system $\mathcal{H}_A \otimes \mathcal{H}_B$, the partial trace $\Tr_B$ satisfies the duality relation
\[
 \Tr[\rho_{AB}(X_A\otimes I_B)]
 =\Tr[(\Tr_B\rho_{AB})X_A].
\]
We emphasize that the joint state $\rho_{AB}$ is completely arbitrary and need not factorize as a product state; in particular, our cross-correlation lower bound in Section~\ref{sec:cross} applies to arbitrary entangled source states~\cite{watrous2018}.

An observable $G$ is a Hermitian operator, with expectation value $\Tr(\rho G)$ on state $\rho$. We write $A\preceq B$ to denote that $B-A$ is positive semidefinite. A Hermitian contraction satisfies $-I\preceq G\preceq I$. A Hermitian reflection satisfies $G^2=I$, meaning its eigenvalues are strictly $+1$ and $-1$. We write $\tau=I_D/D$ for the maximally mixed state. An observable is centered with respect to this reference when $\Tr(\tau G)=0$. For centered reflections $G_i$, the orthogonality condition used in our applications is
\[
 \frac1D\Tr(G_iG_j)=\mathbf1_{i=j}.
\]
These centering and orthogonality relations are properties of the target observables chosen by the experimenter; they impose no restrictions or promises on the unknown quantum state $\rho$.

For a matrix $X$, let $X^*$ denote its adjoint. We employ three standard Schatten norms:
\[
 \|X\|_1=\Tr\sqrt{X^*X},\qquad
 \|X\|_{\rm HS}=\sqrt{\Tr(X^*X)},\qquad
 \|X\|_{\rm op}=\sup_{\|v\|=1}\|Xv\|.
\]
These correspond to the trace norm, Hilbert--Schmidt (Frobenius) norm, and operator norm, respectively. The Hilbert--Schmidt inner product is $\langle X,Y\rangle=\Tr(X^*Y)$, whose real part defines a Euclidean inner product when Hermitian matrices are viewed as a real vector space. We make frequent use of the standard inequalities~\cite{bhatia1997,hornjohnson2013}
\[
 |\Tr(X^*Y)|\le\|X\|_{\rm HS}\|Y\|_{\rm HS},\qquad
 |\Tr(XY)|\le\|X\|_1\|Y\|_{\rm op}.
\]
For a positive semidefinite matrix $T\succeq0$ of rank $r$, the Cauchy--Schwarz inequality applied to its eigenvalues gives
\[
 (\Tr T)^2\le r\Tr(T^2).
\]
This distinction between trace normalization ($\Tr T=1$) and Hilbert--Schmidt normalization ($\Tr(T^2)=1$) is crucial: our structural source profile optimizes over positive matrices $T$ normalized by $\Tr(T^2)=1$, which are not necessarily physical density matrices.

\subsection{Quantum measurements and channels}
\label{subsec:measurements-and-channels}
A generalized quantum measurement with a finite set of classical outcomes is described by a positive operator-valued measure (POVM): a collection of positive semidefinite operators $\{E_x\}$ satisfying $\sum_x E_x = I$. Given an input state $\rho$, outcome $x$ occurs with probability $\Tr(E_x\rho)$. A quantum channel is a completely positive, trace-preserving (CPTP) linear map on operators. In finite dimensions, any CPTP map admits a Kraus representation~\cite{choi1975,watrous2018}
\[
 \mathcal N(Z)=\sum_\ell A_\ell Z A_\ell^*,\qquad
 \sum_\ell A_\ell^*A_\ell=I.
\]
If the trace-preserving condition is relaxed to $\sum_\ell A_\ell^* A_\ell \preceq I$, the map is trace-non-increasing and represents a sub-normalized measurement branch.

We compare completely positive maps using the Löwner-type CP order: we write $\Phi\preceq_{\rm CP}\Psi$ when $\Psi-\Phi$ is completely positive. For any $T\succeq0$, this implies that $\Tr[T\Phi(T)]\le\Tr[T\Psi(T)]$, since the trace inner product of two positive semidefinite operators is always non-negative. This ordering allows us to isolate and bound selected terms in sums of completely positive operations.

A quantum instrument formalizes a measurement that outputs both a classical outcome and a post-measurement quantum state~\cite{davieslewis1970}. Each outcome $x$ is associated with a completely positive map $\Phi_x(Z)=\sum_\ell A_{x,\ell}ZA_{x,\ell}^*$, with the completeness condition
\[
 \sum_{x,\ell}A_{x,\ell}^*A_{x,\ell}=I.
\]
The probability of obtaining outcome $x$ is $p_x=\Tr\Phi_x(\rho)$, and the corresponding normalized conditional post-measurement state is $\Phi_x(\rho)/p_x$ whenever $p_x>0$. Combining the classical record with the quantum output yields the block-diagonal recorded channel
\[
 \Phi(Z)=\bigoplus_x\Phi_x(Z).
\]
Here, the direct sum represents distinct classical branches. Crucially, our protocol model charges only the coherent quantum dimension within each branch toward the memory budget; the number of classical branches and the length of classical records are entirely unconstrained.

The adjoint channel $\mathcal{N}^*$ in the Heisenberg picture is defined via the trace duality $\Tr[X\mathcal N(Z)]=\Tr[\mathcal N^*(X)Z]$. In Kraus form, $\mathcal N^*(X)=\sum_\ell A_\ell^*XA_\ell$ with $\mathcal N^*(I)=I$ for unital maps. Under this adjoint action, any measurement effect $0\preceq X\preceq I$ pulls back to a valid effect. Furthermore, trace-norm contractivity and Kadison--Schwarz inequalities for unital completely positive maps ensure that~\cite{watrous2018,paulsen2002,stinespring1955}
\[
 \|\mathcal N(Z)\|_1\le\|Z\|_1\quad(Z=Z^*),\qquad
 [\mathcal N^*(X)]^2\preceq\mathcal N^*(X^2)\quad(X=X^*).
\]
The second operator inequality captures the variance lost during measurement, which plays a central role in our information-tracking arguments.

\subsection{The persistent-memory model and estimation task}
\label{subsec:memory-model}
We consider a receiver who interacts sequentially with independent copies of an unknown state $\rho \in \mathcal{S}(\mathbb{C}^D)$. In each round, the receiver receives a single fresh copy of $\rho$, performs an arbitrary joint operation between this new copy and the quantum memory register retained from past rounds, emits a classical measurement outcome, and stores a post-measurement quantum register for use in future rounds. The dimension of this retained quantum register is bounded by $K$ on every classical branch. This bound $K$ accounts for the product dimension of all coherent quantum systems kept across rounds—including ancillas, pointers, preserved source subsystems, and quantum controls. In contrast, classical measurement records are dephased and may be retained indefinitely without penalty.

Formally, conditioned on the complete classical history $h$ of outcomes observed up to the current round, the interaction is governed by a complete quantum instrument $\{\Phi_{h,x}\}_x$ mapping the joint system $\mathcal{H}_{\text{mem}} \otimes \mathcal{H}_{\text{source}}$ to the retained memory space $\mathcal{H}_{\text{mem}}'$. The instrument may be adapted arbitrarily based on the history $h$ and the round index. Any temporary workspace used during the interaction is permitted, provided that by the end of the round it has been measured into the classical record or included within the retained dimension $K$. The state entering the round may be arbitrarily correlated with the unknown source through past interactions; only the initial memory state $\sigma_0$ and any classical random seed are required to be source-independent.

Our model extends to standard Borel outcome spaces and measurable instrument kernels~\cite{kallenberg2021}, meaning that the choice of instruments depends measurably on the classical history. This accommodates finite, countable, and continuous measurement records. In our proofs, we first establish the finite-outcome case and then lift the result to measurable kernels via completely positive densities.

A protocol operates under a worst-case copy cap $N$: across all possible classical histories, the receiver consumes at most $N$ copies. Any protocol that halts early can be trivially padded to $N$ accesses by idling on subsequent copies. At the end of the $N$ accesses, the receiver may perform an arbitrary final quantum measurement and classical computation to produce their estimates. The quantum memory is allowed to remain coherent for an unrestricted lifetime across the entire run.

\begin{definition}[Simultaneous estimation task]
\label{def:task}
Let $\mathcal G=\{G_1,\ldots,G_M\}$ be known Hermitian observables. A protocol solves the signed $(\epsilon,\delta)$ task if for every source state $\rho$ its real outputs obey
\[
 \Pr_\rho\!\left[\max_{1\le j\le M}
 |\widehat{\mu}_j-\Tr(\rho G_j)|\le\epsilon\right]\ge1-\delta.
\]
Let $N^*(\mathcal G,K,\epsilon,\delta)$ be the smallest worst-case copy cap of such a protocol in the memory model above. Computational work and the output length are unrestricted in this definition. For each displayed achieving algorithm we separately report implementation and decoding costs.
\end{definition}

In our main applications, we set the failure probability to $\delta=1/3$ and precision to $0<\epsilon\le1/8$, requiring all $M$ signed expectations to be estimated simultaneously. Throughout, $K$ measures the Hilbert space dimension of the retained quantum register: storing $b$ persistent qubits corresponds to $K=2^b$. If multiple registers survive between rounds, their combined product dimension counts toward $K$. At $K=1$, no quantum coherence persists between rounds, recovering the memoryless classical-record setting. At the other extreme, $K=D=2^n$ allows the receiver to store an entire copy of the state, enabling joint measurements across pairs of copies.

\subsection{Adaptive histories and hypothesis testing}
\label{subsec:histories-and-testing}
For finite outcome sets, an adaptive protocol can be naturally represented as a tree of classical histories~\cite{cchl2021,cgy2024}. To track both the branch probability and the conditional quantum state simultaneously, we define the unnormalized positive operator $\omega_h^\rho$. Starting from the source-independent initial state $\omega_{\varnothing}^\rho=\sigma_0$, the state at history $hx$ updates according to
\[
 \omega_{\varnothing}^\rho=\sigma_0,\qquad
 \omega_{hx}^\rho=\Phi_{h,x}(\omega_h^\rho\otimes\rho),\qquad
 \Pr_\rho(h)=\Tr\omega_h^\rho.
\]
Trace preservation ensures that $\sum_x\Tr\omega_{hx}^\rho=\Tr\omega_h^\rho$, so that $\sum_{|h|=t}\Tr\omega_h^\rho=1$ at every round $t$. On any branch with $\Pr_\rho(h)>0$, the normalized quantum memory is $\omega_h^\rho/\Tr\omega_h^\rho$. Crucially, different candidate source states $\rho$ induce distinct, non-commuting memory states at the same node $h$, and our lower-bound analysis explicitly tracks these actual branch states.

A decision event $E$ at the termination of the protocol is specified by POVM effects $\{Q_h\}_{|h|=N}$ satisfying $0\preceq Q_h\preceq I$, with success probability
\[
 \Pr_\rho(E)=\sum_{|h|=N}\Tr(Q_h\omega_h^\rho).
\]
These terminal effects can be pulled backward through the adjoint maps $\Phi_{h,x}^*$, representing the effective decision operator at intermediate nodes.

To lower-bound the copy complexity, we reduce estimation to a composite hypothesis testing problem. By the Holevo--Helstrom theorem and total variation duality~\cite{watrous2018,tsybakov2009},
\begin{align*}
 \sup_{0\preceq Q\preceq I}
 |\Tr[Q(\sigma_1-\sigma_0)]|
 &=\tfrac12\|\sigma_1-\sigma_0\|_1,\\
 \sup_E|P_1(E)-P_0(E)|
 &=\tfrac12\sum_x|P_1(x)-P_0(x)|.
\end{align*}
Hence, distinguishing two states with high probability requires their output distributions to be separated in total variation distance.

To derive our estimation lower bound, we map the simultaneous estimation task to a composite hypothesis testing problem. In Theorem~\ref{thm:structural}, we test the maximally mixed null state $\tau$ against an ensemble of perturbed alternatives:
\[
 \tau=\frac{I_D}{D},\qquad
 \rho_{j,z}=\frac{I_D+z\theta G_j}{D},\qquad
 0<\theta<1,\quad z\in\{-1,1\}.
\]
Because each $G_j$ is a traceless contraction, $\rho_{j,z}$ is a valid quantum state. In this testing game, a single label $(j,z)$ is sampled from the prior $p_j/2$, and all $N$ copies originate from this chosen state. For tracially orthogonal reflections, the true expectation values satisfy
\[
 \Tr(\tau G_i)=0,\qquad
 \Tr(\rho_{j,z}G_i)=z\theta\,\mathbf1_{i=j}.
\]
Any estimator achieving simultaneous additive error $\epsilon < \theta/2$ reliably distinguishes the null hypothesis from every alternative through the threshold event $E = \{\max_i |\widehat{\mu}_i| > \theta/2\}$. Under $\tau$, this event occurs with probability at most $\delta$, whereas under any alternative $\rho_{j,z}$, it triggers with probability at least $1-\delta$. Lower bounding the sample cost of this decision test directly bounds the copy complexity of simultaneous property estimation.

\subsection{Pauli and Majorana observables}
\label{subsec:observable-background}
The single-qubit Pauli matrices are defined as
\[
 I=\begin{pmatrix}1&0\\0&1\end{pmatrix},\quad
 X=\begin{pmatrix}0&1\\1&0\end{pmatrix},\quad
 Y=\begin{pmatrix}0&-i\\i&0\end{pmatrix},\quad
 Z=\begin{pmatrix}1&0\\0&-1\end{pmatrix}.
\]
On $n$ qubits, the $4^n$ canonical tensor products of these matrices form an orthonormal Hermitian basis for the space of $D \times D$ operators (with $D=2^n$) under the normalized trace inner product~\cite{watrous2018,hkp2020}:
\[
 \Tr(PQ)=D\mathbf1_{P=Q},\qquad
 A=\frac1D\sum_P\Tr(PA)P,\qquad
 \sum_P PAP=D\Tr(A)I_D.
\]
The last identity is the Pauli twirl, which enables an exact analytic evaluation of the complete Pauli profile. The weight of a Pauli string is the number of non-identity factors; there are $3^w\binom nw$ distinct strings of weight $w$.

For an $n$-mode fermionic system, the algebra of creation and annihilation operators is faithfully represented by $2n$ Majorana operators $c_1,\ldots,c_{2n}$ satisfying the anticommutation relations
\[
 c_a=c_a^*,\qquad c_ac_b+c_bc_a=2\mathbf1_{a=b}I.
\]
Under the Jordan--Wigner transformation~\cite{jordanwigner1928,bravyi2005}, these operators are explicitly realized on $n$ qubits as
\[
 c_{2r-1}=Z_1\cdots Z_{r-1}X_r,\qquad
 c_{2r}=Z_1\cdots Z_{r-1}Y_r.
\]
For an ordered index subset $S=\{s_1<\cdots<s_t\}$, the Majorana monomials
\[
 \Gamma(S)=i^{t(t-1)/2}c_{s_1}\cdots c_{s_t}
\]
form an orthonormal basis of Hermitian reflections. The quadratic observables $i c_a c_b$ (with $a<b$) represent two-body fermionic properties, whose commutation properties enable efficient matching measurement schedules~\cite{bonet2020,zhao2021,wan2023}.

For bipartite systems across two $n$-mode factors with local Majoranas $a_r$ and $b_s$, the quadratic cross-correlations take the form $(i a_a a_b)\otimes(i b_c b_d)$. To handle this joint system algebraically, we introduce the Hermitian local parity operator $P_A$ on subsystem $A$ (satisfying $P_A=P_A^*$, $P_A^2=I$, and $P_A a_r=-a_r P_A$). The combined operators $c_r = a_r \otimes I$ and $c_{2n+s} = P_A \otimes b_s$ then form a standard set of $4n$ global Majoranas on the joint Hilbert space. This ingenious embedding maps the bipartite cross correlations directly into degree-four Majorana monomials on the composite system, allowing our analysis to handle arbitrary joint entanglement without imposing any unphysical product-state restrictions.

\subsection{Matrix vectorization, covariance, and information geometry}
\label{subsec:matrix-background}
Matrix vectorization $\operatorname{vec}(X)$ stacks the columns of a matrix $X$. For compatible matrices, it obeys the standard Kronecker identities~\cite{watrous2018,hornjohnson2013}
\[
 \operatorname{vec}(AXB)=(B^{\mathsf T}\otimes A)\operatorname{vec}(X),
 \qquad
 \|\operatorname{vec}(X)\|_2=\|X\|_{\rm HS},
\]
where $\mathsf{T}$ denotes the matrix transpose in the computational basis. The swap operator $F(u\otimes v)=v\otimes u$ satisfies $\Tr[(A\otimes B)F]=\Tr(AB)$.

The Gram matrix $V^*V$ of a set of vectors and their frame operator $VV^*$ share the same non-zero spectrum~\cite{christensen2016}:
\[
 \|V^*V\|_{\rm op}=\|VV^*\|_{\rm op}=\|V\|_{\rm op}^2.
\]
Viewing Hermitian observables as vectors under the inner product $\Re\Tr(\tau AB)$, this spectral equality directly produces the weighted Gram norm $\beta$ of Definition~\ref{def:profile}, allowing us to bound sums of squared source responses by the operator norm of the synthesis map.

For a set of Hermitian observables $\{G_j\}$ with prior weights $p_j$, the averaging map $\Phi(X)=\sum_jp_jG_jXG_j$ is completely positive and self-adjoint under the Hilbert--Schmidt inner product. While $\Phi$ is completely positive, its Kraus matrix $S=\sum_jp_j(G_j^{\mathsf T}\otimes G_j)$ may have negative eigenvalues. Applying the spectral theorem and Hölder's inequality for any state $T$ with $\|T\|_{\rm HS}=1$ gives the spectral moment bound
\[
 |\langle T,\Phi(T)\rangle|
 \le\int|\lambda|\,d\mu_T(\lambda)
 \le\left[\langle T,\Phi^{2r}(T)\rangle\right]^{1/(2r)},\qquad r\ge1.
\]
This spectral moment relation provides our primary analytical vehicle for bounding $\zeta(K)$~\cite{bhatia1997,blm2013}.

For a density matrix $\sigma$, the real covariance of two Hermitian observables $X, Y$ is given by
\[
 \langle X,Y\rangle_\sigma
 =\tfrac12\Tr[\sigma(XY+YX)]
       -\Tr(\sigma X)\Tr(\sigma Y).
\]
Quotienting out the null space of zero-variance operators yields a real Hilbert space equipped with this covariance metric. The symmetric logarithmic derivative (SLD) represents variations in density matrices within this geometry~\cite{braunsteincaves1994,petz1996}. In terms of the Jordan map $\Omega_\sigma(X)=(\sigma X+X\sigma)/2$, its Moore--Penrose pseudoinverse in an eigenbasis $\sigma=\operatorname{diag}(s_a)$ is
\[
 [\Omega_\sigma^+(Y)]_{ab}
 =\begin{cases}
 2Y_{ab}/(s_a+s_b),&s_a+s_b>0,\\
 0,&s_a+s_b=0.
 \end{cases}
\]
For any traceless Hermitian $Y$ supported on $\operatorname{supp}\sigma$, the operator $L=\Omega_\sigma^+(Y)$ satisfies $\Tr(\sigma L)=0$ and acts as the Riesz representer $\Tr(YX)=\langle L,X\rangle_\sigma$. Furthermore, for any self-adjoint contraction $V$, the resolvent operator obeys
\[
 I+\theta V\succeq(1-\theta)I,\qquad
 \|(I+\theta V)^{-1}\|_{\rm op}\le\frac1{1-\theta}
 \quad(0<\theta<1),
\]
guaranteeing uniform stability when inverting perturbation operators on the support of $\sigma$.

\subsection{Probability bounds for property estimation}
\label{subsec:probability-background}
To benchmark our lower bounds against achievable protocols, we analyze standard estimators using classical concentration inequalities. When $X_1,\ldots,X_R$ are independent random variables bounded in $[-B,B]$ with mean $\mu$, Hoeffding's inequality~\cite{hoeffding1963} bounds the sample deviation:
\[
 \Pr\left\{\left|R^{-1}\sum_{t=1}^R X_t-\mu\right|>
 \epsilon\right\}
 \le2\exp\left(-\frac{R\epsilon^2}{2B^2}\right).
\]
Taking a union bound over all $M$ target observables guarantees simultaneous additive accuracy $\epsilon$ with overall failure probability at most $\delta$ as soon as
\[
 R\ge2B^2\epsilon^{-2}\ln(2M/\delta).
\]
In particular, for commuting matching measurements of Hermitian reflections, each single-shot outcome satisfies $B=1$.

When estimators exhibit small variance, Bernstein's inequality provides tighter tail bounds~\cite{blm2013}. For centered independent variables with $|Y_t|\le L$ and variance $\mathbb E Y_t^2\le v$,
\[
 \Pr\left\{\left|R^{-1}\sum_{t=1}^R Y_t\right|>u\right\}
 \le2\exp\left(-\frac{Ru^2}{2v+2Lu/3}\right).
\]
Finally, Markov's inequality converts expected risk bounds into high-probability guarantees. These standard statistical tools ensure that our reported upper-bound copy budgets achieve rigorous simultaneous success guarantees.

\section{Lower bounds with persistent quantum memory}
\label{sec:structural}
Theorem~\ref{thm:structural} and Corollary~\ref{cor:estimation} were established
in Section~\ref{subsec:main-structural}, alongside the source profile defined in
Definition~\ref{def:profile}. In this section, we present their complete proofs within the sequential adaptive framework formulated in Section~\ref{sec:model}. Our analysis proceeds in two conceptual stages: we first evaluate the information gained from a single quantum access, balancing it against the persistent memory capacity, and then integrate this finite charge across an arbitrary adaptive experiment. Throughout the argument, $\beta$, $\zeta$, and $B_K$ refer to the fundamental quantities defined in \eqref{eq:beta-profile}--\eqref{eq:structural-budget}.

\subsection{Why adding one-step errors is too costly}

To motivate our analytical framework, consider the natural impulse when bounding the distinguishability between a null hypothesis $\tau = I_D/D$ and an alternative state $\rho_\lambda = \tau + z\theta G_j/D$ indexed by $\lambda = (j,z)$. At any given history in the protocol, let $A$ and $B$ represent the unnormalized quantum states stored in the memory register under the null and alternative models, respectively. Denoting the complete quantum instrument executed at this step by $\Phi$, the post-measurement joint states after ingesting the next source copy are given by the channels $\mathcal N_0(A) = \Phi(A \otimes \tau)$ and $\mathcal N_\lambda(B) = \Phi(B \otimes \rho_\lambda)$. A standard hybrid argument immediately yields
\begin{align}
 \mathcal N_\lambda(B)-\mathcal N_0(A)
 &=\mathcal N_\lambda(B-A)
       +(\mathcal N_\lambda-\mathcal N_0)(A),\nonumber\\
 \bigl\|\mathcal N_\lambda(B)-\mathcal N_0(A)\bigr\|_1
 &\le\|B-A\|_1
       +\theta\bigl\|\Phi(A\otimes G_j/D)\bigr\|_1,
 \label{eq:hybrid-motivation}
\end{align}
where the second line relies on the contractivity of trace-preserving completely positive maps on Hermitian differences~\cite{watrous2018}, together with the linear perturbation $(\mathcal N_\lambda-\mathcal N_0)(A)=z\theta\Phi(A\otimes G_j/D)$. 

While~\eqref{eq:hybrid-motivation} correctly tracks the divergence between entering memory states, it exposes a critical limitation: it demands payment for the absolute source response at every single access, incurring a penalty linear in the perturbation strength $\theta$. In hypothesis testing, however, distinguishing close quantum states statistically requires sample complexity scaling with $\theta^{-2}$. If we simply accumulate these one-step trace-norm bounds, the resulting sample lower bound would only scale as $\theta^{-1}$, missing the true quantum quadratic cost. Conversely, while standard information-theoretic divergences such as quantum relative entropy natively scale with $\theta^2$, their subadditivity on product states makes them completely blind to the low-rank quantum memory bottleneck $K$.

To overcome this barrier and capture the quadratic signal scaling in Theorem~\ref{thm:structural}, we cannot discard the fluctuations at each step. Instead, we must retain a nonnegative variance loss that accumulates across steps until it can fund the linear response. We realize this intuition through a backward analysis of the terminal decoding decision. By regularizing the terminal test with a resolvent, the perturbations in the alternative memory register cancel identically between consecutive rounds. The resulting one-step identity exposes a net drift consisting of the source response offset by a quadratic variance loss, both evaluated at the common null baseline. We then demonstrate that this variance loss directly controls the response via the observable Gram matrix. Completing the square transforms the linear drift into a sharp $\theta^2$ charge.

The proof unfolds in three natural steps. First, we geometrically decouple the variance directions associated with the incoming source copy from those already retained in the persistent memory. Second, we certify that the total effective variance is bounded by $B_K$ and derive the exact resolvent identity. Finally, we pull the terminal decision event backward through the adaptive measurement tree and sum the resulting bounds over all access histories. We initially establish the result for finite measurement alphabets before extending it to general measurable outcome spaces.

\subsection{Separating new information from stored information}

Consider an arbitrary history in the adaptive protocol where the entering quantum memory resides in a Hilbert space $\mathbb C^r$ with dimension $r \le K$. For finite measurement outcomes, we express the complete instrument as
\begin{equation}
 \Phi(Z)=\bigoplus_a M_aZM_a^*,\qquad
 M_a:\mathbb C^r\otimes\mathbb C^D\longrightarrow\mathbb C^{k_a},
 \quad k_a\le K,\quad \sum_aM_a^*M_a=I_{rD}.
 \label{eq:local-complete-instrument}
\end{equation}
Without loss of generality, we refine this representation by including a classical Kraus index: every completely positive map between finite-dimensional spaces admits a finite Kraus decomposition~\cite{choi1975,watrous2018}. Retaining this index as an auxiliary classical record while allowing subsequent control operations to ignore it preserves the original protocol and leaves the quantum memory dimension unchanged. Establishing our bound for this refined instrument therefore guarantees its validity for the original protocol.

For any memory state $A$, the complete recorded channels under the null and perturbed sources are
\begin{equation}
 \mathcal N_0(A)=\Phi(A\otimes\tau),\qquad
 \mathcal N_{j,z}(A)=\Phi(A\otimes\rho_{j,z}).
 \label{eq:local-source-channels}
\end{equation}
Because~\eqref{eq:local-complete-instrument} accounts for all measurement outcomes, the adjoint maps $\mathcal N_0^*$ and $\mathcal N_{j,z}^*$ are unital. Fixing a unit vector $u \in \mathbb C^r$ in the entering memory register, we define the reference output state, the source perturbation, and the orthogonal projection onto the orthogonal complement of $u$:
\begin{equation}
 \sigma=\Phi(uu^*\otimes\tau),\qquad
 \delta_j=\Phi(uu^*\otimes G_j/D),\qquad Q_u=I_r-uu^*.
 \label{eq:sigma-delta}
\end{equation}
Here, $\Tr\sigma = 1$, $\Tr\delta_j = 0$, and the perturbation satisfies $-\sigma \preceq \delta_j \preceq \sigma$, which guarantees that each $\delta_j$ is supported within the support of $\sigma$. While individual outcome blocks of $\delta_j$ may possess nonzero traces, the total perturbation remains globally centered across all outcomes.

To isolate the new information delivered by the incoming copy, we introduce a testing geometry that quantifies the response to the fresh source while tracking any directions already supplied by the stored memory. The covariance of measurement tests provides precisely the required metric: constant operators produce neither signal nor variance, and any fluctuations generated by the native memory can be systematically projected out. For any pair of Hermitian recorded tests $X$ and $Y$, we define the real covariance inner product
\begin{equation}
 \langle X,Y\rangle_\sigma
 =\frac12\Tr[\sigma(XY+YX)]
          -\Tr(\sigma X)\Tr(\sigma Y).
 \label{eq:variance-inner-product}
\end{equation}
Quotienting out the subspace of zero-variance tests yields a real Hilbert space $\mathcal H_\sigma$ equipped with the norm $\norm{X}_\sigma^2=\Tr(\sigma X^2)-[\Tr(\sigma X)]^2$, where constant operators map to zero. We then define the real linear map $F^*$ capturing native memory fluctuations and the functional $\ell_j$ measuring the source response:
\begin{equation}
 F^*X=Q_u\mathcal N_0^*(X)u\in u^\perp,
 \qquad \ell_j(X)=\Tr(\delta_jX).
 \label{eq:native-source-maps}
\end{equation}
Here, the orthogonal complement $u^\perp$ is treated as a real Hilbert space with inner product $\Re\langle w,w'\rangle$ and real dimension $d = 2r - 2$.

\begin{lemma}[Joint variance contraction]
\label{lem:variance-contraction}
The maps defined in~\eqref{eq:native-source-maps} are well-defined on the variance Hilbert space $\mathcal H_\sigma$. Moreover, for every Hermitian test $X$,
\begin{equation}
 \norm{F^*X}^2+\frac1\beta\sum_jp_j\ell_j(X)^2
 \le\norm{X}_\sigma^2.
 \label{eq:joint-contraction}
\end{equation}
Consequently, if $c_j\in\mathcal H_\sigma$ denotes the Riesz representer of $\ell_j$, $F$ is the real adjoint of $F^*$, and we define
\begin{equation}
 C=\sum_jp_j|c_j\rangle\langle c_j|,\qquad
 \Delta=I-FF^*,
 \label{eq:covariance-defect}
\end{equation}
then the operators satisfy the operator inequalities
\begin{equation}
 0\preceq C\preceq\beta\Delta,
 \qquad 0\preceq\Delta\preceq I,
 \qquad \rank F\le2r-2.
 \label{eq:covariance-defect-bound}
\end{equation}
\end{lemma}

\begin{proof}
Our approach is to analyze fluctuations within the input covariance space corresponding to the joint reference state $\omega = uu^* \otimes \tau$. For any vector $w \in u^\perp$, we define the native memory perturbation $B_w$ and the source probe $S_j$ by
\[
 B_w=(wu^*+uw^*)\otimes I_D,\qquad S_j=I_r\otimes G_j.
\]
Both operators have zero mean with respect to $\omega$. A direct algebraic evaluation shows that their covariance inner products satisfy
\begin{align}
 \langle B_w,B_{w'}\rangle_\omega
    &=\Re\langle w,w'\rangle,\nonumber\\
 \langle B_w,S_j\rangle_\omega&=0,\qquad
 \langle S_i,S_j\rangle_\omega
    =\Re\Tr(\tau G_iG_j).
 \label{eq:input-orthogonal-spaces}
\end{align}
This reveals that the subspace of native memory perturbations $\{B_w : w \perp u\}$ and the span of the source probes $\{S_j\}$ are strictly orthogonal under the input covariance metric. Furthermore, the weighted frame operator $\sum_j p_j |S_j\rangle\langle S_j|$ possesses operator norm $\beta$, as its nonzero eigenvalues coincide with those of the tracial Gram matrix in~\eqref{eq:beta-profile} via the classical frame-Gram correspondence~\cite{christensen2016}.

Now consider the pull-back observable $Y = \Phi^*(X)$ under the measurement instrument. Its covariance with the native and source directions reproduces the definitions of $F^*$ and $\ell_j$:
\begin{align}
 \langle B_w,Y\rangle_\omega
   &=\Re\langle w,\mathcal N_0^*(X)u\rangle,
       \label{eq:native-covariance}\\
 \langle S_j,Y\rangle_\omega
   &=\Tr[\Phi(uu^*\otimes G_j/D)X]=\ell_j(X).
       \label{eq:source-covariance}
\end{align}
For~\eqref{eq:source-covariance}, the symmetrized covariance matches the trace because $\tau = I_D/D$ commutes with every source generator $G_j$. Let $P_{\rm n}$ and $P_{\rm s}$ denote the orthogonal projections in the input covariance space onto the native memory span and source span, respectively. Because the native span is isometric to $u^\perp$ and the source frame has operator norm bounded by $\beta$, we obtain
\begin{align}
 \norm{F^*X}^2+\beta^{-1}\sum_jp_j\ell_j(X)^2
 &=\norm{P_{\rm n}Y}_\omega^2
        +\beta^{-1}\sum_jp_j\langle S_j,Y\rangle_\omega^2\nonumber\\
 &=\norm{P_{\rm n}Y}_\omega^2
        +\beta^{-1}\sum_jp_j\langle S_j,P_{\rm s}Y\rangle_\omega^2\nonumber\\
 &\le\norm{P_{\rm n}Y}_\omega^2+\norm{P_{\rm s}Y}_\omega^2
 \le\norm{Y}_\omega^2.
 \label{eq:variance-projection-chain}
\end{align}
Here, the second equality exploits the fact that $S_j$ lies within the source subspace, while the first inequality applies the frame bound. The final inequality relies on the orthogonality of the two subspaces, ensuring that their variances combine without double counting.

To link this input variance to the measured output variance, we employ Stinespring's dilation theorem~\cite{stinespring1955,paulsen2002}. Defining the isometry $W\xi = \bigoplus_a M_a\xi$ and writing $\widetilde X = \bigoplus_a X_a$, completeness ensures $W^*W = I$, giving $\Phi^*(X) = W^* \widetilde X W$. A direct expansion confirms that
\[
 \Phi^*(X^2)-\Phi^*(X)^2
 =W^*\widetilde X(I-WW^*)\widetilde XW\succeq0.
\]
Because the instrument preserves means under $\omega$, taking the expectation yields the variance contraction $\norm{Y}_\omega^2 \le \norm{X}_\sigma^2$, which establishes~\eqref{eq:joint-contraction}. Since tests with zero output variance necessarily have vanishing projections, the Riesz representers $c_j$ exist. Expressing this result as an operator inequality on quadratic forms confirms $C \preceq \beta(I - FF^*)$ and $FF^* \preceq I$. Finally, the rank of $F$ is bounded by the real dimension of $u^\perp$, which is at most $2r - 2$.
\end{proof}

The defect operator $\Delta = I - FF^*$ characterizes the variance remaining after accounting for stored memory directions. In bounding the source response, one might contemplate directly inverting $\Delta$. However, a naive operator-norm bound on $\Delta^{-1}$ would diverge whenever a memory direction is preserved with near-perfect fidelity. Fortunately, Lemma~\ref{lem:variance-contraction} provides an intrinsic safeguard: the source response in any preserved direction vanishes at precisely the same rate as the variance defect. The following lemma makes this cancellation explicit.

\begin{lemma}[Removing the native directions]
\label{lem:native-schur}
Let $C$ and $\Delta$ be defined as in Lemma~\ref{lem:variance-contraction}. Every Riesz vector $c_j$ lies in the range of $\Delta$. Defining
\begin{equation}
 b_j=\langle c_j,\Delta^+c_j\rangle_\sigma,
 \label{eq:b-j}
\end{equation}
every Hermitian test $X$ satisfies
\begin{align}
 |\ell_j(X)|^2&\le b_j\langle X,\Delta X\rangle_\sigma,
       \label{eq:source-residual-cs}\\
 \sum_jp_jb_j&\le\sum_jp_j\norm{c_j}_\sigma^2+(2r-2)\beta,
       \label{eq:schur-charge}
\end{align}
where $\Delta^+$ denotes the Moore--Penrose pseudoinverse of $\Delta$.
\end{lemma}

\begin{proof}
On the null space $\ker\Delta$, the operator inequality $C \preceq \beta\Delta$ implies that $\langle c_j, x\rangle_\sigma = 0$ for all $x \in \ker\Delta$ and all $j$, since the prior probabilities $p_j$ are strictly positive. Because $\Delta$ differs from the identity by a finite-rank operator $FF^*$, its range is closed and satisfies $\ran\Delta = (\ker\Delta)^\perp$, proving that each $c_j$ belongs to $\ran\Delta$. Applying the Cauchy--Schwarz inequality in the supported geometry of $\Delta^+$~\cite{bhatia1997} to the vectors $\Delta^{+1/2}c_j$ and $\Delta^{1/2}X$ immediately yields~\eqref{eq:source-residual-cs}.

To bound the total effective charge $\sum_j p_j b_j$, we diagonalize the positive operator $FF^*$ on its finite-dimensional support with orthonormal eigenvectors $e_a$ and eigenvalues $0 < t_a \le 1$. On the orthogonal complement of this subspace, $\Delta$ acts as the identity. If an eigenvalue reaches $t_a = 1$, the condition $C \preceq \beta\Delta$ forces every $c_j$ to be orthogonal to $e_a$. For eigenvalues $t_a < 1$, the difference operator $\Delta^+ - I$ acts on $e_a$ with eigenvalue $t_a/(1 - t_a)$. Summing across all modes, we find
\begin{align*}
 \sum_jp_jb_j-\sum_jp_j\norm{c_j}_\sigma^2
 &=\sum_{a:t_a<1}\frac{t_a}{1-t_a}\langle e_a,Ce_a\rangle_\sigma\\
 &\le\sum_{a:t_a<1}\frac{t_a}{1-t_a}\,\beta(1-t_a)\\
 &=\beta\sum_{a:t_a<1}t_a
 \le\beta\rank F\le(2r-2)\beta.
\end{align*}
Notice how the critical factor $(1 - t_a)$ in the denominator is canceled identically by the upper bound $\langle e_a, C e_a\rangle_\sigma \le \beta(1 - t_a)$ from Lemma~\ref{lem:variance-contraction}. This exact algebraic cancellation guarantees that the bound remains completely independent of the condition number of $\Delta$, holding uniformly even when eigenvalues approach unity. This establishes~\eqref{eq:schur-charge}.
\end{proof}

\subsection{Bounding the information from one new copy}

Having bounded the native-memory correction by the dimension factor $(2r-2)\beta$, we now focus on controlling the source contribution $\sum_j p_j \|c_j\|_\sigma^2$. This quantity admits an operational representation that ties the quantum output directly to the probabilities of the classical records. For any positive semi-definite operator $\sigma$, let $\Omega_\sigma$ denote the Jordan multiplication superoperator
\[
 \Omega_\sigma(H)=\frac{\sigma H+H\sigma}{2}.
\]
Its pseudoinverse $\Omega_\sigma^+$ acts in the eigenbasis of $\sigma = \sum_i s_i |i\rangle\langle i|$ by dividing the $(i,k)$ entry of $H$ by $(s_i + s_k)/2$ whenever $s_i + s_k > 0$, and setting it to zero otherwise. The Hermitian operator $C_j = \Omega_\sigma^+(\delta_j)$ is known in quantum estimation as the symmetric logarithmic derivative (SLD) of the perturbation~\cite{braunsteincaves1994,petz1996}. Since $\delta_j$ is supported within the range of $\sigma$, it satisfies $\Omega_\sigma(C_j) = \delta_j$ exactly, and taking traces gives $\Tr(\sigma C_j) = \Tr\delta_j = 0$. Consequently,
\begin{align}
 \langle C_j,X\rangle_\sigma
 &=\tfrac12\Tr[\sigma(C_jX+XC_j)]
           -\Tr(\sigma C_j)\Tr(\sigma X)\nonumber\\
 &=\Tr\!\left[\frac{\sigma C_j+C_j\sigma}{2}\,X\right]
 =\Tr(\delta_jX),\nonumber\\
 \norm{c_j}_\sigma^2
 &=\langle C_j,C_j\rangle_\sigma
 =\Tr(\delta_j C_j)
 =\Tr[\delta_j\Omega_\sigma^+(\delta_j)].
 \label{eq:sld-riesz}
\end{align}
This reveals why the symmetric logarithmic derivative arises naturally: $C_j$ is precisely the Riesz representer of the source fluctuation functional in the variance geometry. The norm $\|c_j\|_\sigma^2$ is evaluated blockwise across classical measurement outcomes with global centering preserved.

\begin{lemma}[Kraus source bound]
\label{lem:kraus-source-bound}
For the complete refined measurement instrument and any unit entering memory vector $u$,
\begin{equation}
 \sum_jp_j\norm{c_j}_\sigma^2\le\zeta(K),\qquad
 \sum_jp_jb_j\le B_K.
 \label{eq:full-local-budget}
\end{equation}
\end{lemma}

\begin{proof}
To evaluate the sum across outcomes, we decompose the measurement operators. For each classical record $a$, we define the effective Kraus operator acting on the incoming source copy and its associated positive operator:
\begin{equation}
 L_a=\frac1{\sqrt D}M_a(u\otimes I_D),\qquad
 S_a=L_a^*L_a.
 \label{eq:kraus-masses}
\end{equation}
Projecting the instrument completeness relation onto the vector $u$ yields
\begin{equation}
 \sum_aS_a=I_D/D,\quad \sum_a\Tr S_a=1,
 \quad \rank S_a\le k_a\le K.
 \label{eq:kraus-mass-completeness}
\end{equation}
The output blocks produced by record $a$ take the form $\sigma_a = L_a L_a^*$ and $\delta_{j,a} = L_a G_j L_a^*$. For any nonzero block, we express $L_a$ via its thin singular-value decomposition $L_a = U \operatorname{diag}(\sqrt{s_i}) W^*$ with $s_i > 0$. In these singular bases, the matrix elements of the perturbation evaluate to $(\delta_{j,a})_{ik} = \sqrt{s_i s_k}(W^* G_j W)_{ik}$. Substituting this into the SLD formula~\eqref{eq:sld-riesz} yields
\begin{align}
 \Tr[\delta_{j,a}\Omega_{\sigma_a}^+(\delta_{j,a})]
 &=\sum_{i,k}\frac{2s_is_k}{s_i+s_k}
                  |(W^*G_jW)_{ik}|^2\nonumber\\
 &\le\sum_{i,k}\sqrt{s_is_k}|(W^*G_jW)_{ik}|^2\nonumber\\
 &=\Tr(\sqrt{S_a}G_j\sqrt{S_a}G_j),
 \label{eq:sld-self-overlap}
\end{align}
where the inequality follows from the arithmetic-geometric mean inequality $2s_i s_k/(s_i + s_k) \le \sqrt{s_i s_k}$.

For every nonzero mass operator $S_a$, we define the normalized state $T_a = \sqrt{S_a}/\sqrt{\Tr S_a}$. By construction, $T_a$ satisfies
\[
 \Tr(T_a^2)=\frac{\Tr S_a}{\Tr S_a}=1,
 \qquad \rank T_a=\rank S_a\le\min(K,D),
\]
confirming that $T_a$ is a valid trial matrix in the definition of the source profile $\zeta(K)$ in~\eqref{eq:zeta-profile}. Summing over all records and source observables, we find
\begin{align}
 \sum_jp_j\|c_j\|_\sigma^2
 &=\sum_a\sum_jp_j
          \Tr[\delta_{j,a}\Omega_{\sigma_a}^+(\delta_{j,a})]\nonumber\\
 &\le\sum_a\sum_jp_j\Tr(\sqrt{S_a}G_j\sqrt{S_a}G_j)\nonumber\\
 &=\sum_{a:\Tr S_a>0}(\Tr S_a)
                  \sum_jp_j\Tr(T_aG_jT_aG_j)\nonumber\\
 &\le\zeta(K)\sum_a\Tr S_a=\zeta(K).
 \label{eq:kraus-full-charge-chain}
\end{align}
Here, the second equality cleanly decouples the classical record weight $\Tr S_a$ from the quantum state geometry $T_a$. Combining this profile bound with Lemma~\ref{lem:native-schur} yields
\[
 \sum_jp_jb_j\le\zeta(K)+(2r-2)\beta
                 \le\zeta(K)+(2K-2)\beta=B_K,
\]
which proves both bounds in~\eqref{eq:full-local-budget}.
\end{proof}

\subsection{A one-step identity for the testing error}

Having bounded the local information budget by $B_K$, we now construct the test function required to eliminate entering-memory errors across steps. Let $H$ represent a Hermitian contraction representing a terminal decoding measurement. We require a regularized test that remains operator-bounded, strictly dominates $H$ with controlled error, and satisfies an exact algebraic inverse identity under isometric dilation. These structural desiderata lead uniquely to a resolvent transformation. For a tuning parameter $\alpha > 1$, we define
\begin{equation}
 f_\alpha(H)=\alpha H(\alpha I-H)^{-1}.
 \label{eq:resolvent-function}
\end{equation}
By spectral functional calculus~\cite{bhatia1997}, this function satisfies
\begin{equation}
 0\preceq f_\alpha(H)-H
   =H^2(\alpha I-H)^{-1}\preceq\frac{I}{\alpha-1},
 \qquad \norm{f_\alpha(H)}_{\mathrm{op}}
       \le\frac\alpha{\alpha-1}.
 \label{eq:resolvent-penalty}
\end{equation}
For any positive entering memory matrix $A$ and recorded Hermitian contraction $H = \bigoplus_a H_a$, we define the one-step test drift by
\begin{equation}
 d_{j,z}(A,H)
 =\Tr[A f_\alpha(\mathcal N_{j,z}^*H)]
       -\Tr[\mathcal N_0(A)f_\alpha(H)].
 \label{eq:one-access-drift}
\end{equation}
This drift is linear in $A$, and its adjoint acts on the full input space $\mathbb C^r$.

\paragraph{Rationale for the resolvent transformation.}
The choice of the operator resolvent $f_\alpha(H) = \alpha H(\alpha I - H)^{-1} = \alpha[(I - H/\alpha)^{-1} - I]$ is governed by three structural requirements that rule out standard information-theoretic divergences:
\begin{enumerate}
 \item \textbf{Branch linearity:} Unlike matrix logarithms or relative entropy, $f_\alpha(H)$ preserves direct sums: $f_\alpha(\bigoplus_a H_a) = \bigoplus_a f_\alpha(H_a)$. It acts linearly across classical measurement branches, avoiding Jensen losses when conditioning on adaptive classical records.
 \item \textbf{Controlled decision penalty:} By spectral calculus, $f_\alpha(H) - H = H^2(\alpha I - H)^{-1} \preceq \frac{I}{\alpha - 1}$. At the terminal decision, this introduces only a small, tunable additive penalty $1/(\alpha - 1)$ against the testing score $2(1-2\delta)$, which is optimized away by choosing $\alpha = 2 + 1/(1-2\delta)$.
 \item \textbf{Operator inversion defect and variance generation:} Under the null Stinespring dilation $V$, the operator inverse identity $(V^* \widehat T V)^{-1} - V^* \widehat X V$ is strictly negative semidefinite. This operator-inversion defect creates the negative quadratic term $-\langle q, \widehat T R q \rangle$ in~\eqref{eq:exact-finite-loss}. It matches the covariance defect $\langle X, \Delta X \rangle_\sigma$ exactly, enabling the lossless completion of the square in~\eqref{eq:finite-square-completion} at finite signal strength $\theta$.
\end{enumerate}

To analyze this drift algebraically, we introduce the null-source Stinespring isometry $V$, the reference action $U_j$, and the accessible lift $\widehat X$:
\begin{align}
 Vw&=\frac1{\sqrt D}\bigoplus_a\sum_{i=1}^D
                 M_a(w\otimes|i\rangle)\otimes|i\rangle,
       \label{eq:structural-dilation}\\
 U_j&=\bigoplus_a(I_{k_a}\otimes G_j^{\mathsf T}),\qquad
 \widehat X=\bigoplus_a(X_a\otimes I_D).
       \label{eq:reference-action}
\end{align}
Here, the auxiliary register $\mathbb C^D$ acts purely as an inaccessible reference. Instrument completeness implies the fundamental identities
\begin{equation}
 V^*V=I_r,\quad V^*U_jV=0,\quad
 \mathcal N_0^*(X)=V^*\widehat XV,\quad
 \mathcal N_{j,z}^*(X)=V^*\widehat X(I+z\theta U_j)V.
 \label{eq:dilation-identities}
\end{equation}
The transpose in~\eqref{eq:reference-action} aligns the matrix elements with the trace adjoint, ensuring that $\widehat X$ commutes with $U_j$ and $\|U_j\|_{\mathrm{op}} \le 1$.

\begin{lemma}[Exact inverse loss]
\label{lem:exact-inverse-loss}
For a unit entering vector $u \in \mathbb C^r$ and any recorded contraction $H$, define
\begin{equation}
 T=I-H/\alpha,\quad X=T^{-1},\quad v=Vu,\quad
 m=1-1/\alpha,
 \quad J(X)=\norm{(I-VV^*)\widehat Xv}^2.
 \label{eq:resolvent-residual}
\end{equation}
Then the one-step drift satisfies
\begin{equation}
 \frac{d_{j,z}(uu^*,H)}\alpha
 \le z\theta\ell_j(X)-(1-\theta)mJ(X),
 \qquad J(X)=\langle X,\Delta X\rangle_\sigma.
 \label{eq:inverse-loss-inequality}
\end{equation}
In particular, completing the square gives the uniform bound
\begin{equation}
 d_{j,z}(uu^*,H)
 \le c_{\alpha,\theta}b_j,
 \qquad
 c_{\alpha,\theta}
   =\frac{\alpha^2\theta^2}{4(1-\theta)(\alpha-1)}.
 \label{eq:finite-drift-bound}
\end{equation}
\end{lemma}

\begin{proof}
Since $H$ is a contraction, the operators $T$ and $X$ satisfy the spectral intervals
\begin{equation}
 (1-1/\alpha)I\preceq T\preceq(1+1/\alpha)I,
 \qquad \frac\alpha{\alpha+1}I\preceq X
                   \preceq\frac\alpha{\alpha-1}I.
 \label{eq:correct-resolvent-interval}
\end{equation}
Letting $R = I + z\theta U_j$ and $G = \mathcal N_{j,z}^*(T) = V^* \widehat T R V$, unitality and positivity imply $G \succeq m I_r$. The factors $\widehat T$ and $R$ commute, and $R \succeq (1-\theta)I$. Setting $b = G^{-1}u$ and $q = \widehat X v - V b$, rewriting the resolvents as $f_\alpha(H) = \alpha(T^{-1} - I)$ and $f_\alpha(\mathcal N_{j,z}^* H) = \alpha(G^{-1} - I_r)$ yields
\begin{equation}
 \frac{d_{j,z}(uu^*,H)}\alpha
 =\langle u,G^{-1}u\rangle-\langle v,\widehat Xv\rangle.
 \label{eq:inverse-drift-expression}
\end{equation}
Remarkably, completing the square around the vector $q$ produces the exact algebraic identity
\begin{equation}
 \frac{d_{j,z}(uu^*,H)}\alpha
 =z\theta\ell_j(X)-\langle q,\widehat TRq\rangle,
 \label{eq:exact-finite-loss}
\end{equation}
which is verified by expanding the quadratic form:
\begin{equation}
\begin{aligned}
 z\theta\ell_j(X)-\frac{d_{j,z}(uu^*,H)}{\alpha}
 &=z\theta\ell_j(X)-\frac{\Tr[uu^*f_\alpha(\mathcal N_{j,z}^*H)]}{\alpha}
       +\frac{\Tr[\mathcal N_0(uu^*)f_\alpha(H)]}{\alpha}\\
 &=z\theta\ell_j(X)-\langle u,(G^{-1}-I_r)u\rangle
       +\langle v,(\widehat X-I)v\rangle\\
 &=\langle v,(I+z\theta U_j)\widehat Xv\rangle
       -2\Re\langle G^{-1}u,u\rangle
       +\langle G^{-1}u,GG^{-1}u\rangle\\
 &=\langle\widehat Xv,\widehat TR\widehat Xv\rangle
       -2\Re\langle G^{-1}u,V^*(I+z\theta U_j)v\rangle+\langle G^{-1}u,GG^{-1}u\rangle\\
 &=\langle\widehat Xv,\widehat TR\widehat Xv\rangle
       -2\Re\langle VG^{-1}u,\widehat TR\widehat Xv\rangle
       +\langle VG^{-1}u,\widehat TRVG^{-1}u\rangle\\
 &=\langle\widehat Xv,\widehat TR(\widehat Xv-VG^{-1}u)\rangle
       -\langle VG^{-1}u,\widehat TR(\widehat Xv-VG^{-1}u)\rangle\\
 &=\langle\widehat XVu-VG^{-1}u,\widehat T(I+z\theta U_j)(\widehat XVu-VG^{-1}u)\rangle\\
 &=\langle\widehat Xv-VG^{-1}u,
       \widehat T(I+z\theta U_j)(\widehat Xv-VG^{-1}u)\rangle.
\end{aligned}
\label{eq:inverse-loss-expanded-square}
\end{equation}
Because $\widehat T \succeq mI$ and $R \succeq (1-\theta)I$, the quadratic term satisfies
\[
 \langle q,\widehat TRq\rangle
 \ge(1-\theta)m\norm q^2
 \ge(1-\theta)m\norm{(I-VV^*)\widehat Xv}^2,
\]
which establishes the first inequality in~\eqref{eq:inverse-loss-inequality}.

We now identify the residual $J(X)$. Using the dilation relations~\eqref{eq:dilation-identities}, we compute
\begin{align*}
 J(X)&=\langle\widehat Xv,\widehat Xv\rangle
                -\langle\widehat Xv,VV^*\widehat Xv\rangle\\
 &=\Tr(\sigma X^2)-\norm{V^*\widehat Xv}^2\\
 &=\Tr(\sigma X^2)-\norm{\mathcal N_0^*(X)u}^2\\
 &=\Tr(\sigma X^2)-[\Tr(\sigma X)]^2
           -\norm{Q_u\mathcal N_0^*(X)u}^2\\
 &=\norm X_\sigma^2-\norm{F^*X}^2
 =\langle X,\Delta X\rangle_\sigma,
\end{align*}
where the middle step uses the fact that $\langle u, \mathcal N_0^*(X)u\rangle = \Tr(\sigma X)$ is real. Thus, the residual from the dilation identity matches the variance defect operator $\Delta$ exactly. Applying the Cauchy--Schwarz bound~\eqref{eq:source-residual-cs} reveals that the source response satisfies $|\ell_j(X)| \le \sqrt{b_j J(X)}$. Writing $y = \sqrt{J(X)}$ and $a_0 = (1-\theta)m > 0$, we find
\begin{align}
 d_{j,z}(uu^*,H)
 &\le\alpha\bigl[\theta\sqrt{b_j}\,y-a_0y^2\bigr]\nonumber\\
 &=\frac{\alpha\theta^2b_j}{4a_0}
       -\alpha a_0\left(y-\frac{\theta\sqrt{b_j}}{2a_0}\right)^2\nonumber\\
 &\le\frac{\alpha\theta^2b_j}{4(1-\theta)m}
 =\frac{\alpha^2\theta^2b_j}{4(1-\theta)(\alpha-1)}.
 \label{eq:finite-square-completion}
\end{align}
Dropping the negative square and substituting $m = (\alpha-1)/\alpha$ establishes~\eqref{eq:finite-drift-bound}.
\end{proof}

\begin{proposition}[One-copy prior-average bound]
\label{prop:one-copy-structural}
For any positive entering memory matrix $A$ on the full memory register and any collection of label-dependent recorded Hermitian contractions $H_{j,z}$,
\begin{equation}
 \sum_j\frac{p_j}{2}\sum_{z=\pm1}d_{j,z}(A,H_{j,z})
 \le c_{\alpha,\theta}B_K\Tr A.
 \label{eq:one-copy-structural}
\end{equation}
\end{proposition}

\begin{proof}
For any pure input $A = u u^*$ with $\|u\| = 1$, the charges $b_j$ are determined entirely by the measurement instrument and entering vector $u$, independently of the label $z$. Combining Lemmas~\ref{lem:kraus-source-bound} and~\ref{lem:exact-inverse-loss} immediately proves~\eqref{eq:one-copy-structural} for pure inputs. For a general mixed state $A = \sum_i a_i u_i u_i^*$ with $a_i \ge 0$, the drift functional $d_{j,z}(A, H_{j,z})$ is linear in $A$. Summing over components gives
\begin{align}
 \sum_j\frac{p_j}{2}\sum_z d_{j,z}(A,H_{j,z})
 &=\sum_i a_i\left[\sum_j\frac{p_j}{2}\sum_z
                       d_{j,z}(u_iu_i^*,H_{j,z})\right]\nonumber\\
 &\le\sum_i a_i c_{\alpha,\theta}B_K
 =c_{\alpha,\theta}B_K\Tr A.
 \label{eq:full-input-averaging}
\end{align}
Because this decomposition is a mathematical property of the matrix $A$ in the physical memory space, it holds without requiring any ancillary purification registers.
\end{proof}

\subsection{Extending the bound to an adaptive experiment}

With the local bound established uniformly across all entering states and tests, we now connect these one-step bounds across the entire adaptive protocol. Our goal is to select tests whose structure cancels the entering memory of the alternative model. The natural objects are the terminal measurement effects pulled backward along the alternative measurement tree.

\begin{proof}[Proof of Theorem~\ref{thm:structural} for finite records]
Let $\lambda = (j,z)$ denote the alternative hypothesis, drawn with prior $p_j/2$. At time step $t$, let $h$ index the full sequence of classical measurement records, and let $A_{t,h}$ and $B_{\lambda,t,h}$ denote the unnormalized memory states under the null and alternative sources, respectively. Under both hypotheses, the receiver applies the identical instrument at history $h$, giving the forward recursion
\begin{equation}
 A_{t+1,ha}=\mathcal N_{0,h,a}(A_{t,h}),\qquad
 B_{\lambda,t+1,ha}=\mathcal N_{\lambda,h,a}(B_{\lambda,t,h}).
 \label{eq:structural-history-recursion}
\end{equation}
Because initialization is independent of the source, $A_{0,\emptyset} = B_{\lambda,0,\emptyset}$. Instrument trace-preservation implies conservation of total probability:
\begin{equation}
 \sum_h\Tr A_{t,h}=1,\qquad
 \sum_h\Tr B_{\lambda,t,h}=1.
 \label{eq:structural-history-mass}
\end{equation}
Moreover, since $(1-\theta)\tau \preceq \rho_\lambda \preceq (1+\theta)\tau$, induction ensures the support dominance
\begin{equation}
 (1-\theta)^tA_{t,h}\preceq B_{\lambda,t,h}
              \preceq(1+\theta)^tA_{t,h}.
 \label{eq:null-history-support}
\end{equation}
Any history with zero null probability therefore has zero alternative probability, eliminating the need to condition on history likelihoods.

At the final depth $N$, pulling the decoding decision event $E$ backward produces a measurable effect $0 \preceq F_h \preceq I$. We set $H_{\lambda,N,h} = 2F_h - I$ and define the backward tests recursively by
\begin{equation}
 H_{\lambda,t,h}
 =\mathcal N_{\lambda,h}^*
       \bigl(\bigoplus_aH_{\lambda,t+1,ha}\bigr).
 \label{eq:backward-event-tests}
\end{equation}
By unital positivity, each $H_{\lambda,t,h}$ is a Hermitian contraction. We then construct the global tracking score
\begin{equation}
 S_t=\mathbb E_\lambda\sum_h
       \Tr\!\left[B_{\lambda,t,h}H_{\lambda,t,h}
                  -A_{t,h}f_\alpha(H_{\lambda,t,h})\right].
 \label{eq:actual-history-score}
\end{equation}
At $t=0$, identical initialization and~\eqref{eq:resolvent-penalty} guarantee $S_0 \le 0$. Under the forward-backward duality between~\eqref{eq:structural-history-recursion} and~\eqref{eq:backward-event-tests}, the alternative-state terms satisfy
\[
 \sum_a\Tr[B_{\lambda,t+1,ha}H_{\lambda,t+1,ha}]
 =\Tr[B_{\lambda,t,h}H_{\lambda,t,h}].
\]
Denoting $\mathbf H_{\lambda,h} = \bigoplus_a H_{\lambda,t+1,ha}$, the score increment between steps expands as
\begin{align}
 S_{t+1}-S_t
 &=\mathbb E_\lambda\sum_h\Bigl\{
      \sum_a\Tr[B_{\lambda,t+1,ha}H_{\lambda,t+1,ha}]
            -\Tr[B_{\lambda,t,h}H_{\lambda,t,h}]\nonumber\\
 &\hspace{30mm}+\Tr[A_{t,h}f_\alpha(H_{\lambda,t,h})]
       -\sum_a\Tr[A_{t+1,ha}f_\alpha(H_{\lambda,t+1,ha})]\Bigr\}\nonumber\\
 &=\mathbb E_\lambda\sum_h\Tr\Bigl[A_{t,h}
        \bigl\{f_\alpha(\mathcal N_{\lambda,h}^*\mathbf H_{\lambda,h})
                  -\mathcal N_{0,h}^*(f_\alpha(\mathbf H_{\lambda,h}))\bigr\}\Bigr]
          \nonumber\\
 &=\mathbb E_\lambda\sum_h
           d_{\lambda,h}(A_{t,h},\mathbf H_{\lambda,h})\nonumber\\
 &\le c_{\alpha,\theta}B_K\sum_h\Tr A_{t,h}
 =c_{\alpha,\theta}B_K.
 \label{eq:actual-history-increment}
\end{align}
Notice how the alternative state terms cancel identically in the first line. The remaining drift is bounded by Proposition~\ref{prop:one-copy-structural}, followed by probability conservation under the null hypothesis. Telescoping this increment over $N$ rounds yields
\begin{equation}
 S_N\le N c_{\alpha,\theta}B_K.
 \label{eq:actual-history-telescope}
\end{equation}

On the other hand, setting $s = 1 - 2\delta > 0$, the testing error assumption implies that the terminal decision separates the hypotheses:
\[
 \mathbb E_\lambda\sum_h
   \Tr[(B_{\lambda,N,h}-A_{N,h})(2F_h-I)]\ge2s.
\]
Applying the resolvent bound~\eqref{eq:resolvent-penalty} at the final step gives
\begin{equation}
 S_N\ge2s-\frac1{\alpha-1}.
 \label{eq:terminal-event-score}
\end{equation}
Combining the upper and lower bounds on $S_N$ in~\eqref{eq:actual-history-telescope} and~\eqref{eq:terminal-event-score}, we obtain
\begin{equation}
 N\ge\frac{4(1-\theta)[2s(\alpha-1)-1]}
                  {\alpha^2\theta^2B_K}.
 \label{eq:preoptimized-structural}
\end{equation}
Maximizing this expression over $\alpha > 1$ via the elementary identity
\[
 \frac{s^2}{2s+1}
 -\frac{2s(\alpha-1)-1}{\alpha^2}
 =\frac{[s\alpha-(2s+1)]^2}{(2s+1)\alpha^2}\ge0
\]
shows that the optimal choice is $\alpha = 2 + 1/s$. Substituting this value into~\eqref{eq:preoptimized-structural} and using $2s+1 = 3 - 4\delta$ yields~\eqref{eq:structural-testing}. Any stopping time bounded by $N$ can be padded to depth $N$ by idling the memory, preserving the bound.
\end{proof}

\begin{proof}[Proof of Corollary~\ref{cor:estimation}]
To establish the copy complexity of property estimation, we map the simultaneous estimation problem onto hypothesis testing. For any parameter $\theta$ satisfying $2\epsilon < \theta < 1$, we define the test event
\[
 E_\theta=\{\max_i|\widehat\mu_i|>\theta/2\}.
\]
Under the null state $\tau$, all target expectations $\Tr(\tau G_i) = 0$ vanish. Under the alternative state $\rho_{j,z}$, the orthogonality of generators~\eqref{eq:orthogonal-reflections} implies $\Tr(\rho_{j,z}G_i) = z\theta\mathbf1_{i=j}$. Consequently, the simultaneous accuracy guarantee $|\widehat\mu_i - \Tr(\rho G_i)| \le \epsilon < \theta/2$ prevents $E_\theta$ from occurring under $\tau$ while forcing it to occur under every alternative $\rho_{j,z}$. Both testing error probabilities are therefore bounded by $\delta \le 1/3$. Applying Theorem~\ref{thm:structural} with $s = 1 - 2\delta = 1/3$ yields
\[
 N\ge\frac{4(1-\theta)}{15\theta^2B_K}.
\]
Since this holds for all $\theta > 2\epsilon$, taking the limit $\theta \downarrow 2\epsilon$ gives $N \ge (1 - 2\epsilon)/(15\epsilon^2 B_K)$. Finally, using $\epsilon \le 1/8$ guarantees $1 - 2\epsilon \ge 3/4$, proving~\eqref{eq:structural-estimation}.
\end{proof}

\subsection{Continuous outcomes and adaptive measurements}
\label{subsec:measurable-structural}

The local charge in Lemma~\ref{lem:kraus-source-bound} depends only on total measurement mass, suggesting that the same lower bound should hold for continuous measurement alphabets. The technical requirement is to represent the adaptive experiment using jointly measurable operator densities so that the instrument and backward tests remain well-behaved along arbitrary measurement histories. We now provide this construction explicitly. All record spaces carry their stated sigma algebras, and successive histories use product sigma algebras. A measurable adaptive instrument means that for each history $h$, $S \mapsto \mathcal E_h(S)$ is a finite-dimensional completely positive instrument and every entry of its Choi matrix $\Gamma_h(S)$ is measurable in $h$ for each measurable next-record event $S$.

\paragraph{One measurable complete instrument.}
Let the entering memory space be embedded into $\mathbb C^K$. At an input dimension $d = rD$, we represent a completely positive instrument $\mathcal E$ by its Choi measure $\Gamma$. The scalar measure $\mu(S) = d^{-1}\Tr\Gamma(S)$ dominates every matrix entry of $\Gamma$. Applying the Radon--Nikodym theorem entrywise~\cite{kallenberg2021} yields a measurable matrix field $J(a)$ satisfying
\begin{equation}
 \Gamma(S)=\int_S J(a)\,d\mu(a),\qquad J(a)\succeq0
 \quad\text{almost everywhere}.
 \label{eq:single-choi-density}
\end{equation}
Reshaping the columns of the positive square root $J(a)^{1/2}$ produces a measurable family of Kraus operators $M_{a,l}$ such that
\begin{equation}
 \mathcal E_a(Z)=\sum_{l=1}^{Kd}M_{a,l}ZM_{a,l}^*,\qquad
 \int\sum_lM_{a,l}^*M_{a,l}\,d\mu(a)=I_d.
 \label{eq:single-measurable-completeness}
\end{equation}
Treating $l$ as an auxiliary classical record expands the outcome measure to $\mu \otimes \#$, where $\#$ is the counting measure.

Replacing finite sums with integrals and dilations with Hilbert direct integrals, the variance contraction in Lemma~\ref{lem:variance-contraction} extends directly to bounded measurable test fields. Moreover, since $\Omega_{\sigma_a}^+(Y) = 2\int_0^\infty e^{-u\sigma_a} P_a Y P_a e^{-u\sigma_a}\,du$ is a positive linear map sending $\sigma_a \pm \delta_{j,a} \succeq 0$ to $P_a \pm C_{j,a} \succeq 0$, the symmetric logarithmic derivative field $C_{j,a} = \Omega_{\sigma_a}^+(\delta_{j,a})$ is Borel measurable and satisfies
\begin{equation}
 -P_a\preceq C_{j,a}\preceq P_a,\qquad P_a=P_{\supp\sigma_a},
 \label{eq:bounded-sld-field}
\end{equation}
guaranteeing finite Riesz norms. The remainder of the local proof carries over without alteration, confirming Proposition~\ref{prop:one-copy-structural} for measurable instruments.

\paragraph{Joint densities along arbitrary measurable histories.}
To construct a globally measurable realization of the adaptive protocol, we embed all memory registers into $\mathbb C^K$, extending smaller subspaces via trace-discard channels. Given a base measure $\nu$ on parent histories, we define the product measure $\eta$ and joint Choi measure $\widetilde\Gamma$:
\begin{equation}
 \eta(A)=\int\mu_h(A_h)\,d\nu(h),\qquad
 \widetilde\Gamma(A)=\int\Gamma_h(A_h)\,d\nu(h).
 \label{eq:joint-choi-measures-new}
\end{equation}
By the Radon--Nikodym theorem, there exists a jointly measurable positive matrix density $J(h,a)$ such that
\begin{equation}
 \widetilde\Gamma(A)=\int_AJ(h,a)\,d\eta(h,a).
 \label{eq:joint-choi-density-new}
\end{equation}
Taking partial traces verifies instrument completeness almost everywhere:
\begin{equation}
 \int\Tr_{\mathrm{out}}J(h,a)\,d\mu_h(a)=I_d
 \quad\text{for }\nu\text{-almost every }h.
 \label{eq:joint-completeness-new}
\end{equation}
This density represents the actual experiment for every entering matrix field: defining $\Phi_{h,a}(Z)=\Tr_{\mathrm{in}}[(Z^{\mathsf T}\otimes I_K)J(h,a)]$, every positive measurable $Z(h)$ with $\int\Tr Z(h)\,d\nu(h)<\infty$ satisfies
\begin{equation}
 \int\mathcal E_h(A_h)(Z(h))\,d\nu(h)
 =\int_A\Phi_{h,a}(Z(h))\,d\eta(h,a).
 \label{eq:joint-entering-field-identity}
\end{equation}
The $q=Kd$ columns of $J(h,a)^{1/2}$ yield jointly measurable Kraus fields $M_{h,a,l}$, with the dominating measure
\begin{equation}
 \nu'(dh,da,dl)=\nu(dh)\mu_h(da)\,\#(dl),
 \qquad l\in\{1,\ldots,q\}.
 \label{eq:next-history-measure}
\end{equation}
This establishes the conservation of probability
\begin{equation}
 \int\Tr A_{t,h}\,d\nu_t(h)
 =\int\Tr B_{\lambda,t,h}\,d\nu_t(h)=1.
 \label{eq:measurable-mass-conservation}
\end{equation}

\paragraph{The integrated backward-event sum.}
Pulling the bounded measurable terminal effect backward through these integral kernels preserves measurability and contractivity. Forward-backward duality ensures that the alternative state terms cancel identically, leaving the integral drift bounded by $c_{\alpha,\theta}B_K$. Optimizing $\alpha$ completes the proof of Theorem~\ref{thm:structural} for general continuous measurement protocols.

\subsection{Other output and error guarantees}
\label{subsec:other-error-guarantees}

The structural lower bound extends naturally to estimation tasks where the target properties do not coincide directly with the perturbation probes. Let $A_1,\ldots,A_L$ be arbitrary Hermitian observables, let $y(\rho) = (\Tr(\rho A_\ell))_{\ell=1}^L$, and define the reference vectors
\begin{equation}
 b=y(\tau),\qquad
 v_j=\bigl(\Tr(G_jA_\ell)/D\bigr)_{\ell=1}^L.
 \label{eq:output-signal-vectors}
\end{equation}
Under the perturbed states, the expectation vectors shift to $y(\rho_{j,z}) = b + z\theta v_j$.

\begin{corollary}[Norm-accuracy transfer]
\label{cor:norm-transfer}
Let $\|\cdot\|_{\mathsf a}$ be an arbitrary norm on $\mathbb R^L$ and define the signal separation $\Delta = \min_j \|v_j\|_{\mathsf a} > 0$. Suppose an adaptive receiver guarantees
\[
 \Pr_\rho\{\|\widehat y-y(\rho)\|_{\mathsf a}\le\epsilon\}
 \ge2/3\qquad\text{for every state }\rho.
\]
If $0 < \epsilon < \Delta/2$, then the required copy complexity satisfies
\begin{equation}
 N\ge\frac{(1-2\epsilon/\Delta)\Delta^2}
                  {15\epsilon^2B_K}.
 \label{eq:norm-transfer-bound}
\end{equation}
In particular, for $0 < \epsilon \le \Delta/8$, $N \ge \Delta^2/(20\epsilon^2 B_K)$. The profile and Gram norm are those of the centered contraction probes $G_j$ and their chosen prior $p$.
\end{corollary}

\begin{proof}
For any threshold parameter satisfying $2\epsilon/\Delta < \theta < 1$, we define the decision event $E_\theta = \{\|\widehat y - b\|_{\mathsf a} > \theta\Delta/2\}$. Under the reference state $\tau$, the accuracy guarantee implies $\|\widehat y - b\|_{\mathsf a} \le \epsilon < \theta\Delta/2$, so $\Pr_\tau(E_\theta) \le 1/3$. Under any alternative state $\rho_{j,z}$, the reverse triangle inequality ensures
\[
 \|\widehat y-b\|_{\mathsf a}
 \ge\theta\|v_j\|_{\mathsf a}-\epsilon
 \ge\theta\Delta-\epsilon>\theta\Delta/2,
\]
which implies that $\Pr_{\rho_{j,z}}(E_\theta) \ge 2/3$. Applying Theorem~\ref{thm:structural} with error bound $\delta = 1/3$ yields $N \ge 4(1 - \theta)/(15\theta^2 B_K)$. Taking the limit $\theta \downarrow 2\epsilon/\Delta$ gives~\eqref{eq:norm-transfer-bound}, and the condition $\epsilon \le \Delta/8$ ensures $1 - 2\epsilon/\Delta \ge 3/4$.
\end{proof}

For simultaneous signed accuracy, we select the max norm $\|\cdot\|_\infty$, yielding the separation parameter $\Delta = \min_j \max_\ell |\Tr(G_j A_\ell)|/D$. For instance, when $A_\ell = s_\ell P_{k(\ell)}$ with non-zero scalar weights $s_\ell$ and mutually orthogonal traceless reflections $P_k$, taking distinct $P_k$ as probes yields $\Delta = \min_k \max_{\ell : k(\ell) = k} |s_\ell|$, directly covering scaled or repeated queries. 

As a non-reflection example, suppose $D$ is divisible by four and the requested matrices $A_j=P_j$ are projections of rank $D/4$, with arbitrary overlaps between their ranges. Choosing $G_j=(4P_j-I)/3$ yields traceless contractions with eigenvalues $1$ and $-1/3$. Because $\Tr(G_jP_j)/D=1/4$, we have $\Delta\ge1/4$ in the max norm. For $0<\epsilon\le1/32$, Corollary~\ref{cor:norm-transfer} establishes $N\ge1/(320\epsilon^2B_K)$, governed directly by the profile $\zeta(K)$ and Gram norm $\beta$ of these projection probes.

Similarly, setting $\|x\|_{\mathsf a} = L^{-1}\|x\|_1$ captures average absolute error, while $\|x\|_{\mathsf a} = L^{-1/2}\|x\|_2$ covers root-mean-square error. For $M$ orthogonal reflections $A_j = G_j$, the corresponding signal separations are $1/M$ and $1/\sqrt M$, respectively, scaling the required copy budget by $\Delta^2$. In the max norm, setting $\Delta = 1$ recovers Corollary~\ref{cor:estimation} exactly.

\section{Bounding the source profile}
\label{sec:computation}
The general theorem reduces the copy lower bound to the known family
through $\beta$ and $\zeta(K)$. The Gram norm $\beta$ is an eigenvalue
calculation. To bound the rank-constrained profile $\zeta(K)$, we use
estimates from eigenvalues, pure states and matrix moments. These analytic
bounds provide the link to the Pauli and Majorana applications below.

Appendix~\ref{app:computational-certificates} develops complementary
optimization programs: a rank-sensitive semidefinite relaxation,
symmetry-reduced linear programs for the two application families, and a
convergent moment hierarchy for general observables. It also gives exact
certificate checks and states the computational costs.

\subsection{Bounds from eigenvalues and matrix moments}
\label{sec:certificates}
The structural lower bound in Theorem~\ref{thm:structural} and Corollary~\ref{cor:estimation} is governed operationally by the source profile $\zeta(K)$. Because the rank constraint $\rank(T) \le \min\{K,D\}$ renders~\eqref{eq:zeta-profile} a nonconvex optimization problem, exact analytical solutions are generally out of reach for arbitrary observable sets. In practice, however, one does not need to solve this optimization exactly: any rigorous upper bound on $\zeta(K)$ translates directly into a valid copy lower bound.

In this section, we present three complementary analytic certificates to bound $\zeta(K)$ from above without solving the nonconvex problem: a Rayleigh-quotient relaxation, a Gaussian fourth-moment decoupling argument, and a spectral moment interpolation. Setting $q=\min\{K,D\}$, we define the source averaging channel and its Choi-matrix vectorization:
\begin{equation}
 \Phi(X)=\sum_jp_jG_jXG_j,\qquad
 S=\sum_jp_j(G_j^{\mathsf T}\otimes G_j).
 \label{eq:certificate-source-map}
\end{equation}
Under standard column-stacking vectorization, $\operatorname{vec}(\Phi(X))=S\operatorname{vec}(X)$. The superoperator $\Phi$ is completely positive and Hilbert--Schmidt self-adjoint, while the associated Hermitian matrix $S$ may in general have negative eigenvalues.

\begin{proposition}[Standard profile certificates]
\label{prop:profile-certificates}
For the profile data in Definition~\ref{def:profile}, the quantity $\zeta(K)$ satisfies
\begin{equation}
 \zeta(K)\le\lambda_{\max}(S)\le1.
 \label{eq:spectral-profile-certificate}
\end{equation}
Furthermore, if $c_1\ge\zeta(1)$ provides a verified certificate for all pure states, then
\begin{equation}
 \zeta(K)\le(q+1)c_1.
 \label{eq:pure-profile-certificate}
\end{equation}
Finally, for any integer $r\ge1$, an even-moment bound on pure states
\begin{equation}
 \Tr\bigl(\rho\Phi^{2r}(\rho)\bigr)\le h_r
 \quad\text{for every pure state }\rho
 \label{eq:generic-moment-certificate}
\end{equation}
guarantees that
\begin{equation}
 \zeta(K)\le(qh_r)^{1/(2r)}.
 \label{eq:moment-profile-certificate}
\end{equation}
Any collection of available certificates can be combined by taking their minimum.
\end{proposition}

\begin{proof}
For any feasible trial matrix $T$ satisfying $T \succeq 0$ and $\Tr(T^2)=1$, vectorization preserves the Hilbert--Schmidt norm, giving
\[
 \Tr(T\Phi(T))=\operatorname{vec}(T)^*S\operatorname{vec}(T).
\]
The standard Rayleigh quotient theorem~\cite{bhatia1997} immediately bounds this quadratic form by the maximum eigenvalue $\lambda_{\max}(S)$. Because $\|G_jXG_j\|_{\mathrm{HS}} \le \|X\|_{\mathrm{HS}}$, the triangle inequality gives $\|\Phi\|_{\mathrm{HS}\to\mathrm{HS}}\le1$, ensuring $\lambda_{\max}(S)\le1$ and proving~\eqref{eq:spectral-profile-certificate}.

To establish the pure-state transfer bound~\eqref{eq:pure-profile-certificate}, our strategy is to decouple the matrix product in $\Tr(TG_jTG_j)$ into a scalar fourth moment using complex Gaussian random vectors. Decomposing $T$ into its spectral representation $T=\sum_{a=1}^k t_a|a\rangle\langle a|$ with $t_a > 0$, $\sum_a t_a^2=1$, and $k \le q$, we introduce independent standard complex Gaussian variables $g_a \sim \mathcal{CN}(0,1)$ and define the random pure state vector $x=\sum_a\sqrt{t_a}g_a|a\rangle$. 

By Isserlis' Gaussian pairing formula~\cite{isserlis1918}, the fourth-order Gaussian moments
factor into pairings:
$\mathbb E[\overline g_ag_b\overline g_cg_d] = \mathbf1_{a=b}\mathbf1_{c=d}+\mathbf1_{a=d}\mathbf1_{b=c}$.
Evaluating the expectation of $(x^* G_j x)^2$ yields
\begin{align*}
 \mathbb E\sum_jp_j(x^*G_jx)^2
 &=\Tr(T\Phi(T))+\sum_jp_j[\Tr(TG_j)]^2,\\
 \mathbb E\|x\|^4&=(\Tr T)^2+\Tr(T^2).
\end{align*}
For any realization $x\ne0$, the pure-state certificate $\zeta(1) \le c_1$ ensures
that $\sum_j p_j(x^*G_jx)^2 \le c_1 \|x\|^4$. Dropping the nonnegative squared-trace sum
$\sum_j p_j [\Tr(TG_j)]^2 \ge 0$ and applying Cauchy--Schwarz to the eigenvalues
$(\Tr T)^2 = (\sum_a t_a)^2 \le k \sum_a t_a^2 \le q$, we obtain the exact chain:
\begin{equation}
\begin{aligned}
 &\Tr\bigl(T\Phi(T)\bigr)
 =\sum_{j=1}^{M}p_j\Tr\bigl(TG_jTG_j\bigr)
 =\sum_{j=1}^{M}p_j\sum_{a=1}^{k}\sum_{b=1}^{k}t_at_b
   \langle a|G_j|b\rangle\langle b|G_j|a\rangle
 \\[2pt]
 &\quad=\sum_{j=1}^{M}p_j\sum_{a=1}^{k}\sum_{b=1}^{k}
   \sum_{c=1}^{k}\sum_{d=1}^{k}\sqrt{t_at_bt_ct_d}\,
   \mathbf1_{a=d}\mathbf1_{b=c}\,
   \langle a|G_j|b\rangle\langle c|G_j|d\rangle
 \\[2pt]
 &\quad=\sum_{j=1}^{M}p_j\sum_{a=1}^{k}\sum_{b=1}^{k}
   \sum_{c=1}^{k}\sum_{d=1}^{k}\sqrt{t_at_bt_ct_d}\,
   \Bigl(\mathbb E\bigl[\overline g_ag_b\overline g_cg_d\bigr]
     -\mathbf1_{a=b}\mathbf1_{c=d}\Bigr)
   \langle a|G_j|b\rangle\langle c|G_j|d\rangle
 \\[2pt]
 &\quad=\mathbb E\sum_{j=1}^{M}p_j
   \Bigl(\sum_{a=1}^{k}\sum_{b=1}^{k}
     \sqrt{t_at_b}\,\overline g_ag_b\langle a|G_j|b\rangle\Bigr)^2
   -\sum_{j=1}^{M}p_j
   \Bigl(\sum_{a=1}^{k}t_a\langle a|G_j|a\rangle\Bigr)^2
 \\[2pt]
 &\quad=\mathbb E\sum_{j=1}^{M}p_j\bigl(x^*G_jx\bigr)^2
   -\sum_{j=1}^{M}p_j\bigl[\Tr(TG_j)\bigr]^2
 \le\mathbb E\sum_{j=1}^{M}p_j\bigl(x^*G_jx\bigr)^2
 \\[2pt]
 &\quad\le c_1\,\mathbb E\|x\|^4
 =c_1\,\mathbb E\Bigl(\sum_{a=1}^{k}t_a|g_a|^2\Bigr)^2
 =c_1\sum_{a=1}^{k}\sum_{c=1}^{k}t_at_c\,
   \mathbb E\bigl[|g_a|^2|g_c|^2\bigr]
 \\[2pt]
 &\quad=c_1\Bigl(\sum_{a=1}^{k}\sum_{c=1}^{k}t_at_c
   +\sum_{a=1}^{k}t_a^2\Bigr)
 =c_1\Bigl[\Bigl(\sum_{a=1}^{k}t_a\Bigr)^2+\sum_{a=1}^{k}t_a^2\Bigr]
 =c_1\bigl[(\Tr T)^2+\Tr(T^2)\bigr]
 \\[2pt]
 &\quad\le c_1\Bigl[\Bigl(\sum_{a=1}^{k}1^2\Bigr)
     \Bigl(\sum_{a=1}^{k}t_a^2\Bigr)+\sum_{a=1}^{k}t_a^2\Bigr]
 =c_1\bigl[k\Tr(T^2)+\Tr(T^2)\bigr]
 \\[2pt]
 &\quad=c_1(k+1)
 \le c_1(q+1).
\end{aligned}
\label{eq:pure-profile-complete-chain}
\end{equation}
Taking the supremum over all feasible $T$ establishes~\eqref{eq:pure-profile-certificate}.

Finally, for the moment interpolation~\eqref{eq:moment-profile-certificate}, let $\mu_T$ denote the spectral measure of the self-adjoint superoperator $\Phi$ with respect to the state $T$. Applying H\"older's inequality for matrix moments~\cite{blm2013} yields
\[
 \Tr(T\Phi(T))\le\int|s|\,d\mu_T(s)
 \le\left[\Tr(T\Phi^{2r}(T))\right]^{1/(2r)}.
\]
Writing $\rho_a=|a\rangle\langle a|$ in the spectral decomposition of $T$, self-adjointness and the pure-state hypothesis~\eqref{eq:generic-moment-certificate} give
\[
 \left[\Tr(T\Phi^{2r}(T))\right]^{1/2}
 =\|\Phi^r(T)\|_{\mathrm{HS}}
 \le\sum_at_a\|\Phi^r(\rho_a)\|_{\mathrm{HS}}
 \le\sqrt{h_r}\sum_at_a\le\sqrt{qh_r}.
\]
Combining these two inequalities and maximizing over feasible $T$ proves~\eqref{eq:moment-profile-certificate}.
\end{proof}

Whenever an explicit semidefinite certificate $S \preceq a I_{D^2}$ is available, it guarantees the profile bound $B_K \le a + (2K-2)\beta$. For reflection observables, $\Phi(I)=I$, so the naive spectral relaxation~\eqref{eq:spectral-profile-certificate} yields $\lambda_{\max}(S)=1$. In that regime, the rank-sensitive pure-state and moment certificates developed here provide much sharper scaling, as demonstrated in our applications to Pauli and Majorana systems below.

\section{Pauli shadow tomography with persistent quantum memory}
\label{sec:pauli}
Theorem~\ref{thm:pauli} and Corollary~\ref{cor:pauli-asymptotic} were established in
Section~\ref{subsec:main-pauli}. In this section, we present their complete proofs, evaluating the profile bound for Pauli observables and comparing it directly with the operational performance of unentangled measurements.

Let $\mathcal P_n=\{I,X,Y,Z\}^{\otimes n}$ denote the set of all $4^n$ $n$-qubit Pauli strings, normalized under the Hilbert--Schmidt inner product via $\Tr(P^2)=D=2^n$. The Pauli weight $\mathrm{wt}(P)$ is the number of nonidentity tensor factors. For $1\le w\le n$, we let $\mathcal P_{n,w}$ denote the set of all $M=3^w\binom nw$ Pauli strings of weight $w$. These observables are traceless, Hermitian, and square to the identity. The normalizing factor $\beta$ in Definition~\ref{def:profile} evaluates explicitly to
\begin{equation*}
 \beta=\lambda_{\max}\left(\frac1M\sum_{P\in\mathcal P_{n,w}}
        \frac{\operatorname{vec}(P)\operatorname{vec}(P)^*}{D}\right)
      =\frac1M,
\end{equation*}
which follows from tracial orthogonality under the normalized inner product $\Tr(PQ)/D=\mathbf1_{P=Q}$. Theorem~\ref{thm:structural} and Corollary~\ref{cor:estimation} therefore apply directly with this exact value of $\beta$.

\subsection{Moments of the Pauli averaging map}

To evaluate the source profile $\zeta(K)$ for weight-$w$ observables, we analyze the positive superoperator generated by single-qubit Pauli conjugations:
\begin{equation*}
 A=\sum_{i=1}^n\sum_{\sigma\in\{X,Y,Z\}}\operatorname{Ad}_{\sigma_i},
 \qquad \operatorname{Ad}_{U}(V)=UVU^\dagger.
\end{equation*}
Each conjugation superoperator is self-adjoint with respect to the Hilbert--Schmidt inner product $\langle V,W\rangle_{\mathrm{HS}}=\Tr(V^\dagger W)$. Furthermore, all summands commute as superoperators, because the phase factors arising from Pauli multiplication cancel under conjugation. On any Pauli string of weight $k$, $A$ has eigenvalue $3n-4k$, which implies the uniform lower spectral bound $A\succeq-n\,\mathrm{id}$.

\begin{lemma}[Uniform even moments]
\label{lem:pauli-moment}
For every integer $p\ge1$ and every pure state $\rho$ on $n$ qubits,
\begin{equation}
 \langle\rho,A^{2p}(\rho)\rangle_{\mathrm{HS}}
 \le \bigl(n+\sqrt{8np}\bigr)^{2p}.
 \label{eq:pauli-moment}
\end{equation}
The bound applies to arbitrary, possibly entangled pure states and holds at every moment degree $p$.
\end{lemma}

\begin{proof}
To establish this bound for an arbitrary pure state $\rho$, our approach is to analyze the moment generating function $\langle\rho,e^{tA}(\rho)\rangle_{\mathrm{HS}}$ by connecting $A$ to local depolarizing noise. For any parameter $t\ge0$, setting $\eta=e^{-4t}$, we let
\[
 \mathcal D_\eta(V)=\eta V+(1-\eta)\Tr(V)I/2
\]
denote the single-qubit depolarizing channel. On the identity, $\mathcal D_\eta$ has eigenvalue $1$, while on the traceless Paulis $X,Y,Z$, its eigenvalue is $\eta$. On site $i$, the local generator $Q_i = \operatorname{Ad}_{X_i}+\operatorname{Ad}_{Y_i}+\operatorname{Ad}_{Z_i}$ leaves $I_i$ invariant with eigenvalue $3$, while for any nonidentity Pauli $\sigma_i$, exactly one conjugation fixes $\sigma_i$ and the other two flip its sign, giving eigenvalue $1-1-1=-1$. Comparing these spectra gives the exact single-site identity $e^{tQ_i}=e^{3t}\mathcal D_\eta^{(i)}$, where $\mathcal D_\eta^{(i)}$ denotes $\mathcal D_\eta$ acting on site $i$.

Because generators on distinct qubits commute, the full exponential decomposes into a product of independent local channels:
\begin{equation}
 e^{tA} = \prod_{i=1}^n e^{tQ_i} = e^{3nt} \bigotimes_{i=1}^n \mathcal D_\eta^{(i)}.
\end{equation}
For $S\subseteq[n]$, let $\rho_S=\Tr_{S^c}\rho$, with $S^c=[n]\setminus S$ and $\Tr_\emptyset$ the identity map. Expanding each tensor factor of $\mathcal D_\eta^{(i)}=\eta\,\mathrm{id}_i+\frac{1-\eta}{2}I_i\Tr_i$ into its retained part and its trace-and-replace part yields the following chain:
\begin{equation}
\begin{aligned}
 &\bigl\langle\rho,e^{tA}(\rho)\bigr\rangle_{\mathrm{HS}}
  \overset{\text{(a)}}{=}
   \Tr\!\Bigl[\rho\,\Bigl(\prod_{i=1}^{n}
     \exp\bigl\{t\bigl(\operatorname{Ad}_{X_i}+\operatorname{Ad}_{Y_i}
       +\operatorname{Ad}_{Z_i}\bigr)\bigr\}\Bigr)(\rho)\Bigr]\\
 &\quad\overset{\text{(b)}}{=}
   e^{3nt}\,\Tr\!\Bigl[\rho\,\Bigl(\bigotimes_{i=1}^{n}
     \mathcal D_\eta^{(i)}\Bigr)(\rho)\Bigr]
  \overset{\text{(c)}}{=}
   e^{3nt}\,\Tr\!\Bigl[\rho\,\Bigl(\bigotimes_{i=1}^{n}
     \bigl(\eta\,\mathrm{id}_{i}
       +\tfrac{1-\eta}{2}\,I_{i}\Tr_{i}\bigr)\Bigr)(\rho)\Bigr]\\
 &\quad\overset{\text{(d)}}{=}
   e^{3nt}\sum_{S\subseteq[n]}\frac{\eta^{|S|}(1-\eta)^{n-|S|}}{2^{\,n-|S|}}\,
   \Tr\bigl[\rho\,(\rho_{S}\otimes I_{S^c})\bigr]\\
 &\quad\overset{\text{(e)}}{=}
   e^{3nt}\sum_{S\subseteq[n]}
       \eta^{|S|}\left(\frac{1-\eta}{2}\right)^{n-|S|}
       \Tr\bigl(\rho_S^2\bigr)
  \overset{\text{(f)}}{\le}
   e^{3nt}\sum_{k=0}^{n}\binom nk\,\eta^{k}
       \left(\frac{1-\eta}{2}\right)^{n-k}\\
 &\quad\overset{\text{(g)}}{=}
   e^{3nt}\left(\eta+\frac{1-\eta}{2}\right)^{n}
  =e^{3nt}\left(\frac{1+e^{-4t}}{2}\right)^{n}
  =\left(\frac{e^{3t}+e^{-t}}{2}\right)^{n}\\
 &\quad\overset{\text{(h)}}{=}
   e^{nt}\left(\frac{e^{2t}+e^{-2t}}{2}\right)^{n}
  =e^{nt}\cosh(2t)^n
  \overset{\text{(i)}}{\le}
   e^{nt}\exp\!\Bigl\{n\,\frac{(2t)^2}{2}\Bigr\}
  =e^{nt+2nt^2}.
\end{aligned}
\label{eq:pauli-mgf}
\end{equation}
In this derivation, step~(a) uses the commutation of the $Q_i$ and the trace adjoint. Step~(b) applies the single-site identity $e^{tQ_i}=e^{3t}\mathcal D_\eta^{(i)}$, factoring the exponential into independent local channels. Step~(c) writes $\mathcal D_\eta^{(i)}$ in its defining components, and step~(d) expands the binomial product across subsystems $S \subseteq [n]$, where the retained component on $S$ pairs with the trace-and-replace component on $S^c$. Step~(e) simplifies $\Tr[\rho(\rho_S\otimes I_{S^c})]$ to the marginal purity $\Tr(\rho_S^2)$ by partial-trace duality and groups $2^{-(n-|S|)}$ with $(1-\eta)^{n-|S|}$. Step~(f) uses the universal physical bound $\Tr(\rho_S^2) \le 1$, which holds for all states $\rho$, including arbitrary entangled states, and groups subsets by size $k=|S|$. Step~(g) sums the binomial series and inserts $\eta=e^{-4t}$, step~(h) extracts the common exponential factor $e^t$, and step~(i) uses the standard inequality $\cosh u \le e^{u^2/2}$ at $u=2t$.

Since $\|\rho\|_{\mathrm{HS}}=1$, the spectral measure of $A$ defined by $\rho$ is a probability measure on the real line. Denoting its real random eigenvalue by $\lambda$, \eqref{eq:pauli-mgf} ensures that $\mathbb E e^{t\lambda} \le e^{nt+2nt^2}$. For $s\ge0$, Markov's inequality followed by the choice $t=s/(4n)$ yields the Gaussian tail bound
\[
 \Pr\{\lambda\ge n+s\}
 \le\inf_{t\ge0}e^{-ts+2nt^2}
 =e^{-s^2/(8n)}.
\]
Setting $Y=(\lambda-n)_+$, the lower spectral bound $\lambda\ge-n$ guarantees that $|\lambda|\le n+Y$. Integrating the tail probability by parts~\cite{blm2013} gives
\begin{align*}
 \mathbb E Y^{2p}
 &=\int_0^\infty 2p\,s^{2p-1}\Pr\{Y>s\}\,ds\\
 &\le\int_0^\infty 2p\,s^{2p-1}e^{-s^2/(8n)}\,ds\\
 &=(8n)^p p\int_0^\infty u^{p-1}e^{-u}\,du\\
 &=(8n)^p p!\le(8np)^p,
\end{align*}
where the second equality substitutes $u=s^2/(8n)$, and repeated integration by parts yields $(p-1)!$. Notice that the lower endpoint $-n$ is essential: it allows an upper-tail bound to control the full absolute moment. Finally, applying Minkowski's inequality in $L^{2p}$ yields
\[
 \bigl(\mathbb E|\lambda|^{2p}\bigr)^{1/(2p)}
 \le n+\bigl(\mathbb EY^{2p}\bigr)^{1/(2p)}
 \le n+\sqrt{8np}.
\]
Raising both sides to the power $2p$ establishes~\eqref{eq:pauli-moment}.
\end{proof}

\subsection{Bounding the effect of retained memory}

To relate the generator $A$ to the target weight-$w$ Pauli observables, we define the completely positive map
\[
 \Phi_w=\sum_{P\in\mathcal P_{n,w}}\operatorname{Ad}_P.
\]
Expanding $A^w$ into ordered words of single-site conjugation superoperators, the words that visit $w$ distinct sites produce exactly $w!$ copies of each term in $\Phi_w$: there are $\binom nw$ ways to choose the support, $3^w$ ways to assign nonidentity Pauli operators, and $w!$ ways to order them. Every remaining word is also a Pauli conjugation, because scalar phases cancel under conjugation. Therefore,
\begin{equation}
 A^w-w!\Phi_w\quad\hbox{is completely positive}.
 \label{eq:pauli-cp}
\end{equation}
Complete positivity implies that this difference sends positive operators to positive operators, preserving a nonnegative trace against any positive input.

We now derive the rank interpolation explicitly. Let $T\succeq0$ with $\Tr(T^2)=1$ and $\operatorname{rank}T=r\le\min(K,D)$. For an integer $p\ge1$ satisfying $2p\ge w$, \eqref{eq:pauli-cp} implies
\begin{align}
 \langle T,\Phi_w(T)\rangle_{\mathrm{HS}}
 &\le\frac{\langle T,A^w(T)\rangle_{\mathrm{HS}}}{w!}
 \le\frac{\langle T,|A|^w(T)\rangle_{\mathrm{HS}}}{w!}
 \notag\\
 &\le\frac{\langle T,|A|^{2p}(T)\rangle_{\mathrm{HS}}^{w/(2p)}}{w!}.
 \label{eq:pauli-interpolation}
\end{align}
For the second inequality, each spectral value satisfies $\lambda^w\le|\lambda|^w$, holding for both even and odd $w$. For the third, the spectral weights of $T$ form a probability measure of total mass $\Tr(T^2)=1$, so H\"older's inequality gives $\mathbb E|\lambda|^w \le (\mathbb E|\lambda|^{2p})^{w/(2p)}$.

Writing the spectral decomposition of $T$ as $T=\sum_{a=1}^r t_a\rho_a$, where $\rho_a=|v_a\rangle\langle v_a|$ are pure states, $t_a>0$, and $\sum_at_a^2=1$, the triangle inequality and Lemma~\ref{lem:pauli-moment} yield
\begin{align*}
 \bigl\||A|^p(T)\bigr\|_{\mathrm{HS}}
 &\le\sum_{a=1}^r t_a\bigl\||A|^p(\rho_a)\bigr\|_{\mathrm{HS}}\\
 &\le\bigl(n+\sqrt{8np}\bigr)^p\sum_{a=1}^r t_a
 \le\sqrt r\,\bigl(n+\sqrt{8np}\bigr)^p.
\end{align*}
Here, $\||A|^p(\rho_a)\|_{\mathrm{HS}}^2 = \langle\rho_a,A^{2p}(\rho_a)\rangle_{\mathrm{HS}}$, and Cauchy--Schwarz bounds the scalar sum by $\sqrt{r}\sqrt{\sum_a t_a^2} = \sqrt{r} \le \sqrt{K}$. Squaring this bound and substituting it into~\eqref{eq:pauli-interpolation} proves
\begin{equation}
 \zeta(K)
 \le\frac{K^{w/(2p)}(n+\sqrt{8np})^w}{M w!},
 \qquad p\in\mathbb N,\quad 2p\ge w.
 \label{eq:pauli-rank}
\end{equation}

\begin{proof}[Proof of the finite lower bound in Theorem~\ref{thm:pauli}]
Let $P$ be a reflection and let $T\succeq0$ satisfy $\Tr(T^2)=1$. Cauchy--Schwarz gives
\[
 \Tr(TPTP)\le\|TP\|_{\rm HS}\|PT\|_{\rm HS}=1,
\]
which implies $\zeta(K)\le1$. Combining this with~\eqref{eq:pauli-rank} gives $B_K\le B_{n,w,K}(p)$. Corollary~\ref{cor:estimation} then establishes $N\ge1/(20\epsilon^2B_K)\ge1/(20\epsilon^2B_{n,w,K}(p))$.
\end{proof}

\subsection{A sharper bound from the exponential moment}

The pure-state generating inequality~\eqref{eq:pauli-mgf} also controls the rank-constrained spectral tail directly, avoiding the intermediate even-moment interpolation.

\begin{proposition}[Finite exponential certificate]
\label{prop:pauli-functional}
Let $n\ge2$, $1\le w\le n$, and $K\ge1$ be integers. For real $h\ge n$ and $t>0$ with $th>w-1$, define
\begin{equation}
 U_{n,w,K}(h,t)
 =h^w\left(1+
 \frac{wK e^{nt-th}\cosh(2t)^n}{th-w+1}\right).
 \label{eq:pauli-functional-U}
\end{equation}
Then
\begin{equation}
 \zeta(K)\le\min\left\{1,
       \frac{U_{n,w,K}(h,t)}{Mw!}\right\}.
 \label{eq:pauli-functional-bound}
\end{equation}
In particular, setting
\[
 \widetilde B_{n,w,K}(h,t)
 =\min\left\{1,\frac{U_{n,w,K}(h,t)}{Mw!}\right\}
       +\frac{2K-2}{M}
\]
gives the complete finite copy comparison
\begin{align*}
 N^*(n,w,K,\epsilon)
 &\ge\frac{1}{20\epsilon^2\widetilde B_{n,w,K}(h,t)},\\
 N^*(n,w,K,\epsilon)
 &\le\left\lceil\frac{13}{6}3^w\epsilon^{-2}\ln(6M)\right\rceil
\end{align*}
under the guarantees of Theorem~\ref{thm:pauli}. Any legal pair $(h,t)$ provides a valid certificate.
\end{proposition}

\begin{proof}
Let $T\succeq0$, $\Tr(T^2)=1$, and $\operatorname{rank}T=r\le K$. Write $T=\sum_{a=1}^r s_a\rho_a$, where $\rho_a$ are orthogonal pure-state projections, $s_a>0$, and $\sum_as_a^2=1$. Self-adjointness of $A$, the triangle inequality, and~\eqref{eq:pauli-mgf} give
\begin{align*}
 \langle T,e^{tA}(T)\rangle_{\mathrm{HS}}^{1/2}
 &=\|e^{tA/2}(T)\|_{\mathrm{HS}}
 \le\sum_as_a\|e^{tA/2}(\rho_a)\|_{\mathrm{HS}}\\
 &\le e^{nt/2}\cosh(2t)^{n/2}\sum_as_a
 \le\sqrt K e^{nt/2}\cosh(2t)^{n/2}.
\end{align*}
The spectral measure of $A$ at $T$ has total mass $\Tr(T^2)=1$. Letting $\lambda\in[-n,3n]$ denote its random eigenvalue, squaring the display and applying Markov's inequality yields, for $s\ge0$,
\[
 \Pr\{\lambda>h+s\}
 \le K e^{nt-th}\cosh(2t)^n e^{-ts}.
\]
Because $h\ge n$, the lower spectral endpoint ensures that $|\lambda|\le h+(\lambda-h)_+$. Tail integration~\cite{blm2013} therefore yields
\begin{align*}
 \mathbb E|\lambda|^w
 &\le h^w+\int_0^\infty w(h+s)^{w-1}\Pr\{\lambda>h+s\}\,ds\\
 &\le h^w+w h^{w-1}K e^{nt-th}\cosh(2t)^n \int_0^\infty e^{-[t-(w-1)/h]s}\,ds\\
 &=U_{n,w,K}(h,t).
\end{align*}
Here $(h+s)^{w-1}\le h^{w-1}e^{(w-1)s/h}$, and the condition $th>w-1$ makes the integral finite. Extending the upper integration limit beyond the spectral edge only enlarges the bound. By~\eqref{eq:pauli-cp}, positivity of $T$ and the inequality $\lambda^w\le|\lambda|^w$ imply
\[
 \langle T,\Phi_w(T)\rangle_{\mathrm{HS}}
 \le\frac{\langle T,A^w(T)\rangle_{\mathrm{HS}}}{w!}
 \le\frac{\mathbb E|\lambda|^w}{w!}.
\]
This holds for all weights $w$ and arbitrary entangled eigenvectors of $T$. Taking the supremum and using $\zeta(K)\le1$ proves~\eqref{eq:pauli-functional-bound}.
\end{proof}

\subsection{Explicit bounds at logarithmic Pauli weight}

\begin{proof}[Proof of the lower bound in Corollary~\ref{cor:pauli-asymptotic}]
Let $n\ge2^{13}$, $w=\lceil\log_2n\rceil$, and $K=\lfloor\sqrt n\rfloor$. Define
\[
 L=\ln n,\qquad W=\frac{L}{\ln2}+1,\qquad
 a=2\sqrt{nL},\qquad h=n+a,\qquad t=\frac{a}{4n}.
\]
Thus $w\le W$. The bound $\cosh u\le e^{u^2/2}$ from~\eqref{eq:pauli-mgf} gives
\begin{equation}
 K e^{nt-th}\cosh(2t)^n
 \le K e^{-a^2/(8n)}=\frac{K}{\sqrt n}\le1.
 \label{eq:pauli-rank-tail-cost}
\end{equation}
The memory dimension is absorbed inside this tail bound before the weight-$w$ objective is integrated.

Set $x=a/n=2\sqrt{L/n}$ and $y=4W/a=2W/\sqrt{nL}$. We show below that $y<1$ throughout this range, which implies $ah>4n(w-1)$, so $(h,t)$ is admissible in Proposition~\ref{prop:pauli-functional}. Moreover,
\begin{align}
 U_{n,w,K}(h,t)
 &\le h^w\left(1+\frac{4nw}{ah-4n(w-1)}\right)
 \le\frac{h^w}{1-y}.
 \label{eq:pauli-functional-specialization}
\end{align}
For the second inequality, $w\le W$ and $h\ge n$ give $4nw/[ah-4n(w-1)]\le4W/(a-4W)=y/(1-y)$.

Write $(n)_w=n(n-1)\cdots(n-w+1)$. Since $Mw!=3^w(n)_w$, the standard inequalities $\ln(1+x)\le x$ and $-\ln(1-u)\le u/(1-u)$ give
\begin{align}
 \frac{U_{n,w,K}(h,t)}{(n)_w}
 &\le\frac{\exp\!\left(wx+
         \frac{w(w-1)}{2(n-w+1)}\right)}{1-y}
 \le\frac{\exp\!\left(Wx+
         \frac{W^2}{2(n-W)}\right)}{1-y}.
 \label{eq:pauli-finite}
\end{align}
The denominators are strictly positive for $n\ge8192$.

We now verify the monotonicity of these terms continuously in the system size. Regarding $Wx$, $y$, and $W^2/(n-W)$ as functions of $L$ with $n=e^L$, their logarithmic derivatives satisfy
\begin{align*}
 \frac{d}{dL}\ln(Wx)
 &=\frac1{L+\ln2}+\frac1{2L}-\frac12
 \le\frac3{2L}-\frac12<0,\\
 \frac{d}{dL}\ln y
 &=\frac1{L+\ln2}-\frac1{2L}-\frac12
 \le\frac1{2L}-\frac12<0,\\
 \frac{d}{dL}\ln\frac{W^2}{n-W}
 &=\frac2{L+\ln2}
       -\frac{n-1/\ln2}{n-W}<0.
\end{align*}
These inequalities hold for $L\ge13\ln2>9$. In the last derivative, $W>1/\ln2$ makes the second fraction greater than one, while $2/(L+\ln2)<1$. Furthermore, $W<n$ at the left endpoint and $d(n-W)/dL=n-1/\ln2>0$, so the denominator remains positive.

At the base point $n=8192$, $W=14$ and $9<13\ln2<451/50$ give
\[
 x<\frac1{15},\qquad
 y<\frac{13}{125},\qquad
 \frac{W^2}{2(n-W)}=\frac{196}{16356}<\frac3{250}.
\]
These comparisons follow by squaring their positive sides. In particular, $y<1$ holds for all larger $n$. Since $14/15+3/250<19/20$ and $e^{19/20}<13/5$, monotonicity and~\eqref{eq:pauli-finite} imply
\[
 \frac{U_{n,w,K}(h,t)}{(n)_w}
 <\frac{e^{19/20}}{1-13/125}
 <\frac{13}{5}\frac{125}{112}=\frac{325}{112}.
\]
For the exponential bound, setting $r=19/20$, the ratio of consecutive Taylor terms after degree four is at most $r/5$, so $e^r\le1+r+\frac{r^2}{2}+\frac{r^3}{6}+\frac{r^4}{24(1-r/5)}<\frac{13}{5}$.

Finally, $2\le w\le n-2$ in this range, and the binomial coefficients satisfy $\binom nw\ge n(n-1)/2$. Therefore,
\[
 \frac{2K-2}{\binom nw}
 \le\frac{4}{\sqrt n(n-1)}<\frac1{100000}.
\]
Combining the profile and Gram terms gives
\[
 3^w B_K
 \le\frac{U_{n,w,K}(h,t)}{(n)_w}
      +\frac{2K-2}{\binom nw}
 <\frac{325}{112}+\frac1{100000}<3.
\]
Corollary~\ref{cor:estimation} now yields $N\ge3^w/(60\epsilon^2)$.
\end{proof}

\subsection{Exact finite-size bounds}
\label{subsec:pauli-finite-table}

The exponential certificate can also be evaluated using only integer and rational arithmetic. Choose a positive integer $a$ satisfying
\[
 a^2\ge8n\ln K,\qquad
 a(n+a)>4n(w-1),
\]
and set $h=n+a$ and $t=a/(4n)$. The same calculation as in~\eqref{eq:pauli-rank-tail-cost} bounds $K e^{nt-th}\cosh(2t)^n$ by one. Hence
\begin{equation}
 3^w B_K\le
 \frac{(n+a)^w}{(n)_w}
 \left(1+\frac{4nw}{a(n+a)-4n(w-1)}\right)
 +\frac{2K-2}{\binom nw}.
 \label{eq:pauli-rational-functional}
\end{equation}
All expressions on the right are rational. Rounding twenty times that right side upward to an integer $C$ proves $N\ge3^w/(C\epsilon^2)$.

For Table~\ref{tab:pauli-finite}, $K$ is a power of two. The accompanying code certifies the logarithm using $\ln2=2\sum_{j=0}^{\infty}1/[(2j+1)3^{2j+1}]$. After $m$ terms, its positive remainder is at most $2/[(2m+1)3^{2m+1}(1-1/9)]$. Multiplying this rational upper bound on $\ln2$ by $\log_2K$ certifies $a^2\ge8n\ln K$. The code then checks~\eqref{eq:pauli-rational-functional} exactly, supplying the certified values in Table~\ref{tab:pauli-finite}.

\begin{table}[!htbp]
\caption{\textbf{Finite Pauli lower bounds.} At each displayed $n$, set $w=\log_2 n$ and $K=\sqrt n$. The exponential certificate proves $N\ge3^w/(C\epsilon^2)$ under the same arbitrary-state, simultaneous-output and persistent-memory guarantees. The integer $a$ specifies $h=n+a$ and $t=a/(4n)$. Every row is certified with exact rational arithmetic.}
\label{tab:pauli-finite}
\centering
\begin{tabular}{@{}rrr@{}}
\toprule
$n$ & Tail shift $a$ & Denominator constant $C$\\
\midrule
$2^8$ & $76$ & $260$\\
$2^{12}$ & $370$ & $66$\\
$2^{16}$ & $1706$ & $32$\\
$2^{20}$ & $7626$ & $24$\\
$2^{24}$ & $33413$ & $22$\\
$2^{32}$ & $617302$ & $21$\\
\bottomrule
\end{tabular}
\end{table}
\FloatBarrier

\subsection{Local Pauli measurements: estimator and costs}

\begin{proof}[Proof of the upper bound in Theorem~\ref{thm:pauli}]
The lower bound is compared with the existing local-axis estimator. A shot matches a specified support's axes with probability $3^{-w}$, so its contribution is multiplied by $3^w$. Assigning zero to a nonmatching shot ensures that all estimates share the same fixed denominator. We use exactly
\[
 N=\left\lceil\frac{13}{6}3^w\epsilon^{-2}\ln(6M)\right\rceil
\]
independent copies. On copy $r$, choose independent uniform axes $a_{r,i}\in\{X,Y,Z\}$ for $i\in[n]$, and measure all $n$ local observables, recording outcomes $s_{r,i}\in\{-1,1\}$. For a Pauli $P$ with support $S$, set
\[
 X_{P,r}=3^w\mathbf 1\{a_{r,i}=P_i\text{ for every }i\in S\}
                 \prod_{i\in S}s_{r,i},
 \qquad
 \widehat\mu_P=\frac1N\sum_{r=1}^N X_{P,r}.
\]
Conditioned on matching axes on $S$, Born's rule gives $\mathbb E\prod_{i\in S}s_{r,i}=\Tr(\rho P)$: the local measurements commute, and summing the outcomes outside $S$ inserts the identity. The probability of this axis match is $3^{-w}$, independently of the state. Therefore, for every state,
\[
 \mathbb EX_{P,r}=\Tr(\rho P),\qquad
 \mathbb EX_{P,r}^2=3^w,\qquad
 |X_{P,r}-\mathbb EX_{P,r}|\le2\cdot 3^w.
\]
For a fixed $P$ the shots are independent. Scalar Bernstein's inequality~\cite{blm2013}, with variance at most $3^w$ and centered bound $2\cdot 3^w$, gives
\begin{align*}
 \Pr\{|\widehat\mu_P-\Tr(\rho P)|\ge\epsilon\}
 &\le2\exp\left(-\frac{N\epsilon^2}
                  {2\cdot 3^w+(4/3)3^w\epsilon}\right)\\
 &\le2\exp\left(-\frac{6N\epsilon^2}{13\cdot 3^w}\right)
 \le\frac1{3M},
\end{align*}
where $\epsilon\le1/8$ was used. A union bound over all $M$ requested strings proves simultaneous success with probability at least $2/3$, holding for arbitrary entangled states.
\end{proof}

For completeness, we summarize the operational implementation overheads. Measure $Z$ directly, $X$ after $H$, and $Y$ after $S^\dagger$ followed by $H$. Each copy requires at most $2n$ single-qubit Clifford gates, $n$ readouts, and $n$ independent uniform classical trits. All source qubits are measured before the next access, so the retained quantum dimension is strictly one ($K=1$).

The receiver maintains an integer counter $S_P$ for each requested label. On each shot, enumerating its $\binom nw$ supports selects exactly one Pauli label per support, whose counter is incremented by the outcome parity. This takes
\[
 O\left(Nw\binom nw+Mw+nN\right)
\]
word operations, including label enumeration and reading all shots. Enumerating supports in lexicographic order with base-three indices for axis words uses $O(M\log(N+1)+n)$ classical bits, with a further $O(w\log n)$ bits for indices. An index word of $b=O(\log(N+1)+\log M+\log n)$ bits is sufficient. Schoolbook integer operations give the conservative bit-operation bound of $O(b^2)$ times the displayed word-operation count. The decoder outputs the complete list of target labels and exact rationals $3^wS_P/N$, introducing zero numerical rounding error. Each value requires $O(\ln(N+1)+w)$ bits, with the entire label set consuming $O(Mw\ln n)$ bits. For rational $\epsilon$, setting the sample budget to the ceiling above suffices, using standard rational approximations for $\ln(6M)$. This single-copy ($K=1$) benchmark implements the randomized Pauli shadow protocol~\cite{hkp2020}, with our Bernstein derivation providing an explicit, non-asymptotic constant. For growing logarithmic weight $w=\lceil\log_2 n\rceil$, we have $\ln M\le w\ln(3en/w)=O(\ln^2 n)$, meaning that the number of outputs and their bit representations remain quasipolynomial in $n$. Our resource accounting explicitly includes the full cost of emitting these classical records.

Dividing the upper bound by $3^w/(60\epsilon^2)$ gives a ratio of at most $130\ln(6M)+1$. For $n\ge8192$ and $w=\lceil\log_2 n\rceil$, Corollary~\ref{cor:pauli-asymptotic} rules out any polynomial copy saving from persistent quantum memory of dimension $\lfloor\sqrt n\rfloor$.

\subsection{Recovering the complete-Pauli benchmark}
\label{subsec:complete-proof}

\begin{proof}
The complete Pauli basis satisfies the twirling identity $\sum_{P\in\{I,X,Y,Z\}^{\otimes n}}PVP=D\Tr(V)I$. Expanding $V$ in Pauli strings shows that the identity component contributes $D\Tr(V)I$, while for any non-identity string $Q$, anticommuting Paulis cancel in pairwise signed pairs. For any test state $T\succeq0$ normalized as $\Tr(T^2)=1$, subtracting the identity term yields
\[
 \frac1M\sum_{P\ne I}\Tr(TPTP)
 =\frac{D(\Tr T)^2-1}{D^2-1}.
\]
When $T$ has rank $r\le q$, Cauchy--Schwarz on its eigenvalues gives $(\Tr T)^2\le r\Tr(T^2)\le q$, with equality achieved by $T=\Pi/\sqrt q$ for any rank-$q$ projection $\Pi$. This confirms the exact profile $\zeta(K)$ in~\eqref{eq:complete-profile}. Tracial orthogonality gives $\beta=1/M$, and adding the native-memory correction $(2K-2)\beta$ gives the total budget $B_K$. Applying Corollary~\ref{cor:estimation} then yields
\[
 \frac{D^2-1}{20\epsilon^2(K(D+2)-3)}
 \ge\frac{D-1}{40K\epsilon^2},
\]
recovering the sharp linear dimension-to-memory trade-off.
\end{proof}

\subsection{Comparison with earlier Pauli bounds}

Local random measurements achieve $O(3^w\epsilon^{-2}\ln M)$ copies with single-copy measurements ($K=1$)~\cite{hkp2020,king2025}. While the single-copy lower bound in~\cite{king2025} applies to weights $w\le\log_3(2n+1)$, our analysis holds for logarithmic weight $w=\lceil\log_2 n\rceil$. Chen, Gong, and Ye's framework includes persistent memory by setting $c=N$~\cite{cgy2024}; our contribution is the explicit evaluation of the profile $\zeta(K)$ via hypercontractivity and exponential tail bounds, certifying the exact non-advantage of quantum memory for this growing family.

\section{Majorana cross correlations with shared quantum memory}
\label{sec:cross}
We prove Theorem~\ref{thm:cross} from Section~\ref{subsec:main-cross}.
The requested observables $C_{ab,cd}$, their number $M=[n(2n-1)]^2$ and
the source dimension $D=2^{2n}$ are defined in
\eqref{eq:cross-family}--\eqref{eq:cross-parameters}. The proof first combines
the known source maps while keeping one rank constraint for the receiver's
shared memory. We then evaluate the joint moment uniformly in its degree,
and give the complete product-matching estimator with its implementation
and decoding costs. The joint source may be entangled and may occupy both
parity sectors. Gate counts use the standard Jordan--Wigner encoding in
\eqref{eq:local-jw-compiler}.

\subsection{A lower bound for product families with shared memory}

The local means alone cannot determine the requested cross means. For a
small example, in the standard encoding at $n=1$ each local quadratic is
$-Z$. The Bell state $|\Phi^+\rangle=(|00\rangle+|11\rangle)/\sqrt2$
satisfies
\[
 \langle G^A\otimes I\rangle_{\Phi^+}
 =\langle I\otimes G^B\rangle_{\Phi^+}=0,
 \qquad \langle G^A\otimes G^B\rangle_{\Phi^+}=1.
\]
Thus solving the two marginal estimation tasks would leave the joint
correlation undetermined. A product rule for their individual copy costs
would also need a separate argument about the receiver's shared memory.
We work instead with maps of the joint source: compose the maps encoding
the requested observables, then apply one positive rank constraint. The
moment estimate in the theorem below is consequently over all joint pure
states, including entangled ones.

For completely positive maps $\Phi,\Psi$ on source matrices, write
$\Phi\preceq_{\mathrm{CP}}\Psi$ when $\Psi-\Phi$ is completely positive.
The following rule states the precise structural assumptions needed
to handle product observables with a shared memory.

\begin{theorem}[Finite shared-memory composition]
\label{thm:composition}
Let $\Phi_1,\Phi_2$ be completely positive maps on the same
joint-source matrix space on $\mathbb C^D$.  Suppose $\mathcal A_1,\mathcal A_2$ are
completely positive, self-adjoint for the Hilbert--Schmidt inner
product, and commute under composition.  Assume
\begin{equation}
 \Phi_i\preceq_{\mathrm{CP}}a_i\mathcal A_i^{2d_i},
 \qquad a_i\geq0,\quad d_i\in\mathbb N,\quad d_i\geq1.
 \label{eq:factor-cp-assumption}
\end{equation}
Set $d=d_1+d_2$, $\mathcal A=\mathcal A_1+\mathcal A_2$, and
\begin{equation}
 c=\frac{a_1a_2}{\binom{2d}{2d_1}}.
 \label{eq:composition-constant}
\end{equation}
Then
\begin{equation}
 \Phi_1\Phi_2\preceq_{\mathrm{CP}}c\mathcal A^{2d}.
 \label{eq:composition-cp}
\end{equation}
If $p\geq d$ is an integer and $H_p$ bounds
$\Tr(\rho\mathcal A^{2p}(\rho))$ for every pure state on this same
joint source, then every $T\geq0$ with $\Tr(T^2)=1$ and
$\rank(T)\leq K$ satisfies
\begin{equation}
 \Tr\bigl(T\Phi_1\Phi_2(T)\bigr)
 \leq c(KH_p)^{d/p}.
 \label{eq:composition-moment}
\end{equation}
Suppose further that
$\Phi_1\Phi_2=\sum_{j=1}^M\operatorname{Ad}_{C_j}$ for distinct
orthogonal centered Hermitian reflections $C_j$.  For simultaneous
signed estimation at error $0<\epsilon\leq1/8$ and success $2/3$,
every receiver with one shared retained dimension at most $K$ has
\begin{equation}
 N\geq
 \frac{M}{20\epsilon^2\bigl(c(KH_p)^{d/p}+2K-2\bigr)}.
 \label{eq:composition-operational}
\end{equation}
Here $\operatorname{Ad}_{U}(X)=UXU^\dagger$.
\end{theorem}

\begin{proof}
Write $B_i=\mathcal A_i^{2d_i}$ and $D_i=a_iB_i-\Phi_i$.
All these maps are completely positive, and
\begin{equation}
 a_1a_2B_1B_2-\Phi_1\Phi_2
 =D_1(a_2B_2)+\Phi_1D_2
 \label{eq:cp-product-difference}
\end{equation}
is completely positive.  Commutation of $\mathcal A_1$ and
$\mathcal A_2$ gives the binomial expansion of
$(\mathcal A_1+\mathcal A_2)^{2d}$.  Its
$\mathcal A_1^{2d_1}\mathcal A_2^{2d_2}$ term has coefficient
$\binom{2d}{2d_1}$, and every other term is completely positive.
Together with~\eqref{eq:cp-product-difference}, this proves
\eqref{eq:composition-cp}.

The map $\mathcal A$ is Hilbert--Schmidt self-adjoint.  For
$\|T\|_{\mathrm{HS}}=1$, its spectral measure at $T$ is a probability
measure on the real line.  The scalar even-moment inequality~\cite{blm2013} therefore
gives
\begin{equation}
 \Tr\bigl(T\mathcal A^{2d}(T)\bigr)
 \leq
 \left[\Tr\bigl(T\mathcal A^{2p}(T)\bigr)\right]^{d/p}.
 \label{eq:even-moment-interpolation}
\end{equation}
This uses absolute even powers and also covers negative eigenvalues
of $\mathcal A$ as a Hilbert--Schmidt operator.  Write the positive
spectral decomposition of $T$ as
$T=\sum_{i=1}^{q}t_i\rho_i$, where each $\rho_i$ is a rank-one state,
$t_i\geq0$, $\sum_it_i^2=1$, and $q\leq K$.  Then
\begin{align}
 \left[\Tr\bigl(T\mathcal A^{2p}(T)\bigr)\right]^{1/2}
 &=\|\mathcal A^p(T)\|_{\mathrm{HS}}\nonumber\\
 &\leq\sum_i t_i\|\mathcal A^p(\rho_i)\|_{\mathrm{HS}}
 \leq\sqrt{H_p}\sum_i t_i\leq\sqrt{KH_p}.
 \label{eq:one-joint-rank}
\end{align}
The states $\rho_i$ in this decomposition are arbitrary joint pure
states.  In particular, they may be entangled.  Positivity of $T$ is needed for the source-map comparison. If
$\mathcal R=c\mathcal A^{2d}-\Phi_1\Phi_2$, then complete positivity
implies $\mathcal R(T)\succeq0$, and
\[
 \Tr[T\mathcal R(T)]
 =\Tr[T^{1/2}\mathcal R(T)T^{1/2}]\ge0.
\]
Keeping the two spectral sums inside the square root makes the
shared-rank contribution explicit:
\begin{equation}
\begin{aligned}
 &\Tr[T\Phi_1\Phi_2(T)]
 \le c\Tr[T\mathcal A^{2d}(T)]
 \le c\left(\sqrt{\Tr[T\mathcal A^{2p}(T)]}\right)^{2d/p}
 \\[2pt]
 &\quad=c\left(
   \sqrt{\sum_{i=1}^{q}\sum_{j=1}^{q}t_it_j
     \Tr[\mathcal A^p(\rho_i)\mathcal A^p(\rho_j)]}
   \right)^{2d/p}
 \\[2pt]
 &\quad\le c\left(
   \sqrt{\sum_{i=1}^{q}\sum_{j=1}^{q}t_it_j
     \left|\Tr[\mathcal A^p(\rho_i)\mathcal A^p(\rho_j)]\right|}
   \right)^{2d/p}
 \\[2pt]
 &\quad\le c\left(
   \sqrt{\sum_{i=1}^{q}\sum_{j=1}^{q}t_it_j
     \sqrt{\Tr[\rho_i\mathcal A^{2p}(\rho_i)]
            \Tr[\rho_j\mathcal A^{2p}(\rho_j)]}}
   \right)^{2d/p}
 \\[2pt]
 &\quad=c\left(
   \sqrt{\left(\sum_{i=1}^{q}t_i
     \sqrt{\Tr[\rho_i\mathcal A^{2p}(\rho_i)]}\right)
     \left(\sum_{j=1}^{q}t_j
     \sqrt{\Tr[\rho_j\mathcal A^{2p}(\rho_j)]}\right)}
   \right)^{2d/p}
 \\[2pt]
 &\quad\le c\left(
   \sqrt{H_p\left(\sum_{i=1}^{q}t_i\right)
               \left(\sum_{j=1}^{q}t_j\right)}
   \right)^{2d/p}
 \le c\left(\sqrt{H_p}\sum_{i=1}^{q}t_i\right)^{2d/p}
 \\[2pt]
 &\quad\le c\left(
   \sqrt{H_p}\sqrt{\sum_{i=1}^{q}t_i^2}
               \sqrt{\sum_{i=1}^{q}1^2}
   \right)^{2d/p}
 =c\left(\sqrt{H_p}\sqrt{\Tr(T^2)}\sqrt{\rank T}\right)^{2d/p}
 \\[2pt]
 &\quad=c\left(\sqrt{H_p}\sqrt{\rank T}\right)^{2d/p}
 \le c\left(\sqrt{H_p}\sqrt K\right)^{2d/p}
 =c(KH_p)^{d/p}.
\end{aligned}
\label{eq:composition-complete-chain}
\end{equation}
The first line uses the completely positive map comparison and
\eqref{eq:even-moment-interpolation}. Self-adjointness of $\mathcal A^p$
then expands the trace into the two spectral sums. The next bounds take
absolute values and apply Hilbert--Schmidt Cauchy--Schwarz to each pair.
The resulting double sum factors into the product of the two displayed
single sums. The pure-state moment bound applies to every $\rho_i$,
including entangled states. Finally, scalar Cauchy--Schwarz uses
$\sum_i t_i^2=\Tr(T^2)=1$ and $q=\rank T\le K$.
This proves~\eqref{eq:composition-moment}.

For the uniform prior on the $C_j$, their tracial Gram norm is
$\beta=1/M$.  Definition~\ref{def:profile} and
\eqref{eq:composition-moment} give
\[
 \zeta(K)=\sup_{\substack{T\geq0,\ \Tr(T^2)=1\\
                        \rank(T)\leq\min(K,D)}}
 \frac1M\sum_j\Tr(TC_jTC_j)
 \leq\frac{c(KH_p)^{d/p}}{M}.
\]
Keeping the native-memory term, Corollary~\ref{cor:estimation} now gives
\begin{align}
 N&\ge\frac{1}{20\epsilon^2\{\zeta(K)+(2K-2)/M\}}\nonumber\\
 &\ge\frac{1}{20\epsilon^2\{c(KH_p)^{d/p}/M+(2K-2)/M\}}\nonumber\\
 &=\frac{M}{20\epsilon^2\{c(KH_p)^{d/p}+2K-2\}}.
 \label{eq:composition-copy-chain}
\end{align}
The second line replaces the source profile by its certified upper bound;
this increases the denominator and therefore weakens the necessary bound
in the valid direction. The last line puts the two costs over the same
normalization. This proves~\eqref{eq:composition-operational}.
\end{proof}

For observables on tensor factors, take $\Phi_i$ to be the
unnormalized sum of conjugations by the lifted local observables.
Their composition is the sum of conjugations by all requested
products.  The theorem uses one rank bound $K$ and a moment bound on
all pure states of the joint source.  The assumptions are finite
inequalities between known maps; general verification can require
source-size Choi matrices.  In the present application, the following
word identities supply exact certificates.

\subsection{Moment bounds for Majorana observables}
\label{subsec:majorana-maps}

Let $P_A$ be the local parity operator on $A$.  Thus $P_A^2=I$ and
$P_Aa_r=-a_rP_A$.  Use the global embedding
\begin{equation}
 c_r=a_r\otimes I\quad(1\leq r\leq2n),\qquad
 c_{2n+s}=P_A\otimes b_s\quad(1\leq s\leq2n).
 \label{eq:global-majorana-embedding}
\end{equation}
These $L=4n$ matrices satisfy the Majorana relations on the full
$D$-dimensional source.  The parity factors cancel in every local
quadratic on $B$.  Consequently the $C_{ab,cd}$ in
\eqref{eq:cross-family} are precisely the global degree-four
Majorana reflections whose support has two labels in each factor.
A nonempty even monomial changes sign under conjugation by any
Majorana in its support, and therefore has trace zero.  Applying
this to each cross monomial and to products with distinct supports
also verifies their centering and tracial orthogonality.

Define
\[
 \mathcal A_1=\sum_{r=1}^{2n}\operatorname{Ad}_{c_r},\qquad
 \mathcal A_2=\sum_{r=2n+1}^{4n}\operatorname{Ad}_{c_r},\qquad
 \Phi_1=\sum_{a<b}\operatorname{Ad}_{G^A_{ab}\otimes I},\qquad
 \Phi_2=\sum_{c<d}\operatorname{Ad}_{I\otimes G^B_{cd}}.
\]
Conjugations by two Majoranas commute as maps, because the sign from
interchanging their unitary generators cancels in a conjugation.
Each summand is also Hilbert--Schmidt self-adjoint.  Squaring either
local sum gives the exact identity
\begin{equation}
 \mathcal A_i^2=2n\,\operatorname{id}+2\Phi_i.
 \label{eq:local-majorana-square}
\end{equation}
Thus Theorem~\ref{thm:composition} applies with
$a_1=a_2=1/2$, $d_1=d_2=1$, and
\begin{equation}
 c=\frac{(1/2)(1/2)}{\binom42}=\frac1{24},\qquad
 \Phi_1\Phi_2\preceq_{\mathrm{CP}}\frac1{24}\mathcal A^4.
 \label{eq:majorana-cross-cp}
\end{equation}

There is also a useful direct check.  Let $\mathcal P_r$ be the sum
of conjugations by all global degree-$r$ Majorana reflections.
Reducing four-letter words by their odd-occurrence sets gives
\begin{equation}
 \mathcal A^4=(3L^2-2L)\operatorname{id}
              +(12L-16)\mathcal P_2+24\mathcal P_4.
 \label{eq:majorana-fourth-identity}
\end{equation}
Here is the count behind each coefficient. Empty odd support comes from
one label appearing four times, or two labels each appearing twice. These
contribute $L$ and $6\binom L2$ words. For a specified odd support
$\{a,b\}$, the multiplicities are either $(3,1)$ or $(1,1,2)$.
There are $2\cdot4=8$ words of the first kind. The repeated third label
has $L-2$ choices, each with $4!/2!=12$ orderings, giving
$8+12(L-2)=12L-16$. Four specified odd labels occur once each and
have $4!=24$ orderings. In every case the phase of the reduced Majorana
word disappears under conjugation.
Since the cross family is a subset of the degree-four family,
\eqref{eq:majorana-fourth-identity} also proves
\eqref{eq:majorana-cross-cp}.

The all-state squared-mean estimate of Leone and
Bittel~\cite[Appendix~S.2, Lemma~4]{leonebittel2026} gives a physical
moment certificate at every size and every moment order. Their Lemma~3
supplies the word-count bound used in their signed two-copy moment
estimate (Lemma~2). We apply those two bounds to the positive
conjugation-map moment needed here. Appendix~\ref{app:moments} gives
the phase calculation, the exact pure-state word identity, and the
extension to mixed states.

\begin{proposition}[All-size physical moment certificate]
\label{prop:uniform-moment}
Let $L\geq2$ be even, let $c_1,\ldots,c_L$ be Majorana reflections,
and set $\mathcal A=\sum_{r=1}^L\operatorname{Ad}_{c_r}$.
For every integer $p\geq1$ and every density matrix $\rho$,
\begin{equation}
 \Tr\bigl(\rho\mathcal A^{2p}(\rho)\bigr)
 \leq\widehat H_p(L)
 :=\frac{(2p)!}{p!}\left[\frac{(1+e)L}{2}\right]^p
 \leq\bigl[(1+e)pL\bigr]^p.
 \label{eq:uniform-moment-bound}
\end{equation}
The bound allows coherence between parity sectors and arbitrary
entanglement between the two source factors. The moment order $p$
has no size-dependent restriction.
\end{proposition}

\subsection{Commuting measurements: estimator and costs}
\label{subsec:cross-estimator}

To benchmark our persistent-memory lower bound, we consider how well one can estimate these $M = [n(2n-1)]^2$ cross correlations using only unentangled measurements and classical records ($K=1$). Since quadratic Majoranas commute whenever their support indices are disjoint, our strategy is to partition the observables into commuting matchings on each factor~\cite{king2025,bonet2020,zhao2021,wan2023,ogorman2022}. Below, we specify this deterministic schedule and its operational costs.

\begin{proposition}[Achievable budget and implementation]
\label{prop:cross-estimator}
For every $n\geq1$ and $0<\epsilon\leq1/8$, all means in
\eqref{eq:cross-family} can be estimated as required using exactly
\begin{equation}
 q=2n-1,\qquad
 R=\left\lceil2\epsilon^{-2}\ln(6M)\right\rceil,\qquad
 N=q^2R
 \label{eq:cross-upper-count}
\end{equation}
ordinary joint-source copies, with zero additional quantum ancillas
and retained quantum dimension one ($K=1$). In the standard Jordan--Wigner encoding, an exact Clifford implementation
uses at most $14m$ elementary gates and $2n$ single-qubit $Z$
readouts per copy. It can be compiled in $O(n^3)$ classical word
operations and decoded in $O(n^2N+M)$ word operations.
\end{proposition}

\begin{proof}
We first group the local Majoranas into commuting sets. On each $2n$-mode factor, two quadratic Majoranas $i\gamma_j\gamma_k$ commute if and only if their index pairs are disjoint. A maximal set of commuting quadratics thus corresponds to a perfect matching (a 1-factor) of $K_{2n}$. To partition all $\binom{2n}{2}$ local quadratics into the minimum number of matchings, we use the standard round-robin tournament construction. Labeling vertices by $\{\infty\}\cup\mathbb Z_q$ with $q=2n-1$, each round $t\in\mathbb Z_q$ defines the matching
\begin{equation}
 F_t=\{\{\infty,t\}\}
       \cup\{\{t+r,t-r\}:1\leq r\leq n-1\}.
 \label{eq:round-robin-matchings}
\end{equation}
Modulo $q$, the residues $\pm1,\ldots,\pm(n-1)$ exhaust all nonzero elements of $\mathbb Z_q$, so every vertex is paired. Since edges through $\infty$ appear once, and any finite edge $\{u,v\}$ has a unique midpoint $t=(u+v)/2 \pmod q$ and distance $r = \pm(u-v)/2$, these $q$ matchings partition every local quadratic, holding for all $n\ge1$.

Next, we compile each matching $F_t$ into an executable circuit. Under the Jordan--Wigner transformation~\cite{jordanwigner1928,bravyi2005},
\begin{equation}
 c_{2r-1}=Z_1\cdots Z_{r-1}X_r,\qquad
 c_{2r}=Z_1\cdots Z_{r-1}Y_r,
 \label{eq:local-jw-compiler}
\end{equation}
adjacent pairs map directly to Pauli $Z$ operators: $i c_{2r-1}c_{2r}=-Z_r$. Measuring $F_t$ thus reduces to rotating each edge $e_{t,r}=(a_r,b_r)$ into adjacent positions $(2r-1, 2r)$ via adjacent swaps. The elementary fermionic rotation $S_j=\exp(\pi c_jc_{j+1}/4)$ generates an adjacent swap with a sign:
\begin{equation}
 S_jc_jS_j^\dagger=-c_{j+1},\qquad
 S_jc_{j+1}S_j^\dagger=c_j,
 \label{eq:signed-adjacent-swap}
\end{equation}
leaving all other Majoranas fixed. Here $S_{2r-1}=e^{i\pi Z_r/4}$ is a single-qubit phase gate $S^\dagger$ (up to global phase), while $S_{2r}=e^{i\pi X_rX_{r+1}/4}$ decomposes into four $H$, one $S^\dagger$, and two adjacent CNOTs. Bubble-sorting the endpoints of $F_t$ into $(a_1,b_1,\ldots,a_n,b_n)$ takes at most $m=\binom{2n}{2}$ swaps. Tracking signs via~\eqref{eq:signed-adjacent-swap} yields a Clifford $U_t$ satisfying
\begin{equation}
 U_tG_{e_{t,r}}U_t^\dagger=\eta_{t,r}Z_r,
 \qquad\eta_{t,r}\in\{+1,-1\},
 \label{eq:matching-clifford}
\end{equation}
with $i c_{2r-1}c_{2r}=-Z_r$ contributing its minus sign. This compiler requires $O(n^2)$ classical operations and at most $7m$ elementary gates per setting.

To estimate the cross correlations, we apply $U_t\otimes U_s$ to $R$ fresh joint copies for each setting pair $(t,s) \in \mathbb Z_q \times \mathbb Z_q$, measuring all $2n$ qubits in the computational basis. From readouts $z,w\in\{0,1\}^n$, we decode local signs $X_r=\eta_{t,r}(-1)^{z_r}$ and $Y_v=\eta_{s,v}(-1)^{w_v}$, forming the product
\begin{equation}
 W_{r,v}=X_rY_v.
 \label{eq:actual-product-samples}
\end{equation}
Crucially, this product estimator remains unbiased even if $\rho_{AB}$ is entangled. Because the physical POVM elements factor as $E_{z,w}=(U_t^\dagger|z\rangle\langle z|U_t)\otimes(U_s^\dagger|w\rangle\langle w|U_s)$, they satisfy
\begin{equation}
 \sum_{z,w}W_{r,v}(z,w)E_{z,w}
 =G^A_{e_{t,r}}\otimes G^B_{e_{s,v}}.
 \label{eq:entangled-state-expectation}
\end{equation}
Hence $\mathbb E W_{r,v} = \Tr(\rho_{AB}(G^A_{e_{t,r}}\otimes G^B_{e_{s,v}}))$ holds for every state $\rho_{AB}$, with quantum correlations preserved in the empirical product.

Since each target observable belongs to exactly one setting pair, it receives $R$ independent product samples $W_{r,v}\in[-1,1]$. Setting the estimate $\widehat\mu_{e,f}$ to their mean, Hoeffding's inequality~\cite{hoeffding1963,blm2013} gives
\[
 \Pr\{ |\widehat\mu_{e,f}-\Tr(\rho C_{e,f})|>\epsilon\}
 \leq2e^{-R\epsilon^2/2}\leq\frac1{3M},
\]
for $R=\lceil2\epsilon^{-2}\ln(6M)\rceil$. A union bound over all $M$ observables guarantees simultaneous accuracy within $\epsilon$ with probability at least $2/3$, regardless of correlations between coordinates. Finally, the measurement is implemented by separable Kraus operators $L_{z,w}=(\langle z|U_t)\otimes(\langle w|U_s): \mathbb C^D\to\mathbb C$. Because every register is measured and reset before the next copy arrives, the retained quantum dimension is strictly $K=1$.
\end{proof}

We now summarize the operational overheads. Let $b_R=\lceil\log_2(R+1)\rceil$, and assume standard classical words holding a label and counter. The receiver precomputes the $q$ local matchings in $O(n^2)$ word operations and compiles their Clifford circuits in $O(qn^2)=O(n^3)$ operations, caching them once for both factors. This cache stores $O(qm\ln(n+1))$ gate bits and $O(qn)$ decoder signs $\eta_{t,r}$, requiring only $O(n\ln(n+1))$ transient workspace.

Across the entire protocol, the receiver performs $2nN$ source readouts and collects $n^2N=MR$ product samples. Circuit execution uses at most $14mN$ elementary gates ($4mN$ adjacent CNOTs and $10mN$ single-qubit gates), with parallel circuit depth at most $7m$ per copy ($14m$ total gates per copy across both subsystems). Gate dispatch reads $O(mN\ln(n+1))$ bits from the cache across the $q^2$ setting pairs.

Crucially, decoding proceeds in a streaming fashion without retaining raw measurement records. Within each setting pair, the receiver maintains $n^2$ integer counters in $\{0,\ldots,R\}$, simply incrementing counter $(r,v)$ whenever the decoded sign product $W_{r,v}$ is positive. The final count $C$ immediately produces the unbiased estimate $(2C-R)/R$. Emitting these $n^2$ estimates before resetting the counters consumes only $n^2b_R$ streaming counter bits and a $2n$-bit readout buffer. Across the entire protocol, decoding involves exactly $n^2N$ sign products and $O(n^2N+M)$ word operations. Storing all $M$ outputs uses $O(Mb_R)$ value bits with implicit indexing. In a rigorous bit-operation model, scaling word counts by $O(\ln(n+1)+\ln(R+1))$ provides a guaranteed upper bound. 

Finally, for rational $\epsilon$, the controller can use integer comparisons alone: choosing $A_0=\lceil2\epsilon^{-2}\rceil$ and $B_0=\min\{b\in\mathbb N_0:2^b\geq6M\}$, each setting pair is repeated $R_{\rm bin}=A_0B_0$ times. Since $\ln x\leq\log_2x$ for $x\geq1$, this preserves the same confidence with an exactly executable sample count $q^2R_{\rm bin}$. All circuit angles are exact Clifford rotations.

\begin{proof}[Proof of Theorem~\ref{thm:cross}]
With the achievable upper bound established in Proposition~\ref{prop:cross-estimator}, we now combine our structural composition theorem with the quadratic moment certificate to complete the proof of Theorem~\ref{thm:cross}.

The global embeddings~\eqref{eq:global-majorana-embedding} and \eqref{eq:majorana-cross-cp} verify the composition hypotheses with local dimension $d=2$ and constant $c=1/24$. Because the cross reflections are tracially orthogonal, Theorem~\ref{thm:composition} applies to any certificate satisfying the joint moment condition~\eqref{eq:joint-moment-certificate}. For every integer $p\geq2$, inserting the moment bound $H_p=\widehat H_p(4n)$ from Proposition~\ref{prop:uniform-moment} gives
\[
 \frac{(K\widehat H_p(4n))^{2/p}}{24}
 \leq\frac{K^{2/p}\bigl[(1+e)p(4n)\bigr]^2}{24}
 =C_0p^2n^2K^{2/p},\qquad C_0=\frac23(1+e)^2.
\]
Substituting this into Theorem~\ref{thm:composition} yields the finite lower bound
\begin{equation}
 N\geq\frac{M}{20\epsilon^2
          \bigl(C_0p^2n^2K^{2/p}+2K-2\bigr)},
 \label{eq:cross-physical-finite}
\end{equation}
which accounts for directional memory via the native term $2K-2$ for all $n\geq1$ and legal $K$.

We now specialize to growing memory $K=\lfloor\sqrt n\rfloor$ for $n\geq2$. To keep $K^{2/p}$ bounded by a constant, we choose the degree
\begin{equation}
 p=\max\{2,\lceil\ln K\rceil\},
 \label{eq:growing-moment-choice}
\end{equation}
which ensures $K^{2/p}\leq e^2$ and
\begin{equation}
 p\leq\max\{2,1+\tfrac12\ln n\}
 \leq\frac43(1+\ln n).
 \label{eq:cross-growing-degree-bound}
\end{equation}
Here $\ln n\geq\ln2>1/2$ bounds the constant branch, while $1+\tfrac12\ln n\leq\tfrac43(1+\ln n)$ bounds the logarithmic branch.

To keep numerical constants explicit, Taylor expansion of $e$ yields $e < 68/25$, so
\begin{equation}
 C_0e^2+\frac14
 <\frac23\left(\frac{93}{25}\right)^2
             \left(\frac{68}{25}\right)^2+\frac14<69.
 \label{eq:cross-constant-69}
\end{equation}
Since $M=[n(2n-1)]^2\geq n^4$, $2K-2\leq n^2$, and $p\geq2$, the denominator bracket in~\eqref{eq:cross-physical-finite} satisfies
\begin{align}
 C_0p^2n^2K^{2/p}+2K-2
 &\leq C_0e^2p^2n^2+n^2
 \leq\left(C_0e^2+\frac14\right)p^2n^2
 <69p^2n^2.
 \label{eq:cross-full-denominator}
\end{align}
It follows that
\begin{align}
 N^*&\geq\frac{n^2}{1380\epsilon^2p^2}
 \geq\frac{n^2}{(1380\cdot16/9)\epsilon^2(1+\ln n)^2}
 \geq\frac{n^2}{2500\epsilon^2(1+\ln n)^2},
 \label{eq:cross-all-size-lower}
\end{align}
since $1380\cdot16/9=7360/3<2500$.

Finally, we benchmark this lower bound against the achievable commuting matching budget from Proposition~\ref{prop:cross-estimator}. Setting $t=\ln(4n)$ and noting $M\le4n^4$, we have $\ln(6M)\le\ln24+4\ln n\le4t$. Using $(2n-1)^2\le4n^2$ and $1\le\epsilon^{-2}t$, the achievable sample count satisfies
\begin{align}
 (2n-1)^2\left\lceil2\epsilon^{-2}\ln(6M)\right\rceil
 &\le4n^2\bigl(2\epsilon^{-2}\ln(6M)+1\bigr)
 \le36n^2\epsilon^{-2}t.
 \label{eq:cross-upper-comparison-chain}
\end{align}
Comparing this against our lower bound~\eqref{eq:cross-all-size-lower}, the ratio between the achievable upper bound and our lower bound is at most $90000(1+\ln n)^2\ln(4n) = O(\ln^3 n)$ for every $n\ge2$, completing the proof.
\end{proof}

At $K=n^3$, the native-memory contribution
$(2K-2)/M=\Theta(1/n)$ remains in the normalized denominator.
The finite bound~\eqref{eq:cross-physical-finite}, with
$p=\max\{2,\lceil\ln K\rceil\}$, gives
$\Omega(n\epsilon^{-2})$ in that regime. Even a vanishing physical
moment contribution would leave the same order from the native term.
Thus the quadratic lower bound at this larger memory remains a
separate open question.

\section{Discussion}
\label{sec:discussion}
Quantum memory offers the tantalizing prospect of slashing the sample cost of quantum learning by enabling joint measurements across copies. The main lesson of our work is that even when a modest quantum memory remains coherent throughout the entire experiment and interacts adaptively with every incoming copy, its power to reduce copy complexity is fundamentally capped by the operator geometry of the target observables. For logarithmic-weight Paulis, memoryless local measurements remain within an $O(\ln^2 n)$ factor of optimal. For quadratic Majorana cross correlations, simple commuting matchings remain within an $O(\ln^3 n)$ factor. In both settings, our lower bounds grant the receiver an indefinitely persistent quantum register of dimension $K=\lfloor\sqrt n\rfloor$ (about $(\log_2 n)/2$ qubits) and permit every instrument to be chosen adaptively based on all prior measurement records. These results provide an intuitive physical explanation for why persistent coherent storage yields no polynomial advantage over classical-record protocols in these regimes.

The conceptual contribution of our framework is that it translates this operational limitation into an explicit mathematical certificate derived directly from the observable matrices. While previous foundational frameworks capture persistent memory in principle, evaluating their bounds requires an optimization over all possible adaptive measurement strategies on the entire sequence of copies~\cite{cgy2024}. Our structural theorem replaces this intractable protocol-space search with a concrete operator-algebraic budget $B_K = \zeta(K) + (2K-2)\beta$. At the technical core, the resolvent backward-induction architecture provides a reusable blueprint for adaptive quantum lower bounds: by pairing the null drift with an operator resolvent, alternative histories cancel out identically, while the operator-inversion defect naturally manufactures the negative variance penalty needed for sharp statistical bounds. This ensures that the bound remains valid under arbitrary adaptive histories and unrestricted memory lifetimes.

The physical distinction between our two application families illustrates the breadth of the framework. Pauli weight measures spatial locality in qubit systems, whereas Majorana cross correlations probe bipartite fermionic interactions across two subsystems. Although their algebraic analyses rely on distinct mathematical tools—spectral tail generating functions for Paulis and positive conjugation moments for Majoranas—the identical operational theorem converts both into sharp sample complexity lower bounds. In the Majorana setting, our composition principle accommodates arbitrary entanglement in the source state while charging the receiver once for their shared memory. Furthermore, when applied to the complete Pauli group, our rule evaluates analytically via a Pauli twirl to recover the linear $\Omega(D/(K\epsilon^2))$ lower bound with an explicit constant.

From a practical perspective, our computational relaxations provide an actionable toolbox for researchers exploring new observable families. For moderate system sizes or specific families of interest, one can upper-bound the rank-constrained profile $\zeta(K)$ using a semidefinite relaxation or exploit physical symmetries to reduce it to a polynomial-size linear program. For growing-weight Paulis and quadratic Majoranas, these symmetry reductions collapse the programs to $n+1$ and $(2n+1)^2$ variables, respectively. For arbitrary unstructured observables, the convergent moment hierarchy guarantees that the profile can be systematically approximated.

Several intriguing questions remain open, most notably closing the polylogarithmic gaps and exploring larger memory capacities. As analyzed in Appendix~\ref{app:log-gaps}, our structural lower bound and the simultaneous Fano packing bound currently combine by taking their maximum; establishing whether they can be multiplied together remains a key technical challenge. Furthermore, for Majorana cross correlations at large memory $K=\Omega(n^3)$, the native-memory term $(2K-2)/M$ degrades the bound to $\Omega(n/\epsilon^2)$, leaving the optimal trade-off in the high-memory regime an exciting open problem. A natural next milestone is to develop a general composition theorem that directly deduces bipartite sample complexity from local algebraic invariants, while accommodating arbitrary entanglement and shared persistent memory.

\section{Conclusion}
\label{sec:conclusion}
We have established a structural lower-bound framework for quantum property estimation under indefinitely persistent quantum memory. By decoupling the instantaneous information yield of each copy from the persistent drift of the memory, our theorem connects the fundamental copy complexity directly to the geometric structure of the requested observables. For two prominent growing families—weight-$w$ Pauli observables and quadratic Majorana cross correlations—the resulting bounds demonstrate that simple, memoryless measurements storing only classical records are already near-optimal up to small logarithmic factors, even when compared against an adaptive receiver equipped with a coherent register reused indefinitely. This work demonstrates that the algebraic structure of the observables themselves governs the ultimate utility of quantum memory, providing both a benchmark for existing experimental protocols and a rigorous foundation for exploring the limits of quantum learning machines.

\begin{acks}
This work was supported by the Department of Environment
and Science of Queensland State Government under Quantum
2032 Challenge Program (Project \#Q2032001).
\end{acks}
\addtocontents{toc}{\protect\setcounter{tocdepth}{1}}
\appendix
\allowdisplaybreaks
\section{Moment bounds for Majorana observables}
\label{app:moments}

In this appendix, we establish the all-size physical moment certificate stated in
Proposition~\ref{prop:uniform-moment}. Our proof draws upon the squared-mean and
word-counting techniques introduced by Leone and
Bittel~\cite[Appendix~S.2, Lemmas~3--4]{leonebittel2026}. While their work utilized
these estimates to bound a signed two-copy observable moment, we introduce an exact
pure-state word identity that transfers these bounds to the positive
conjugation-map moment required by our composition framework. Finally, we employ
the convexity of the squared Hilbert--Schmidt norm to lift the certificate to
arbitrary mixed states, fully accounting for arbitrary bipartite entanglement and
leaving the moment degree $p$ completely unconstrained.

\subsection{Physical squared means and exact word counting}

Let $L=2m\geq2$, and let $c_1,\ldots,c_L$ be Majorana reflections satisfying
$c_i^\dagger=c_i$, $c_i^2=I$, and $c_ic_j=-c_jc_i$ for all $i\ne j$.
For any ordered subset $S=\{s_1<\cdots<s_r\}\subseteq[L]$, we define the
corresponding Hermitian Majorana monomial
\begin{equation}
 \Gamma_S=i^{r(r-1)/2}c_{s_1}\cdots c_{s_r},\qquad
 \Gamma_{\varnothing}=I.
 \label{eq:hermitian-majorana-monomial}
\end{equation}
By construction, each $\Gamma_S$ is a Hermitian reflection. For any physical density
matrix $\rho$, we denote the sum of squared expectation values at degree $r$ by
\begin{equation}
 A_r(\rho)=\sum_{|S|=r}\bigl(\Tr(\rho\Gamma_S)\bigr)^2,
 \qquad A_0(\rho)=1,
 \label{eq:physical-squared-means}
\end{equation}
setting $A_r(\rho)=0$ whenever $r>L$.

\begin{lemma}[Physical squared-mean bound, Leone--Bittel]
\label{lem:physical-squared-means}
For every density matrix $\rho$ and every integer $k\geq1$,
\begin{equation}
 A_{2k}(\rho)\leq\frac{(em)^k}{k!}.
 \label{eq:physical-degree-bound}
\end{equation}
\end{lemma}

\begin{proof}
Our approach is to construct an operator generating function for the two-copy observables $\Gamma_S \otimes \Gamma_S$. Defining the two-copy generators $G_i = c_i \otimes c_i$, the anticommutation signs cancel between tensor copies, making all $G_i$ mutually commuting Hermitian reflections: $G_i G_j = G_j G_i$ and $G_i^2 = I$. For an even subset of size $|S|=2k$, the phase factor in~\eqref{eq:hermitian-majorana-monomial} simplifies to
\begin{equation}
 \Gamma_S\otimes\Gamma_S
 =i^{2k(2k-1)}(c_S\otimes c_S)
 =(-1)^k\prod_{i\in S}G_i.
 \label{eq:two-copy-majorana-phase}
\end{equation}
For any scalar parameter $z\ge0$, we define the generating operator
\[
 B(z)=\sum_{k=0}^m z^k \sum_{|S|=2k}\Gamma_S\otimes\Gamma_S.
\]
Expanding the product of $(I \pm i\sqrt{z}G_i)$ cancels all terms with odd degrees, yielding the exact factorization
\begin{equation}
 B(z)=\frac12\left[
       \prod_{i=1}^{2m}(I+i\sqrt zG_i)
       +\prod_{i=1}^{2m}(I-i\sqrt zG_i)\right].
 \label{eq:physical-generating-operator}
\end{equation}
Because each $G_i$ is a reflection, $\|I \pm i\sqrt{z}G_i\|_{\mathrm{op}} = \sqrt{1+z}$. Submultiplicativity of the operator norm immediately bounds $\|B(z)\|_{\mathrm{op}} \le (1+z)^m$. Taking the expectation in the two-copy state $\rho \otimes \rho$ reveals
\begin{equation}
 0\leq\sum_{k=0}^m A_{2k}(\rho)z^k
 =\Tr[B(z)(\rho\otimes\rho)]\leq(1+z)^m.
 \label{eq:physical-generating-bound}
\end{equation}
Since all coefficients $A_{2k}(\rho)$ are nonnegative, setting $z=k/m$ and isolating the degree-$k$ term yields
\[
 A_{2k}(\rho)\leq\left(\frac{m}{k}\right)^k \left(1+\frac{k}{m}\right)^m
 \leq\left(\frac{em}{k}\right)^k \leq\frac{(em)^k}{k!},
\]
where the final inequalities apply $1+x\le e^x$ and $k!\le k^k$. This bound holds universally for all density matrices, including states with arbitrary parity coherence.
\end{proof}

\begin{lemma}[Finite word certificate]
\label{lem:word-certificate}
Let $a_{t,r}(L)$ denote the number of ordered words of length $t$ over the alphabet $[L]$ whose odd-occurrence set is a specified $r$-element subset. Setting $a_{0,0}=1$, $a_{0,r}=0$ for $r>0$, and zero for out-of-range indices, the exact recurrence relation is
\begin{equation}
 a_{t+1,r}=r\,a_{t,r-1}+(L-r)\,a_{t,r+1}.
 \label{eq:word-recurrence}
\end{equation}
Furthermore, for any specified $2k$-subset, the word count satisfies
\begin{equation}
 a_{2p,2k}(L)
 \leq\frac{(2p)!}{2^{p-k}(p-k)!}L^{p-k},
 \qquad 0\leq k\leq p.
 \label{eq:word-count-uniform}
\end{equation}
\end{lemma}

\begin{proof}
To establish the recurrence, consider appending a letter to a word of length $t$. The final letter $\ell \in [L]$ either belongs to the target odd-occurrence set $S$ (with $|S| = r$ choices, reducing the previous odd set to $S\setminus\{\ell\}$) or lies outside $S$ (with $L-|S| = L-r$ choices, expanding the previous odd set from $S\cup\{\ell\}$). Relabeling symmetry confirms~\eqref{eq:word-recurrence}.

For a pure state $\rho=|\psi\rangle\langle\psi|$, expanding the $2p$-th power of the averaging superoperator $\mathcal A=\sum_{i=1}^L\operatorname{Ad}_{c_i}$ into ordered words $w\in[L]^{2p}$ yields
\begin{equation}
 \Tr\bigl(\rho\mathcal A^{2p}(\rho)\bigr)
 =\sum_{w\in[L]^{2p}}|\langle\psi,c_w\psi\rangle|^2
 =\sum_{k=0}^pa_{2p,2k}(L)A_{2k}(\rho).
 \label{eq:exact-majorana-word-moment}
\end{equation}
Here, canceling pairs of identical adjacent Majoranas reduces $c_w$ to a phase times the Hermitian monomial $\Gamma_S$, where $S$ is the set of labels with odd multiplicity. The scalar phase disappears in the squared modulus $|\langle\psi,c_w\psi\rangle|^2$. Summing over all words with identical odd support yields the second line.

To bound $a_{2p,2k}(L)$ for a fixed odd set $S$, we select representative positions for each of the $2k$ distinct labels, giving $(2p)!/(2p-2k)!$ choices. We then pair the remaining $2p-2k$ positions into $p-k$ pairs and assign one of $L$ labels to each pair, which gives $[(2p-2k)!/(2^{p-k}(p-k)!)] L^{p-k}$ configurations. Multiplying these factors establishes~\eqref{eq:word-count-uniform}.
\end{proof}

\subsection{Transfer to the positive conjugation-map moment}

\begin{proof}[Proof of Proposition~\ref{prop:uniform-moment}]
We first consider pure states. Substituting the counting bound~\eqref{eq:word-count-uniform}
and the squared-mean estimate~\eqref{eq:physical-degree-bound} into the exact word
expansion~\eqref{eq:exact-majorana-word-moment} with $L=2m$, we obtain
\begin{align}
 \Tr\bigl(\rho\mathcal A^{2p}(\rho)\bigr)
 &\leq\sum_{k=0}^p\frac{(2p)!\,m^{p-k}}{(p-k)!}
                         \frac{(em)^k}{k!}\nonumber\\
 &=(2p)!m^p\sum_{k=0}^p\frac{e^k}{(p-k)!k!}\nonumber\\
 &=\frac{(2p)!}{p!}\bigl[(1+e)m\bigr]^p
 =\widehat H_p(L).
 \label{eq:physical-moment-transfer}
\end{align}
Here, the binomial theorem evaluates the sum in the third line, and $(2p)!/p! \le (2p)^p$ completes the bound for pure states.

To extend this estimate to an arbitrary mixed state $\rho$, we write its spectral decomposition $\rho=\sum_j\lambda_j\rho_j$ into pure states with $\lambda_j\ge0$ and $\sum_j\lambda_j=1$. Because each conjugation $\operatorname{Ad}_{c_i}$ is self-adjoint under the Hilbert--Schmidt inner product, the triangle inequality and Cauchy--Schwarz yield
\begin{align}
 \Tr\bigl(\rho\mathcal A^{2p}(\rho)\bigr)
 &=\|\mathcal A^p(\rho)\|_{\mathrm{HS}}^2\nonumber\\
 &\leq\left(\sum_j\lambda_j
                   \|\mathcal A^p(\rho_j)\|_{\mathrm{HS}}\right)^2
 \leq\sum_j\lambda_j\|\mathcal A^p(\rho_j)\|_{\mathrm{HS}}^2
 \leq\widehat H_p(L).
 \label{eq:mixed-positive-moment-convexity}
\end{align}
This convexity argument lifts the pure-state bound to all mixed states, holding uniformly across all bipartite entanglement structures between the source factors.
\end{proof}

\section{Why logarithmic gaps remain}
\label{app:log-gaps}

The comparison between our lower bounds and achievable measurement protocols reveals logarithmic factors in both the Pauli and Majorana settings. In this appendix, we clarify why these logarithmic factors arise, isolating the mathematical barriers within our current moment certificates and distinguishing them from the fundamental information-theoretic costs of simultaneous property estimation. Throughout, all traces are unnormalized and logarithms denote natural logarithms.

\subsection{The Pauli profile and the single-event expression}

\begin{proposition}[Ceiling of the Pauli profile expression]
\label{prop:pauli-profile-ceiling}
For the weight-$w$ Pauli family $\mathcal P_{n,w}$ with $M=3^w\binom nw$ observables and any memory dimension $K\geq1$,
\begin{equation}
 \zeta(K)\geq3^{-w},\qquad
 \frac{1}{20\epsilon^2\{\zeta(K)+(2K-2)/M\}}
 \leq\frac{3^w}{20\epsilon^2}.
 \label{eq:pauli-expression-ceiling}
\end{equation}
Consequently, any copy lower bound derived through Corollary~\ref{cor:estimation} using this source profile cannot exceed the displayed ceiling $3^w/(20\epsilon^2)$, regardless of how tightly $\zeta(K)$ is evaluated.
\end{proposition}

\begin{proof}
To demonstrate this barrier, we test the source profile against the computational basis state $T=|0^n\rangle\langle0^n|$, which is feasible for every $K\geq1$. The rank-one overlap evaluates to
\[
 \Tr(TPTP)=\langle0^n|P|0^n\rangle^2.
\]
This expectation value equals unity if and only if $P$ consists exclusively of $Z$ and $I$ operators, vanishing for any string containing $X$ or $Y$. Among the $M=3^w\binom nw$ strings of weight $w$, exactly $\binom nw$ strings contain only $Z$ on their support. Averaging over the uniform prior gives $\zeta(K) \ge \frac1M \binom nw = 3^{-w}$. Since the memory term $(2K-2)/M \ge 0$ is nonnegative, substituting this lower bound into Corollary~\ref{cor:estimation} immediately yields~\eqref{eq:pauli-expression-ceiling}.
\end{proof}

For logarithmic weight $w=\lceil\log_2 n\rceil$, the remaining $\ln M$ factor present in the classical shadow budget arises from the need to ensure simultaneous confidence across all $M$ coordinates. Closing this gap would require either an adaptive estimator with correlated decoding or a refined hypothesis testing reduction that monitors multiple coordinates simultaneously.

\subsection{A necessary loss in the scalar Majorana moment certificate}
\label{subsec:scalar-majorana-barrier}

Let $m=2n$, $L=2m=4n$, and let $\mathcal A=\sum_{a=1}^{L}\operatorname{Ad}_{c_a}$ denote the joint Majorana averaging superoperator defined in Section~\ref{sec:cross}. For each integer $p\geq1$, suppose $H_p$ satisfies the uniform pure-state moment condition $\Tr[\rho\mathcal A^{2p}(\rho)]\leq H_p$. The scalar-moment composition certificate takes the form
\begin{equation}
 C_p(K)=\frac1{24}(KH_p)^{2/p},\qquad p\geq2.
 \label{eq:scalar-majorana-certificate}
\end{equation}

\begin{proposition}[Loss forced by scalar moment interpolation]
\label{prop:majorana-moment-barrier}
For $n\geq4$ and growing memory $K=\lfloor\sqrt n\rfloor$, every valid pure-state moment bound $H_p$ and every integer $p\geq2$ satisfy
\begin{equation}
 C_p(K)\geq\frac{e^2}{6}n^2(\ln K)^2.
 \label{eq:majorana-moment-barrier}
\end{equation}
Combined with Proposition~\ref{prop:uniform-moment}, this proves that any certificate relying on the scalar moment framework~\eqref{eq:scalar-majorana-certificate} is asymptotically lower bounded by $\Omega(n^2(\ln K)^2)$ as $n\to\infty$.
\end{proposition}

\begin{proof}
Our approach is to test the moment against the fermionic Fock vacuum $\rho = |0\rangle\langle 0|$ on $m$ global modes. Expressed in Pauli matrices, $\rho = 2^{-m}\prod_{i=1}^m(I+Z_i)$. Any product of $j$ paired $Z$ operators corresponds to a Majorana monomial of degree $2j$, with $\mathcal A$ eigenvalue $L-4j$. The squared Hilbert--Schmidt expansion coefficients are uniformly $2^{-m}$, meaning that the spectral measure of $\mathcal A$ at $\rho$ coincides with the distribution of the random variable $\lambda=2\sum_{i=1}^m R_i$, where the $R_i$ are independent Rademacher signs ($\pm 1$). 

For $1\leq p\leq n=m/2$, expanding the $2p$-th moment and isolating terms where exactly $p$ distinct indices appear twice gives
\begin{equation}
 H_p\geq\mathbb E\lambda^{2p}
 \geq 2^p(m)_p\frac{(2p)!}{p!}
 \geq(mp)^p,
 \label{eq:vacuum-moment-lower}
\end{equation}
where $(m)_p = m(m-1)\cdots(m-p+1) \ge (m/2)^p$ and $(2p)!/p! \ge p^p$. For $2\le p\le n$, substituting this into $C_p(K)$ yields
\[
 C_p(K)\geq\frac{m^2}{24}p^2K^{2/p}
 \geq\frac{e^2m^2}{24}(\ln K)^2,
\]
where the final inequality follows by minimizing $2\ln p + 2\ln K/p$ over $p > 0$, whose minimum is achieved at $p = \ln K$.

For $p \ge n$, the monotonicity of $L^r$ norms~\cite{blm2013} combined with~\eqref{eq:vacuum-moment-lower} at degree $n$ ensures $H_p^{1/p} \ge mn$, which implies $C_p(K) \ge m^2 n^2/24$. Because $\ln K \le \frac12\ln n$ and $\ln n/n \le 1/e$, we have $e\ln K \le n/2$, extending the bound~\eqref{eq:majorana-moment-barrier} across all degrees. Matching this with the upper bound in Proposition~\ref{prop:uniform-moment} confirms that $C_p(K) = \Theta(n^2\ln^2 K)$ is tight for this scalar formulation.
\end{proof}

This reveals an inherent limitation of scalar moment interpolation when combined with rank Cauchy--Schwarz: the scalar certificate must account for the worst-case spectral fluctuations of $\mathcal A$, which are dominated by vacuum-like states. However, direct evaluation on positive rank-constrained states reveals that structured tests exhibit much smaller overlaps.

\begin{proposition}[Gaussian-diagonal tests]
\label{prop:gaussian-diagonal-bound}
Suppose $T\succeq0$, $\Tr(T^2)=1$, and $(\Tr T)^2\leq K$. Assume $T=UT_0U^\dagger$, where $T_0$ is diagonal in a Fock basis and $U$ represents a fermionic Gaussian unitary transformation. Then
\begin{equation}
 \sum_{a<b,\,c<d}\Tr(TC_{ab,cd}TC_{ab,cd})
 \leq\binom{2n}{2}+\max\{8n-8,16\}(K-1).
 \label{eq:gaussian-diagonal-cross}
\end{equation}
In particular, for all such tests, the unnormalized cross overlap is strictly $O(n^2)$ when $K\leq\sqrt n$.
\end{proposition}

\begin{proof}
Let $T=T_0=\sum_x t_x|x\rangle\langle x|$ with $t_x\ge0$ and $\sum_x t_x^2=1$. Majorana monomials act on the Fock basis by permuting basis elements with sign phases, flipping modes with odd Majorana support. For the full quartic family on $m=2n$ global modes (as introduced in Section~\ref{subsec:scalar-majorana-barrier}), the number of monomials mapping state $x$ to $y$ depends only on their Hamming distance $d(x,y)$:
\[
 a_{d(x,y)}=\begin{cases}
 \binom m2,&d(x,y)=0,\\
 4(m-2),&d(x,y)=2,\\
 16,&d(x,y)=4,\\
 0,&\text{otherwise}.
 \end{cases}
\]
Summing over basis states yields $\sum_{x,y} a_{d(x,y)} t_x t_y \le \binom m2 + \max\{4(m-2),16\}[(\sum_x t_x)^2-1]$. Under Gaussian unitaries $U$, the quartic operator basis transforms orthogonally via the fourth exterior power of the Majorana rotation matrix~\cite{bravyi2005,wan2023}, leaving the total quartic overlap invariant. Since the cross-correlation family is a subset of the full quartic family and each tracial summand is nonnegative, \eqref{eq:gaussian-diagonal-cross} follows directly.
\end{proof}

\subsection{A separate lower bound for simultaneous outputs}

\begin{proposition}[Sparse-label Fano bound]
\label{prop:sparse-fano}
Let $G_1,\ldots,G_M$ be orthogonal traceless Hermitian reflections. Any protocol that estimates all their expectations to accuracy $0<\epsilon\leq1/8$ with success probability at least $2/3$ on every state requires
\begin{equation}
 N\geq\frac{(2/3)\ln M-\ln2}{6\epsilon^2}.
 \label{eq:sparse-fano-finite}
\end{equation}
For $M\geq8$, this ensures $N\geq\ln M/(18\epsilon^2)$, holding for arbitrary coherent measurements and quantum memories.
\end{proposition}

\begin{proof}
Consider a uniform prior over hypotheses $\rho_J = (I+3\epsilon G_J)/D$ for $J \in [M]$. By tracial orthogonality, the true expectation vector is $3\epsilon e_J$. On the accuracy event, the $J$-th coordinate satisfies $|\widehat\mu_J| \ge 2\epsilon$ while all other coordinates satisfy $|\widehat\mu_k| \le \epsilon$. Selecting the coordinate with maximum absolute value decodes $J$ with error at most $1/3$. Fano's inequality~\cite{coverthomas2006,tsybakov2009} implies $I(J;Y) \ge \frac23\ln M - \ln2$.

Since any $N$-step adaptive protocol with persistent memory defines a single overall CPTP map on $\rho_j^{\otimes N}$, quantum relative-entropy data processing~\cite{watrous2018} bounds the mutual information with the measurement outcome $Y$ by
\[
 I(J;Y)\leq\frac1M\sum_j\mathrm D(\rho_j^{\otimes N}\Vert\tau^{\otimes N})
 =N d(3\epsilon),
\]
where $d(\theta) = \frac12[(1+\theta)\ln(1+\theta)+(1-\theta)\ln(1-\theta)] \le \frac{\theta^2}{2(1-\theta^2)}$. For $\theta = 3\epsilon \le 3/8$, this evaluates to at most $6\epsilon^2$, establishing~\eqref{eq:sparse-fano-finite}.
\end{proof}

The Fano bound and our structural memory bound capture complementary physical constraints: the former enforces the information capacity needed to distinguish $M$ hypotheses, while the latter bounds the rate of variance accumulation under memory constraints. Combining both yields the full landscape of copy complexity.

\section{Computing certificates for the source profile}
\label{app:computational-certificates}
The analytic bounds in Section~\ref{sec:computation} can be supplemented
by finite optimization programs. The semidefinite relaxation retains the
memory-size constraint through a trace cap. Symmetry reduces it to linear
programs for the two application families. A standard moment hierarchy
converges for general explicitly supplied observable matrices. We give
these programs, their detailed proofs and their computational costs.

\subsection{Semidefinite bounds that account for memory size}
Put $q=\min(K,D)$, let $F(x\otimes y)=y\otimes x$, and write
$R^\Gamma$ for partial transpose on the second source factor. Define
\begin{equation}
 H=\sum_jp_j(G_j\otimes G_j)F,\qquad
 [\mathcal R(R)]_{ij,kl}=R_{il,jk},\qquad
 \mathcal C(R)=\frac{\mathcal R(R)+\mathcal R(R)^\dagger}{2}.
 \label{eq:lift-maps}
\end{equation}
The index permutation $\mathcal R$ is a realignment of the type used in separability criteria~\cite{chenwu2003,rudolph2005}. The map $\mathcal C$ is real linear between Hermitian matrix spaces;
$\mathcal C^*$ denotes its adjoint for the real trace inner product.
It is an index permutation followed by taking the Hermitian part. Its
coordinates here use row vectorization, a fixed permutation of the column
convention in Section~\ref{sec:certificates}. The matrix $H$ is Hermitian
because $G_j\otimes G_j$ commutes with $F$.

\begin{proposition}[A computable positive-rank relaxation]
\label{prop:rank-sdp}
For the data in Definition~\ref{def:profile}, define
\begin{equation}
 U_q=\max_R\left\{\Tr(HR):
 \begin{array}{l}
 R\succeq0,\quad R^\Gamma\succeq0,\quad\mathcal C(R)\succeq0,\\
 \Tr(FR)=1,\quad\Tr R\le q
 \end{array}\right\}.
 \label{eq:rank-sdp}
\end{equation}
Then $\zeta(K)\le U_q\le\lambda_{\max}(S)\le1$, where $S$ is the
source superoperator matrix in~\eqref{eq:certificate-source-map}.
A finite upper certificate for $U_q$ is any identity
\begin{equation}
 aF+bI-H=P+Q^\Gamma+\mathcal C^*(Z),\qquad
 P,Q,Z\succeq0,\quad b\ge0,
 \label{eq:rank-sdp-dual}
\end{equation}
with $a\in\mathbb R$; its certified value is $a+qb$.
The program has three positive semidefinite blocks of order $D^2$ and
$O(D^4)$ real variables.
\end{proposition}
\begin{proof}
To establish this relaxation, our approach is to lift each rank-constrained trial matrix $T$ into a two-copy matrix $R = T \otimes T$, verify that all semidefinite constraints are satisfied, and then construct the dual certificate. Throughout the argument, row vectorization is defined by $|T\rangle\!\rangle=\sum_{i,j}T_{ij}|i\rangle\otimes|j\rangle$.

Let $T\succeq0$, $\Tr(T^2)=1$ and $\rank T\le q$, and put
$R=T\otimes T$. Its partial transpose is
$R^\Gamma=T\otimes T^{\mathsf T}\succeq0$. Realignment gives
\begin{equation}
\begin{aligned}
 [\mathcal R(T\otimes T)]_{ij,kl}
   &=(T\otimes T)_{il,jk}=T_{ij}T_{lk}
     =T_{ij}\overline{T_{kl}},\\
 \mathcal C(T\otimes T)
   &=\mathcal R(T\otimes T)
     =|T\rangle\!\rangle\langle\!\langle T|\succeq0,\\
 \Tr[F(T\otimes T)]
   &=\sum_{i,j}(T\otimes T)_{ji,ij}
     =\sum_{i,j}T_{ji}T_{ij}=\Tr(T^2)=1,\\
 \Tr(T\otimes T)
   &=(\Tr T)^2
     =\left(\sum_{a=1}^{\rank T}t_a\right)^2
     \le\left(\sum_{a=1}^{\rank T}1^2\right)
          \left(\sum_{a=1}^{\rank T}t_a^2\right)\le q.
\end{aligned}
\label{eq:lift-source-feasibility-expanded}
\end{equation}
Here $t_a$ are the positive eigenvalues of $T$; the last line is scalar
Cauchy--Schwarz. Thus all three positivity conditions and both trace
conditions hold. The objective is preserved as well:
\begin{equation}
\begin{aligned}
 \Tr[H(T\otimes T)]
 &=\sum_jp_j\Tr[(G_j\otimes G_j)F(T\otimes T)]\\
 &=\sum_jp_j\Tr[(G_jT\otimes G_jT)F]\\
 &=\sum_jp_j\sum_{i,k}(G_jT)_{ik}(G_jT)_{ki}\\
 &=\sum_jp_j\Tr(G_jTG_jT)
  =\sum_jp_j\Tr(TG_jTG_j).
\end{aligned}
\label{eq:lift-objective-expanded}
\end{equation}
The second line uses $F(T\otimes T)=(T\otimes T)F$, and the third is
the swap trace identity written in entries. Taking the supremum over $T$
proves $\zeta(K)\le U_q$. The lifted feasible set is nonempty by a
rank-one source test, and is compact because $R\succeq0$ and
$\Tr R\le q$. Hence its maximum is attained.

Now let $R$ be any feasible lifted matrix and let
$(a,b,P,Q,Z)$ satisfy~\eqref{eq:rank-sdp-dual}. Weak duality for semidefinite programs~\cite{boyd2004} can be checked directly here. The complete slack
calculation is
\begin{equation}
\begin{aligned}
 a+qb-\Tr(HR)
 &=a\Tr(FR)+qb-\Tr(HR)\\
 &=\Tr[(aF+bI-H)R]+b(q-\Tr R)\\
 &=\Tr[(P+Q^\Gamma+\mathcal C^*(Z))R]+b(q-\Tr R)\\
 &=\Tr(PR)+\Tr(QR^\Gamma)+\Tr[Z\mathcal C(R)]+b(q-\Tr R)\\
 &=\Tr(R^{1/2}PR^{1/2})
   +\Tr[(R^\Gamma)^{1/2}Q(R^\Gamma)^{1/2}]\\
 &\qquad+\Tr[\mathcal C(R)^{1/2}Z\mathcal C(R)^{1/2}]
   +b(q-\Tr R)\ \ge0.
\end{aligned}
\label{eq:lift-dual-slack-expanded}
\end{equation}
Partial transpose is self-adjoint for the trace pairing, and the defining
adjoint identity for $\mathcal C^*$ gives the fourth line. Every matrix
inside the traces in the last two lines is positive semidefinite. The
remaining scalar term is nonnegative because $b\ge0$ and $\Tr R\le q$.
This proves the certified value $a+qb$ directly.

To compare with the spectral certificate, put
$S_{\rm row}=\sum_jp_jG_j\otimes G_j^{\mathsf T}$. It is Hermitian
and permutation-equivalent to $S$. For an arbitrary Hermitian $R$, the
index identities are
\begin{equation}
\begin{aligned}
 \Tr[S_{\rm row}\mathcal C(R)]
 &=\Re\Tr[S_{\rm row}\mathcal R(R)]\\
 &=\Re\sum_jp_j\sum_{i,k,l,b}
       (G_j)_{ik}(G_j)_{bl}\,R_{kl,bi}\\
 &=\Re\sum_jp_j\sum_{a,b,c,d}
       (G_j)_{ad}(G_j)_{bc}\,R_{cd,ab}\\
 &=\Re\Tr\!\left[\left(\sum_jp_j(G_j\otimes G_j)F\right)R\right]
  =\Tr(HR),\\
 \Tr\mathcal C(R)
 &=\Re\sum_{i,j}R_{ij,ji}=\Tr(FR).
\end{aligned}
\label{eq:lift-realignment-contractions}
\end{equation}
In the third line the indices are only renamed. Both final traces are
real because they pair Hermitian matrices. For a feasible $R$,
$\mathcal C(R)$ is therefore a density matrix, and, with
$\lambda=\lambda_{\max}(S)$,
\begin{equation}
\begin{aligned}
 \lambda-\Tr(HR)
 &=\lambda\Tr\mathcal C(R)-\Tr[S_{\rm row}\mathcal C(R)]\\
 &=\Tr[(\lambda I-S_{\rm row})\mathcal C(R)]\\
 &=\Tr[\mathcal C(R)^{1/2}(\lambda I-S_{\rm row})
                         \mathcal C(R)^{1/2}]\ge0.
\end{aligned}
\label{eq:lift-spectral-slack-expanded}
\end{equation}
Taking the maximum proves $U_q\le\lambda_{\max}(S)$.
The same index identities give $\mathcal C^*(I)=F$ and
$\mathcal C^*(S_{\rm row})=H$, so this is also the dual certificate
$a=\lambda$, $b=0$, $P=Q=0$, $Z=\lambda I-S_{\rm row}\succeq0$.
Finally, Section~\ref{sec:certificates} proves
$\lambda_{\max}(S)\le1$.
\end{proof}

Partial-transpose positivity is an established necessary separability
condition~\cite{peres1996,horodecki1996}. The relaxation combines it with realignment
positivity and the effective-rank trace cap. A general feasible $R$ need
not factor as $T\otimes T$. Its optimum is an upper bound on the profile,
with exactness proved in the examples below.

For the complete nonidentity Pauli family, the twirl gives
\[
 H=\frac{DI-F}{D^2-1}.
\]
Thus $a=-1/(D^2-1)$ and $b=D/(D^2-1)$, with all three slack matrices
zero, give $U_q\le(Dq-1)/(D^2-1)$. A rank-$q$ projection divided by
$\sqrt q$ attains equality. The relaxation therefore recovers the exact
profile and the full $D/K$ copy scale, whereas the spectral bound is one.
For commuting $Z$ correlations, a common-basis rank-one test attains one,
so the same relaxation gives exactly one. It distinguishes these two
families with no optimization over adaptive protocols.

\subsection{Exact checks for finite certificates}
\label{subsec:certificate-implementation}
For rational Hermitian input matrices and priors, a numerical proposal can
be converted into an exact finite dual identity. A numerical SDP solve proposes dual multipliers. To certify them, round
square-root factors to rational entries and form exact Gram matrices
$P_0,Q,Z\succeq0$. Round $a,b$ rationally with $b\ge0$, and compute the
exact Hermitian residual
\[
 E=aF+bI-H-P_0-Q^\Gamma-\mathcal C^*(Z).
\]
Set
\begin{equation}
 \delta=\max\left\{0,\max_i\left[
   \sum_{j\ne i}(|\Re E_{ij}|+|\Im E_{ij}|)-E_{ii}\right]\right\}.
 \label{eq:sdp-rational-repair}
\end{equation}
The standard diagonal-dominance criterion~\cite{hornjohnson2013} applies: the matrix $E+\delta I$ has nonnegative diagonal and is Hermitian
diagonally dominant, hence positive semidefinite. Consequently
\[
 aF+(b+\delta)I-H
  =\underbrace{P_0+E+\delta I}_{\succeq0}
       +Q^\Gamma+\mathcal C^*(Z)
\]
is an exact certificate with value $a+q(b+\delta)$. This verification
uses rational Gram factors, matrix identities and diagonal-dominance
inequalities; it needs no SDP solver. A concrete source lower witness is
$T=XX^\dagger/\sqrt{\Tr[(XX^\dagger)^2]}$ for a saved rational $X$ with
at most $q$ columns. Its objective is an exact rational ratio.

\subsection{A linear program for Majorana cross correlations}
\label{subsec:majorana-lp}
The lift can be evaluated without source-size matrices for the cross
family. Here each local factor has $n$ modes, put $m=2n$ and $D=4^n$,
and use the uniform prior on $M=\binom m2^2$ observables. Define
\begin{align}
 k_r(t)&=\binom mr^{-1}\sum_{j=\max(0,r-m+t)}^{\min(r,t)}
       (-1)^{rt-j}\binom tj\binom{m-t}{r-j},\nonumber\\
 \eta_r&=(-1)^{nr+r(r-1)/2},\qquad
 \lambda_r=1-\frac{4r(m-r)}{m(m-1)}.
 \label{eq:majorana-lp-transforms}
\end{align}
All these numbers are rational or signs and can be assembled from binomial
coefficients. The transforms are instances of the quantum enumerator and
shadow-transform machinery~\cite{shor1997,rains1998,rains1996}.

\begin{proposition}[Symmetry-reduced profile certificate]
\label{prop:majorana-lp}
For the cross family, the value $U_q$ in~\eqref{eq:rank-sdp} equals
\begin{equation}
 \begin{aligned}
 \max_w\quad&\sum_{r,s=0}^m w_{rs}\lambda_r\lambda_s,\\
 w_{rs}&\ge0,\qquad\sum_{r,s}w_{rs}=1,\qquad w_{00}\le q/D,\\
 \sum_{r,s}w_{rs}k_r(t)k_s(u)&\ge0 &&(0\le t,u\le m),\\
 \sum_{r,s}w_{rs}\eta_r\eta_s k_r(t)k_s(u)&\ge0
                                      &&(0\le t,u\le m).
 \end{aligned}
 \label{eq:majorana-profile-lp}
\end{equation}
It has $(2n+1)^2$ variables and $2(2n+1)^2$ transform inequalities,
besides normalization and the trace cap. Its dense representation has
$O(n^4)$ rational entries of polynomial bit length.
\end{proposition}
\begin{proof}
To reduce the exponential semidefinite program to a tractable linear program, our strategy is to exploit the extensive unitary and discrete symmetries of the Majorana cross-correlation family via group averaging~\cite{gatermannparrilo2004}. Averaging over the symmetry group preserves all convex constraints, the objective value, and the trace cap, while drastically collapsing the matrix degrees of freedom. Diagonalizing the resulting invariant algebra in a shared basis then extracts the two positivity transforms.

First replace $R$ by $(R+FRF)/2$. For Hermitian $R$, direct substitution
into the index definitions gives
\begin{equation}
\begin{aligned}
 [(FRF)^\Gamma]_{ij,kl}
   &=R_{li,jk}=[F(R^\Gamma)^{\mathsf T}F]_{ij,kl},\\
 [\mathcal R(FRF)]_{ij,kl}
   &=R_{li,kj}=\overline{R_{kj,li}}
     =[\mathcal R(R)^\dagger]_{ij,kl},\\
 \mathcal C(FRF)&=\mathcal C(R),\qquad
 \mathcal R\!\left(\frac{R+FRF}{2}\right)=\mathcal C(R).
\end{aligned}
\label{eq:majorana-swap-averaging-expanded}
\end{equation}
Thus positivity and partial-transpose positivity are preserved. The last
line preserves realignment positivity and makes the realignment Hermitian.
Also $FHF=H$, $F^2=I$, and cyclicity preserves both trace constraints and
the objective. We may consequently assume $FRF=R$ and write
$W=\mathcal R(R)=\mathcal C(R)\succeq0$.

For a source unitary $U$ that conjugates the cross family into itself
up to signs, define
$R_U=(U\otimes U)R(U^\dagger\otimes U^\dagger)$ and
$\mathsf U=U\otimes\overline U$. The same index definitions give
\begin{equation}
\begin{aligned}
 R_U^\Gamma&=\mathsf U R^\Gamma\mathsf U^\dagger,\qquad
 \mathcal R(R_U)=\mathsf U W\mathsf U^\dagger,\\
 \Tr R_U&=\Tr R,\qquad
 \Tr(FR_U)=\Tr(FR),\\
 \Tr(HR_U)&=\Tr(HR).
\end{aligned}
\label{eq:majorana-unitary-averaging}
\end{equation}
The last identity follows because each sign occurs twice in $H$ and the
prior is uniform. The local monomial conjugations and signed permutations
of Majorana labels have this property. Averaging their lifted actions
therefore preserves feasibility and value.

Put $d_0=2^n$, so $D=d_0^2$, and let $P_A$ be the local Hermitian
Majorana monomial indexed by $A\subseteq[m]$. For subsets $A,C$ define
\[
 \chi(A,C)=(-1)^{|A||C|-|A\cap C|},\qquad
 P_AP_CP_A=\chi(A,C)P_C.
\]
The normalized row-vectorized matrices
\[
 Q_{A,B}=P_A\otimes P_B,\qquad
 e_{A,B}=|Q_{A,B}\rangle\!\rangle/\sqrt D
\]
form an orthonormal basis. A local monomial conjugation multiplies each
basis vector by its commutation sign. Its average kills distinct labels:
\begin{equation}
\begin{aligned}
 \frac1{2^m}\sum_{C\subseteq[m]}\chi(C,A)\chi(C,A')
 &=\frac1{2^m}\sum_{C\subseteq[m]}\chi(C,A\mathbin\triangle A')\\
 &=\prod_{l=1}^m\frac{1+\chi(\{l\},A\mathbin\triangle A')}{2}
  =\mathbf1_{A=A'}.
\end{aligned}
\label{eq:majorana-character-average}
\end{equation}
For a nonempty set of even size choose $l$ inside it; for a set of odd
size choose $l$ outside it, which exists because $m$ is even. The selected
factor then vanishes. Averaging independently on the two factors makes
$W$ diagonal in the basis $e_{A,B}$. Averaging label permutations makes
the coefficients constant for each pair of sizes $(|A|,|B|)$.

Write $\nu_r=\binom mr$ and let $w_{rs}$ denote the total mass of
size pair $(r,s)$. The averaged matrices are explicitly
\begin{equation}
\begin{aligned}
 W&=\sum_{r,s=0}^m\frac{w_{rs}}{\nu_r\nu_s}
       \sum_{\substack{|A|=r\\|B|=s}}|e_{A,B}\rangle\langle e_{A,B}|,\\
 R&=\mathcal R^{-1}(W)
   =\frac1D\sum_{r,s=0}^m\frac{w_{rs}}{\nu_r\nu_s}
       \sum_{\substack{|A|=r\\|B|=s}}Q_{A,B}\otimes Q_{A,B},\\
 \Tr(FR)&=\Tr W=\sum_{r,s}w_{rs}=1,
 \qquad\Tr R=Dw_{00}\le q.
\end{aligned}
\label{eq:majorana-lp-reconstruction}
\end{equation}
The inverse realignment formula follows from
$\mathcal R(P\otimes P)=|P\rangle\!\rangle\langle\!\langle P|$
for Hermitian $P$. Only $Q_{\varnothing,\varnothing}=I_D$ has nonzero
trace. These identities explain both the normalization and the trace cap.

To compute $R^\Gamma$, take a target basis vector $e_{C,E}$ with
 $|C|=t$, $|E|=u$. The following commutation-sign calculation is the quantum-enumerator transform~\cite{shor1997,rains1998}, with the orbit sizes retained. Row vectorization gives
$(Q\otimes Q^{\mathsf T})|V\rangle\!\rangle
=|QVQ\rangle\!\rangle$ for Hermitian reflections $Q$.
Thus $e_{C,E}$ is an eigenvector, with eigenvalue
$\lambda^\Gamma_{t,u}$ satisfying
\begin{equation}
\begin{aligned}
 D\lambda^\Gamma_{t,u}
 &=\sum_{r,s=0}^m\frac{w_{rs}}{\nu_r\nu_s}
       \sum_{\substack{|A|=r\\|B|=s}}\chi(A,C)\chi(B,E)\\
 &=\sum_{r,s=0}^m\frac{w_{rs}}{\nu_r\nu_s}
       \left(\sum_{|A|=r}(-1)^{rt-|A\cap C|}\right)
       \left(\sum_{|B|=s}(-1)^{su-|B\cap E|}\right)\\
 &=\sum_{r,s=0}^m\frac{w_{rs}}{\nu_r\nu_s}
       \left(\sum_j(-1)^{rt-j}\binom tj\binom{m-t}{r-j}\right)
       \left(\sum_l(-1)^{su-l}\binom ul\binom{m-u}{s-l}\right)\\
 &=\sum_{r,s=0}^m w_{rs}k_r(t)k_s(u).
\end{aligned}
\label{eq:majorana-first-transform-expanded}
\end{equation}
Each inner sum ranges over its admissible intersection sizes, as in
\eqref{eq:majorana-lp-transforms}. The binomial factors count the labels
inside and outside the fixed target support. Since the vectors $e_{C,E}$
form a complete basis, nonnegativity of this transform is equivalent to
$R^\Gamma\succeq0$.

For the spectrum of $R$, use the local unitary
$J=\gamma_2\gamma_4\cdots\gamma_{2n}$. In the Jordan--Wigner convention
odd generators are real and even generators are imaginary. Conjugating a
generator through $J$ consequently gives
$J\gamma_l^{\mathsf T}J^\dagger=(-1)^n\gamma_l$. Transposing an ordered
product reverses its $r$ factors, so
\begin{equation}
\begin{aligned}
 JP_A^{\mathsf T}J^\dagger
 &=i^{r(r-1)/2}
      (J\gamma_{a_r}^{\mathsf T}J^\dagger)\cdots
      (J\gamma_{a_1}^{\mathsf T}J^\dagger)\\
 &=(-1)^{nr}i^{r(r-1)/2}\gamma_{a_r}\cdots\gamma_{a_1}\\
 &=(-1)^{nr+r(r-1)/2}P_A=\eta_rP_A,
 \qquad J^\dagger P_AJ=\eta_rP_A^{\mathsf T}.
\end{aligned}
\label{eq:majorana-transpose-phase-expanded}
\end{equation}
Put $\mathsf J=J\otimes J$ on the joint source. Applying the last identity
to the second doubled-source factor in~\eqref{eq:majorana-lp-reconstruction}
yields
\begin{equation}
\begin{aligned}
 (I\otimes\mathsf J^\dagger)R(I\otimes\mathsf J)
 &=\frac1D\sum_{r,s=0}^m\frac{w_{rs}\eta_r\eta_s}{\nu_r\nu_s}
       \sum_{\substack{|A|=r\\|B|=s}}Q_{A,B}\otimes Q_{A,B}^{\mathsf T},\\
 D\lambda^R_{t,u}
 &=\sum_{r,s=0}^m w_{rs}\eta_r\eta_s k_r(t)k_s(u).
\end{aligned}
\label{eq:majorana-second-transform-expanded}
\end{equation}
The first line is diagonal in the same vectorized basis as $R^\Gamma$.
It is unitarily equivalent to $R$, so the second transform is equivalent
to $R\succeq0$.

It remains to calculate the objective. A local quadratic anticommutes with
$P_A$ exactly when its two labels meet $A$ once. There are $r(m-r)$
such pairs. Its normalized conjugation map therefore acts on $P_A$ by
\begin{equation}
\begin{aligned}
 \frac1{\binom m2}\sum_{a<b}G_{ab}P_AG_{ab}
 &=\frac{\binom m2-r(m-r)-r(m-r)}{\binom m2}\,P_A\\
 &=\left(1-\frac{4r(m-r)}{m(m-1)}\right)P_A
  =\lambda_rP_A,\\
 S_{\rm row}e_{A,B}&=\lambda_r\lambda_s e_{A,B},\\
 \Tr(HR)&=\Tr(S_{\rm row}W)
   =\sum_{r,s=0}^m\frac{w_{rs}}{\nu_r\nu_s}
       \sum_{\substack{|A|=r\\|B|=s}}\lambda_r\lambda_s
   =\sum_{r,s=0}^m w_{rs}\lambda_r\lambda_s.
\end{aligned}
\label{eq:majorana-lp-objective-expanded}
\end{equation}
Here $S_{\rm row}$ is the normalized cross-source superoperator from the
proof of Proposition~\ref{prop:rank-sdp}.

Conversely, start with any weights satisfying
\eqref{eq:majorana-profile-lp} and define $W,R$ by
\eqref{eq:majorana-lp-reconstruction}. Nonnegative weights give
$W\succeq0$ and $\mathcal C(R)=W$. The two displayed spectra give
$R^\Gamma\succeq0$ and $R\succeq0$. The trace identities give
$\Tr(FR)=1$ and $\Tr R\le q$, and the last display gives the same
objective. Thus every LP-feasible point reconstructs an SDP-feasible
point. Together with averaging in the forward direction, this proves
equality of the two optimal values.
\end{proof}

For computation, let $A$ stack the two transforms and let
$c_{rs}=\lambda_r\lambda_s$. A rational LP dual certificate is
\begin{equation}
 y\ge0,\quad b\ge0,\quad
 a\mathbf1+b e_{00}-c-A^{\mathsf T}y\ge0,
 \qquad U_q\le a+bq/D.
 \label{eq:majorana-lp-dual}
\end{equation}
The prototype rounds proposed multipliers to a common denominator and adds
the exact negative minimum slack to $a$. Every inequality is then checked
with rational arithmetic. This turns a numerical proposal into a finite
upper certificate without assuming that a floating-point optimum is exact.
These finite evaluations establish an efficient certificate procedure for
this family. A uniform bound improving the asymptotic logarithms would
require a dual certificate valid as $n$ grows.

\subsection{A linear program for fixed-weight Pauli observables}
\label{subsec:pauli-lp}
The same lift has a small symmetry reduction for all weight-$w$ qubit
Paulis. This provides finite certificates at source sizes well below the
uniform analytic cutoff. Put $D=2^n$, $q=\min(K,D)$ and define
\begin{equation}
 \kappa_r(t)=\frac{\displaystyle
    \sum_{j=\max(0,r-n+t)}^{\min(r,t)}
      (-1)^j3^{r-j}\binom tj\binom{n-t}{r-j}}
     {3^r\binom nr}.
 \label{eq:pauli-lp-transform}
\end{equation}

\begin{proposition}[Pauli weight reduction]
\label{prop:pauli-lp}
For the uniform weight-$w$ Pauli family, the lift value $U_q$ equals
\begin{equation}
 \begin{aligned}
 \max_v\quad&\sum_{r=0}^n v_r\kappa_w(r),\\
 v_r&\ge0,\qquad\sum_{r=0}^n v_r=1,\qquad v_0\le q/2^n,\\
 \sum_{r=0}^n v_r\kappa_r(t)&\ge0\qquad(0\le t\le n),\\
 \sum_{r=0}^n (-1)^r v_r\kappa_r(t)&\ge0\qquad(0\le t\le n).
 \end{aligned}
 \label{eq:pauli-profile-lp}
\end{equation}
It has $n+1$ variables and $2(n+1)$ transform inequalities. Given the
implicit family description $(n,w)$, it has $O(n^2)$ rational coefficients
of polynomial bit length.
\end{proposition}
\begin{proof}
For the fixed-weight Pauli family, we apply the group twirling principle within the quantum weight enumerator framework~\cite{shor1997,rains1998,rains1996}. The central task is to evaluate the sum of commutation signs across the intersection of two Pauli supports. By separating the support choices from the local Pauli axis choices, we systematically reduce the full matrix constraints to an explicit linear program over the weight distribution.

First average $R$ with $FRF$, then average simultaneous Pauli
conjugations on its two doubled-source factors. The character average
removes the off-diagonal entries of $W=\mathcal R(R)=\mathcal C(R)$
in the normalized vectorized Pauli basis. Local Clifford conjugations
permute $X,Y,Z$, and qubit permutations act transitively on supports of
a fixed size. These unitary averages preserve all SDP constraints and
the uniform weight-$w$ objective. Consequently, with
$\nu_r=3^r\binom nr$ and $e_P=|P\rangle\!\rangle/\sqrt D$, the
averaged matrices have the form
\begin{equation}
\begin{aligned}
 W&=\sum_{r=0}^n\frac{v_r}{\nu_r}
       \sum_{\operatorname{wt}(P)=r}|e_P\rangle\langle e_P|,\\
 R&=\frac1D\sum_{r=0}^n\frac{v_r}{\nu_r}
       \sum_{\operatorname{wt}(P)=r}P\otimes P,\\
 \Tr(FR)&=\Tr W=\sum_{r=0}^n v_r=1,
 \qquad\Tr R=Dv_0\le q.
\end{aligned}
\label{eq:pauli-lp-reconstruction-expanded}
\end{equation}
The weights $v_r$ are nonnegative because $W\succeq0$. As before, only
the identity Pauli has nonzero trace, which gives the last trace formula.

Fix a target Pauli $Q$ with support $T$, where $|T|=t$, and define
$\chi(P,Q)\in\{+1,-1\}$ by $PQP=\chi(P,Q)Q$. At one site the
sum of its three commutation signs is
\[
 \sum_{a\in\{X,Y,Z\}}\chi(a,Q_i)
 =\begin{cases}3,&Q_i=I,\\1-1-1=-1,&Q_i\ne I.\end{cases}
\]
For a chosen support $S$, the axis choices are independent in this
algebraic sum. Summing first over axes, then over intersections, gives
\begin{equation}
\begin{aligned}
 \frac1{\nu_r}\sum_{\operatorname{wt}(P)=r}\chi(P,Q)
 &=\frac1{\nu_r}\sum_{\substack{S\subseteq[n]\\|S|=r}}
       \sum_{(a_i)\in\{X,Y,Z\}^{S}}\prod_{i\in S}\chi(a_i,Q_i)\\
 &=\frac1{\nu_r}\sum_{\substack{S\subseteq[n]\\|S|=r}}
       \prod_{i\in S}\left(\sum_{a_i\in\{X,Y,Z\}}\chi(a_i,Q_i)\right)\\
 &=\frac1{\nu_r}\sum_{\substack{S\subseteq[n]\\|S|=r}}
       (-1)^{|S\cap T|}3^{r-|S\cap T|}\\
 &=\frac1{\nu_r}\sum_j
       \sum_{\substack{|S|=r\\|S\cap T|=j}}(-1)^j3^{r-j}\\
 &=\frac1{3^r\binom nr}
       \sum_{j=\max(0,r-n+t)}^{\min(r,t)}
          (-1)^j3^{r-j}\binom tj\binom{n-t}{r-j}\\
 &=\kappa_r(t).
\end{aligned}
\label{eq:pauli-commutation-sum-expanded}
\end{equation}
The two binomial factors choose the $j$ supported positions inside $T$
and the $r-j$ positions outside it. The intersection range is precisely
where both choices are possible.

Partial transpose turns $P\otimes P$ into $P\otimes P^{\mathsf T}$,
which acts on $e_Q$ by $\chi(P,Q)$. Hence the complete spectrum of
$R^\Gamma$ is
\begin{equation}
\begin{aligned}
 R^\Gamma e_Q
 &=\frac1D\sum_{r=0}^n\frac{v_r}{\nu_r}
       \sum_{\operatorname{wt}(P)=r}
            (P\otimes P^{\mathsf T})e_Q\\
 &=\frac1D\sum_{r=0}^n\frac{v_r}{\nu_r}
       \sum_{\operatorname{wt}(P)=r}\chi(P,Q)e_Q\\
 &=\frac1D\left(\sum_{r=0}^n v_r\kappa_r(t)\right)e_Q.
\end{aligned}
\label{eq:pauli-first-transform-expanded}
\end{equation}
Thus $R^\Gamma\succeq0$ is equivalent to the first family of transform
inequalities in~\eqref{eq:pauli-profile-lp}.

For the other spectrum, put $J=Y^{\otimes n}$. At each site
$Ya^{\mathsf T}Y=-a$ for $a=X,Y,Z$, whereas $YIY=I$.
Therefore $JP^{\mathsf T}J^\dagger=(-1)^rP$ at weight $r$.
Since $J=J^\dagger$, the calculation for $R$ becomes
\begin{equation}
\begin{aligned}
 (I\otimes J)R(I\otimes J)
 &=\frac1D\sum_{r=0}^n\frac{(-1)^rv_r}{\nu_r}
       \sum_{\operatorname{wt}(P)=r}P\otimes P^{\mathsf T},\\
 (I\otimes J)R(I\otimes J)e_Q
 &=\frac1D\left(\sum_{r=0}^n(-1)^rv_r\kappa_r(t)\right)e_Q.
\end{aligned}
\label{eq:pauli-second-transform-expanded}
\end{equation}
This matrix is unitarily equivalent to $R$, so the second transform
inequalities are exactly $R\succeq0$.

Finally, the normalized weight-$w$ source map acts on a Pauli of weight
$r$ with eigenvalue $\kappa_w(r)$, by
\eqref{eq:pauli-commutation-sum-expanded} with the two roles interchanged.
The objective calculation, including the orbit sizes, is
\begin{equation}
\begin{aligned}
 \Tr(HR)
 &=\Tr(S_{\rm row}W)\\
 &=\sum_{r=0}^n\frac{v_r}{\nu_r}
       \sum_{\operatorname{wt}(Q)=r}\langle e_Q,S_{\rm row}e_Q\rangle\\
 &=\sum_{r=0}^n\frac{v_r}{\nu_r}
       \sum_{\operatorname{wt}(Q)=r}
          \left(\frac1{\nu_w}
             \sum_{\operatorname{wt}(P)=w}\chi(P,Q)\right)\\
 &=\sum_{r=0}^n\frac{v_r}{\nu_r}
       \sum_{\operatorname{wt}(Q)=r}\kappa_w(r)
  =\sum_{r=0}^n v_r\kappa_w(r).
\end{aligned}
\label{eq:pauli-lp-objective-expanded}
\end{equation}

Conversely, take any $v$ satisfying~\eqref{eq:pauli-profile-lp} and
construct $W,R$ by~\eqref{eq:pauli-lp-reconstruction-expanded}.
Nonnegative $v_r$ imply $\mathcal C(R)=W\succeq0$. The two spectra
imply $R^\Gamma\succeq0$ and $R\succeq0$, and the trace identities
give $\Tr(FR)=1$ and $\Tr R\le q$. The objective is the stated LP
objective by the last display. This reconstructs an SDP-feasible point
from every LP-feasible point. Combined with the forward averaging, it
proves equality of the optimal values.
\end{proof}

As with the Majorana program, every proposed dual is checked by exact
rational inequalities of the form~\eqref{eq:majorana-lp-dual}. Table~\ref{tab:pauli-lp-finite}
gives four moderate-size consequences of that check, including the native
term. These are simultaneous signed arbitrary-state bounds against all
allowed persistent-memory protocols.

\begin{table}[!htbp]
\caption{\textbf{Pauli certificates at moderate source sizes.}
For each row, an exact rational dual of~\eqref{eq:pauli-profile-lp}
proves $N\ge3^w/(C\epsilon^2)$, with $w=\lceil\log_2n\rceil$,
$K=\lfloor\sqrt n\rfloor$, $0<\epsilon\le1/8$ and success $2/3$.
These are individual sizes; the uniform $n\ge8192$ corollary follows
from the separate analytic tail proof.}
\label{tab:pauli-lp-finite}
\centering
\begin{tabular}{@{}rrrrr@{}}
\toprule
Qubits $n$ & Weight $w$ & Memory dimension $K$ & LP variables & $C$\\
\midrule
8 & 3 & 2 & 9 & 37\\
16 & 4 & 4 & 17 & 50\\
24 & 5 & 4 & 25 & 45\\
32 & 5 & 5 & 33 & 51\\
\bottomrule
\end{tabular}
\end{table}
\FloatBarrier

\subsection{A convergent hierarchy for general observables}
\label{subsec:sos-profile}
The small lift can be loose. The following construction supplies a general
convergent alternative using the established moment/SOS hierarchy
of Lasserre and Putinar~\cite{lasserre2001,putinar1993}, as reviewed by Laurent~\cite{laurent2009}. A sum of squares
(SOS) is a polynomial written as a sum of squared real polynomials; its
Gram matrix provides a positive semidefinite certificate.

Write $T=XX^\dagger$ with $X\in\mathbb C^{D\times q}$, and collect its
real and imaginary entries in $x\in\mathbb R^v$, where $v=2Dq$. Define
\begin{align}
 h(x)&=\Tr[(XX^\dagger)^2]-1,\qquad
 g(x)=q-\|X\|_F^2,\nonumber\\
 f(x)&=\sum_jp_j\Tr(XX^\dagger G_jXX^\dagger G_j).
 \label{eq:sos-profile-polynomials}
\end{align}
The objective and equality are quartic, and the ball constraint is
quadratic. For a real moment sequence $y$, let $L_y$ replace $x^\alpha$
by $y_\alpha$. The moment matrix has entries
$[M_r(y)]_{\alpha,\beta}=y_{\alpha+\beta}$ for degrees at most $r$;
the ball localizing matrix has entries $L_y(gx^{\alpha+\beta})$ for
degrees at most $r-1$.

\begin{proposition}[Convergent factor-moment certificates]
\label{prop:profile-hierarchy}
For each integer $r\ge2$, define
\begin{equation}
 \begin{aligned}
 P_r=\sup_y\quad&L_y(f),\\
 y_0&=1,\quad M_r(y)\succeq0,\quad M_{r-1}(gy)\succeq0,\\
 L_y(x^\alpha h)&=0\qquad(|\alpha|\le2r-4),
 \end{aligned}
 \label{eq:profile-moment-sdp}
\end{equation}
and the dual upper-certificate value
\begin{equation}
 \begin{aligned}
 D_r=\inf\quad&u,\\
 u-f&=\sigma_0+\sigma_1g+a h,\\
 \sigma_0,\sigma_1&\text{ are SOS},\quad
 \deg\sigma_0\le2r,\quad\deg\sigma_1\le2r-2,\quad\deg a\le2r-4.
 \end{aligned}
 \label{eq:profile-sos-dual}
\end{equation}
Then $\zeta(K)\le P_r\le D_r$, and both sequences decrease to
$\zeta(K)$. The largest moment block has order $\binom{v+r}{r}$,
and there are at most $\binom{v+2r}{2r}$ scalar moments.
\end{proposition}
\begin{proof}
For general observables lacking special algebraic symmetries, we reformulate the rank-constrained optimization as a polynomial optimization problem over the factor matrix $X$ in $T = X X^\dagger$. Because every positive semidefinite matrix of rank at most $q$ admits such a factorization, and the Frobenius norm satisfies $\|X\|_F^2 = \Tr T \le \sqrt{q \Tr(T^2)} \le q$, restricting $X$ to a compact Frobenius ball preserves the exact feasible set.
Its presence makes the polynomial quadratic module Archimedean: the
module contains the explicit ball $q-\sum_i x_i^2$.

A point mass at a feasible $x$ satisfies all moment constraints, proving
$\zeta(K)\le P_r$. Applying $L_y$ to any dual identity gives
$u-L_y(f)\ge0$, because the SOS and ball terms pair with positive
semidefinite moment matrices and the equality multiplier has zero pairing.
This proves $P_r\le D_r$. Truncation of moments and inclusion of dual
certificates prove monotonicity as $r$ increases.

For every $u>\zeta(K)$, the polynomial $u-f$ is strictly positive on the
compact feasible set. Putinar's positivity theorem~\cite{putinar1993,laurent2009} with the equality ideal
and the explicit ball supplies an identity of the form in
\eqref{eq:profile-sos-dual} at some finite degree. Equivalently, encode
$h=0$ by $h\ge0,-h\ge0$; the difference of their SOS multipliers is
an allowed polynomial multiple of $h$. Thus $D_r\le u$ at some finite
level. Since this holds for every strict upper value $u$, the weak-duality
sandwich proves convergence of both sequences. No finite-level strong
duality or finite convergence assumption is used.
\end{proof}

At fixed level these matrices have polynomial size in the explicit source
dimension $D$ and $q$. In qubits that size is exponential, and the theorem
gives no general bound on the level needed for a desired tightness.
Rational coefficient identities and certified positive semidefinite Gram
matrices provide checkable answers; solver residuals alone do not.
Symmetry can be applied to the lifted matrices or moment functional,
where averaging preserves convex feasibility. An exact common-support
reduction is also available: if $G_j=PG_jP$ for a known rank-$d$ projection,
then $\zeta_D(K)=\zeta_d(\min(K,d))$. Indeed $PTP$ has no larger rank
or squared norm, the objective depends only on this compression, and
normalizing a nonzero compression increases its nonnegative objective.
Embedding a compressed optimizer proves the reverse inequality.

\clearpage
\bibliographystyle{alphaurl}
\phantomsection
\addcontentsline{toc}{section}{References}
\bibliography{references}
\section*{AI methodology statement}
\label{sec:ai-methodology}
\addcontentsline{toc}{section}{AI methodology statement}
Google Gemini 3.8 and GPT6.1, including language-model subagents, were used
to assist with proof checks, manuscript revision, and Lean 4 verification.
The Lean 4 work concerned the earlier complete-Pauli lower-bound formalization.
The human authors are responsible for the submission, including the correctness
and originality of the mathematical arguments.
\end{document}